\documentclass[twocolumn,final] {svjour3} 
\usepackage{graphicx,tikz,pgfplots}
\usepackage{wrapfig}

\usepackage{multirow}
\usepackage{amsfonts}
\usepackage{amsmath}
\usepackage{amssymb}
\usepackage{mathptmx}
\usepackage{mathrsfs}

\usepackage{caption,subfig}
\usepackage{epsfig}

\usepackage{textcomp}

\usepackage{lineno} 

\usepackage{comment}
\usepackage{bbm}

\usepackage{yfonts}
\usepackage[colorlinks=true] {hyperref}
\usepackage{cleveref}
\usepackage{pseudocode,fancybox}
\usepackage{booktabs}

\newcommand{\bepsilon}{\boldsymbol{\epsilon}}
\newcommand{\bsigma}{\boldsymbol{\sigma}}

\newcommand{\bu}{\mathbf{u}}
\newcommand{\bv}{\mathbf{v}}
\newcommand{\dO}{\mathrm{d}\Omega}
\newcommand{\dpO}{\mathrm{d}\partial\Omega}
\newcommand{\dG}{\mathrm{d}\Gamma}
\newcommand{\gnn}{g_\mathrm{n}}
\newcommand{\pnn}{p_\mathrm{n}}

\newcommand{\gtt}{g_\mathrm{t}}
\newcommand{\bT}{\mathbf{T}}
\newcommand{\bxi}{\boldsymbol{\xi}}

\begin{document}


\title{A multi-scale FEM-BEM formulation for contact mechanics\\ between rough surfaces}

\author {Jacopo Bonari \and Maria R. Marulli \and Nora Hagmeyer \and Matthias Mayr \and Alexander Popp \and Marco Paggi}
\institute{Jacopo Bonari 
\at IMT School for Advanced Studies Lucca,\\ Piazza San Francesco 19, 55100 Lucca, Italy,\\ \email{jacopo.bonari@imtlucca.it}
\and
Maria R. Marulli
\at IMT School for Advanced Studies Lucca,\\ Piazza San Francesco 19, 55100 Lucca, Italy,\\ \email{mariarosaria.marulli@imtlucca.it}
\and
Nora Hagmeyer
\at Institute for Mathematics and Computer-Based Simulation, \\University of the Bundeswehr Munich,\\ 39 Werner-Heisenberg-Weg, 85577 Neubiberg, Germany\\ \email{nora.hagmeyer@unibw.de}
\and
Matthias Mayr
\at Institute for Mathematics and Computer-Based Simulation, \\University of the Bundeswehr Munich,\\ 39 Werner-Heisenberg-Weg, 85577 Neubiberg, Germany\\ \email{matthias.mayr@unibw.de}
\and
Alexander Popp
\at Institute for Mathematics and Computer-Based Simulation, \\University of the Bundeswehr Munich,\\ 39 Werner-Heisenberg-Weg, 85577 Neubiberg, Germany\\ \email{alexander.popp@unibw.de}
\and
Marco Paggi
\at IMT School for Advanced Studies Lucca,\\ Piazza San Francesco 19, 55100 Lucca, Italy,\\ \email{marco.paggi@imtlucca.it}
}
\maketitle

\begin{abstract}
A novel multi-scale finite element formulation for contact mechanics between nominally smooth but microscopically rough surfaces is herein proposed. The approach integrates the interface finite element method (FEM) for modelling interface interactions at the macro-scale with a boundary element method (BEM) for the solution of the contact problem at the micro-scale. The BEM is used at each integration point to determine the normal contact traction and the normal contact stiffness, 
allowing to take into account any desirable kind of rough topology, either real, e.g. obtained from profilometric data, or artificial, evaluated with the most suitable numerical or analytical approach. Different numerical strategies to accelerate coupling between FEM and BEM are discussed in relation to a selected benchmark test.  
\end{abstract}



\section{Introduction}

Due to the technological trend of producing structures down to the micro- and nano-scales, surface-related phenomena become predominant over bulk properties \cite{LR05}. Therefore, local imperfections and deviation from the ideal flatness of surfaces \cite{RMF02}, and especially waviness, roughness and other forms of texturing, have a fundamental effect on surface physics, as for instance for heat and electrical transfer, optical properties, fluid-solid interactions. Similarly, they play a crucial role in tribology as far as stress transfer between interacting surfaces in relative motion, friction, wear, and lubrication are concerned \cite{PH_SI_JMES,PH_SI_JSA,vakis}. The role of mechanics is essential for understanding, modelling and simulating the stress and the deformation fields experienced by rough surfaces in contact, as well as for the description of their evolution over time \cite{rabino,KJ,goriacheva,perssonbook,vlpopov,popov_hess,barber_contact}. 

Even in the simplest case of linear elastic continua, the presence of roughness introduces a nonlinearity, since the effective contact area of the micro-scale now also depends on the applied load level. Therefore, understanding the connection between the geometrical/topological features of roughness and the consequent non-linear constitutive relation at the interface, the relation between the thermal/electrical contact resistance and the contact pressure, or the apparent value of the friction coefficient, just to name a few exemplary problems, is an intriguing research question with many practical technological implications.     

Semi-analytical micromechanical contact theories relying on the statistical distribution of the elevation of the asperities and their radii of curvature have been proposed and widely explored (see \cite{exp4,ZBP04} for comprehensive review articles), following the pioneering approach in \cite{GW66} and extending it to more complex statistical distributions of elevations and curvatures \cite{revitalized,GW2006,PC10}, considering also elastic interactions between asperities \cite{CGP08}. Since the 1990's, research focused on the multi-scale features of roughness, exploiting the use of fractal geometry for the understanding of its role on the contact behaviour \cite{MB90,BCC,CARPINTERI20052901,Persson1}. In all such studies, the primary focus was the characterization of the constitutive behaviour of the rough interface, regardless of the bulk. Hence, the boundary element method (BEM) has been historically preferred over the finite element method (FEM) \cite{A81,M94} for this purpose. This is essentially due to the fact that only the surface must be discretized in the boundary element method, and not the surrounding continuum, as required by the finite element method. Moreover, it is not necessary to adopt surface interpolation techniques, like Bezier curves, to discretize the interface (see,
e.g., the approach in \cite[Ch. 9]{wriggers2}) and make it amenable for the application of contact search algorithms. This avoids an undesired smoothing of the fine scale geometrical features of roughness.

However, standard boundary element formulations are based on the fundamental assumptions of linear elasticity and homogeneity of the materials. Consequently, their generalization to inhomogeneities \cite{nelias} and finite-size geometries \cite{finite} are sometimes possible but are not straightforward.
The finite element method would open new perspectives, even if applied at the micro-scale. Within this approach, in fact, it is possible to take into account any material \cite{PEI,HPMR} or interface constitutive nonlinearity \cite{MPJR}. Moreover, it is prone to be extended for the solution of nonlinear multi-field problems involved in heat transfer or in reaction-diffusion systems \cite{Zava1,Sapora2014,LENARDA2018374}, for which the boundary element method has not been applied so far. Last but not least, new robust contact discretization schemes and solution strategies have been advanced within the framework of the FEM in recent years, including nonlinear thermomechanics and wear \cite{popp2018contact,Seitz2018,Seitz2018bis,HIERMEIER2018532,Farah2017}.

Industrial applications, for which the size-scale of the components is usually much bigger than the microscopical size-scale of roughness, are challenging also for the above finite element techniques designed for micro-scale computations. Hence, multi-scale approaches should be invoked. In this regard, a node-to-segment finite element formulation for contact mechanics was proposed in \cite{Zava1}, where a penalty approach was used to enforce the satisfaction of the unilateral contact constraint. Moreover, the contact force and the penalty stiffness were predicted by a modified nonlinear penalty formulation where the nodal force-nodal gap relation was dictated by a micromechanical contact model. This approach assumes a scale separation between the micro-scale solution of the contact problem, provided in closed form according to the micromechanical contact model, and the macro-scale one, where the finite element method is applied.
As a limitation, this method strongly relies on the micromechanical contact model prediction, which is based on simplified assumptions related to the form of roughness, the statistical distribution of asperity heights and curvatures. 

In this article, a multi-scale finite element formulation is proposed, where any statistically representative microscopically rough surface can be provided as input, also variable with the position along the contact surface of the macro-scale finite element model. Specifically, at the macro-scale, an implicit finite element formulation based on the interface finite element topology is exploited. At each integration point, the statistically representative rough surface height field is stored, and the boundary element method is called by passing the macroscopic relative displacement in the normal direction. 
The micro-scale model based on BEM provides the homogenized normal contact traction (and its linearization) to be used within the nonlinear solver of macro-scale FEM model.
This approach allows testing any surface roughness topology without making assumptions on the surface height distribution. On the other hand, the computation cost associated to this problem is much higher than in \cite{Zava1}. Therefore, some possible acceleration strategies are presented and their effect compared.

This article is structured as follows: in Sec.~\ref{sec:variational} the variational formulation at the macro-scale is presented. Sec.~\ref{sec:multi-scale} details the multi-scale contact formulation and the way coupling between the finite element method at the macro-scale and the boundary element method at the micro-scale is enforced. Sec.~\ref{sec:examples} is devoted to numerical examples and to the comparison of the different solution schemes to accelerate computations. Conclusive remarks and future developments complete the article.


\section{Variational formulation}\label{sec:variational}
In this section, we propose the variational formulation governing the problem of contact between two bodies across a rough interface. Starting from the strong differential form describing the mechanics of the continua and the problem of contact along the interface, we derive the corresponding weak form that provides the basis for the new interface finite element detailed in Sec.~\ref{sec:multi-scale}.

\subsection{Governing equations and strong form}\label{sec:strong_form}

\begin{figure}[h!]
\begin{center}
\includegraphics[width=.45\textwidth,angle=0]{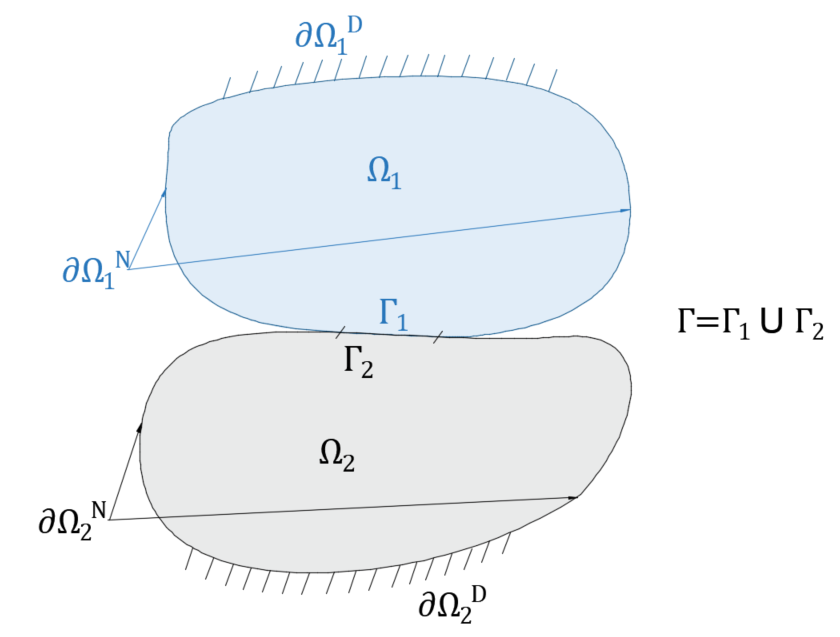}
\caption{Domains $\Omega_i$ $(i=1,2)$, their Dirichlet $(\partial\Omega_i^D)$ and Neumann $(\partial\Omega_i^N)$ boundaries, and the contact interface $\Gamma=\Gamma_1\bigcup \Gamma_2$.}
\label{fig:fig1}
\end{center}
\end{figure}

Let two deformable bodies occupy the domains $\Omega_i \in \mathbb{R}^2$ $(i=1,2)$ in the undeformed configuration defined by the reference system $Oxz$ (see Fig.~\ref{fig:fig1}). The two domains are separated by an interface $\Gamma$ defined by the opposite boundaries $\Gamma_i$ $(i=1,2)$ of the two bodies, viz. $\Gamma=\bigcup_{i=1,2} \Gamma_i$, where contact takes place. The whole boundary of the $i$-th body, $\partial \Omega_i$, is therefore divided into three parts: 
\begin{itemize}
\item a portion where displacements are imposed, i.e., the Dirichlet boundary $\partial \Omega_i^D$;
\item a portion where tractions are specified, i.e., the Neumann boundary $\partial \Omega_i^N$; 
\item the interface $\Gamma_i$ where specific boundary conditions have to be imposed to model contact.
\end{itemize}

Here, we assume that $\Gamma_i$ is nominally smooth but microscopically rough. A smoother representation of each interface $\Gamma_i^*$ is introduced by considering a surface parallel to the average one of the rough surface and passing through its lowest point, i.e. the deepest valley. In the present 2D setting, this surface coincides with a smooth line, discretized by a set of appropriate interface elements. At the initial condition, $\Gamma_1^*$ and $\Gamma_2^*$ are coincident but distinct lines and the two bodies are in contact just in correspondence of a single point given by the  highest peaks of the undeformed surfaces.  

We also associate the tangential and normal unit vectors $\mathbf{t}_i(x, z)$ and $\mathbf{n}_i(x,z)$ at any point of $\Gamma_i^*$, with $\mathbf{n}_i$ pointing outwards from the domain $\Omega_i$. Due to the assumption that the two non-conformal profiles are microscopically rough but nominally smooth, the two coincident smoother lines are parallel to each other and therefore $\mathbf{n}_1(x,z)=-\mathbf{n}_2(x,z)$ and $\mathbf{t}_1(x,z)=-\mathbf{t}_2(x,z)$, $\forall x,z$ on $\Gamma^*$. As a result, we can define a unique tangential and normal unit vectors $\mathbf{n}$ and $\mathbf{t}$ and introduce a zero-thickness interface model for $\Gamma^*$. 

In the most general case, we now postulate the existence of a displacement field for each body, $\bu_i=(u_i,v_i)^\text{T}$, that 
can be used to map the undeformed configuration to the deformed one.
Such functions are thereby assumed to be continuous, invertible and differentiable functions of the position vector $\mathbf{x}=(x,z)^{\text{T}}$ within each body. At the interface $\Gamma^*$, on the other hand, the configuration of the system is described by the relative displacement field $\Delta\bu$, usually denoted as gap field across the interface $\mathbf{g}=(\gtt,\gnn)^{\text{T}}$,  which is mathematically defined as  the projection of the relative displacement $\bu_1-\bu_2$ onto the normal and tangential directions of the interface defined by the unit vectors $\mathbf{n}$ and $\mathbf{t}$, respectively. In components, the vector $\Delta\bu$ collects the relative tangential displacement, $\Delta u_t$, and the relative normal displacement, $\Delta u_n$, i.e., $\Delta\bu=(\Delta u_t,\Delta u_n)^{\text{T}}$.

Inside each deformable material, the small deformation strain tensor $\bepsilon_i$ $(i=1,2)$ is introduced as customary, which is defined as the symmetric part of the displacement gradient: $\bepsilon_i =  \nabla^{s} \bu_i$. In the sequel, the standard Voigt notation will be used and the strain tensor components will be collected in the vector  $\bepsilon_i=(\epsilon_{xx},\epsilon_{zz},\gamma_{xz})_i^{\text{T}}$.

In the absence of body forces, the strong (differential) form of equilibrium for each body is provided by the linear momentum equation along with the Dirichlet and the Neumann boundary conditions on $\partial \Omega_i^D$ and $\partial \Omega_i^N$, respectively $(i=1,2)$, equipped by the conditions for contact on $\Gamma^*$:
\begin{subequations}\label{strong}
\begin{align}
\nabla \cdot \bsigma_i =\mathbf{0} \;\; &\text{in}\, \Omega_i,\\
\bu_i =\hat{\bu} \;\; &\text{on}\, \partial\Omega_i^D,\\
\bsigma_i \cdot \mathbf{n} = \mathbf{T}\;\; &\text{on}\, \partial\Omega_i^N, \\
\gnn \ge 0, ~ \pnn \ge 0\;\; &\text{on}\, \Gamma^*
\end{align}
\end{subequations}
where $\hat{\bu}$ denotes the imposed displacement, and $\mathbf{T}$ the applied traction vector. 

For its solution, the strong form has to be equipped by the constitutive equations for the bulk and for the interface. For the bulk, recalling standard thermodynamics arguments, general  (linear or nonlinear) constitutive stress-strain relations can be postulated without any loss of generality for the $i$-th material domain:  $\bsigma_i := \partial_{\bepsilon_i} \Psi (\bepsilon_i)$ and 
$\mathbb{C}_i : = \partial^{2}_{\bepsilon_i \bepsilon_i} \Psi (\bepsilon_i)$, whereby $\Psi (\bepsilon_i)$ is the Helmholtz free-energy function for body $i$, whereas its corresponding Cauchy stress tensor and the constitutive operator are respectively denoted by  $\bsigma_i$ and $\mathbb{C}_i$. 

Regarding the interface, the constitutive response should be introduced by distinguishing between the normal and the tangential directions. Although the present formulation can encompass any type of loading condition, we restrict our attention in this study to the frictionless normal contact problem and we neglect the influence of adhesion. In general, the constitutive relation in the tangential direction should account for frictional effects and adhesion, and it is left for further investigation.


\subsection{Weak form}

According to the principle of virtual work, the weak form associated with the strong form in Eq.~\eqref{strong} reads:
\begin{align}\label{weak}
    \Pi &= \int_{\Omega_1}\bsigma_1(\bu_1)^{\text{T}}\bepsilon_1(\bv_1)\dO
    + \int_{\Omega_2}\bsigma_2(\bu_2)^{\text{T}}\bepsilon_2(\bv_2)\dO\\
    &- \int_{\partial\Omega_1^{\text{N}}}\bT^{\text{T}}\bv_1 \dpO           \notag
    - \int_{\partial\Omega_2^{\text{N}}}\bT^{\text{T}}\bv_2 \dpO\\
     &- \int_{\Gamma^*} \pnn (\Delta \bu) \gnn (\Delta \bv) \dG = 0          \notag
\end{align}
where $\bv_i$ is the test function (virtual displacement field) and $\gnn(\Delta\bv)$ is the virtual normal gap at the interface $\Gamma^*$. The test function in the $i$-th body fulfills the condition $\bv_i=\mathbf{0}$ on $\partial\Omega_i^D$ and the contact condition on  $\Gamma^*$, which can be formulated as:
\begin{equation}\label{pn}
\pnn(\gnn)=
\begin{cases}
\pnn & \text{if}\; \gnn > 0,\\
0, & \text{if}\; \gnn \le 0.
\end{cases}
\end{equation}
where the nominal pressure $\pnn$ is given by the micro-scale contact interactions predicted by the boundary element method as described in the following section.

The contact conditions on $\Gamma^*$ impose that the corresponding integral is greater or equal to zero everywhere on $\Gamma^*$. Thus, the solution of the problem implies the solution of the following variational inequality:

\begin{multline}
\int_{\Omega_1}\bsigma_1(\bu_1)^{\text{T}}\bepsilon_1(\bv_1)\dO+\int_{\Omega_2}\sigma_2(\bu_2)^{\text{T}}\bepsilon_2(\bv_2)\dO\\
-\int_{\partial\Omega_1^N}\bT^{\text{T}} \bv_1\text{d}\partial\Omega
-\int_{\partial\Omega_2^N}\bT^{\text{T}} \bv_2\text{d}\partial\Omega\ge 0.
\end{multline}

The displacement field $\bu_i$ solution of the weak form in Eq.~\eqref{weak} is such that it corresponds to the minimum of $\Pi$ for any choice of the test functions $\bv_i$.


\section{Multi-scale contact formulation} \label{sec:multi-scale}
For the bulk, standard linear quadrilateral or triangular isoparametric finite elements can be invoked. On the other hand, at the interface, the solution of the presented contact problem is treated at two different levels. At the macro-scale, a zero-thickness interface finite element is employed to model interface interactions. The integral expressions for the stiffness operator and the residual vector are approximated via a Gaussian integration, as explained in detail in Sec.~\ref{sec:FEM_discret}, and the values of the contact pressure are evaluated by solving, for each Gauss point, the contact problem between the elastic half plane and the rough surface, exploiting the boundary element method, see Sec.~\ref{sec:BEM_discret}.

The macro-scale model, analyzed in Sec.~\ref{sec:FEM_discret}, is 2D under plain strain assumption, while the micro-scale is 3D and deals with two surfaces coming into contact. For guaranteeing the consistency between the two scales, the average pressure acting on the surfaces and evaluated using the BEM is multiplied by a unit depth before passing it to the FEM model.


\subsection{Finite element discretization of the interface at the macro-scale} \label{sec:FEM_discret}

\begin{figure}
\begin{center}
\includegraphics[width=.4\textwidth,angle=0]{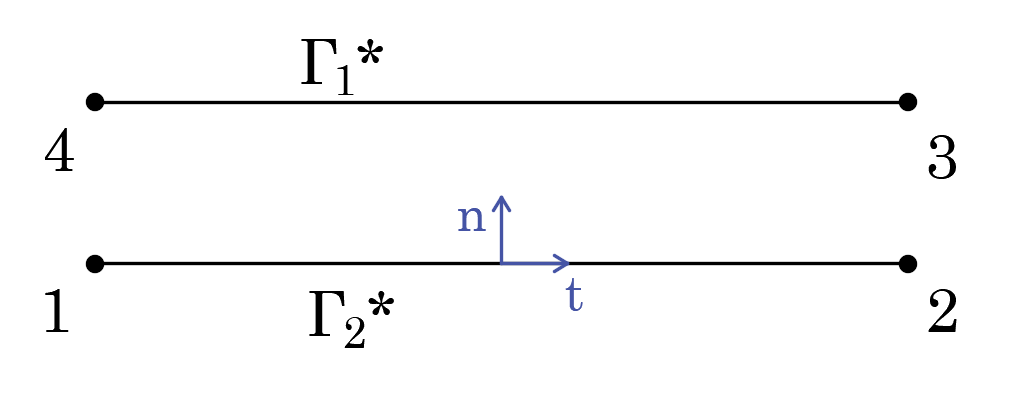}
\caption{Sketch of the interface finite element topology.}\label{fig2}
\end{center}
\end{figure}

At the macro-scale, we introduce a conforming finite element discretization whose kinematics follows from the formulation of interface elements used in non-linear fracture mechanics for cohesive crack growth. This interface element is characterized by nodes $1$ and $2$, belonging to $\Gamma_{2}^*$, and by nodes $3$ and $4$, which belong to $\Gamma_{1}^*$, see Fig.~\ref{fig2}.

The interface integral in Eq.~\eqref{weak} can be evaluated as the sum of the contribution of the whole interface elements. Each element contribution (denoted by the subscript $e$) is herein computed by using the 2 points Gauss quadrature formula which implies the sampling of the integrand at the Gauss points $x_{\mathrm{g}1}$ and $x_{\mathrm{g}2}$:
\begin{equation}
\int_{\Gamma^*_\mathrm{e}}p(\gnn) \gnn\,\mathrm{d}\Gamma=\det J\sum_{i=1,2}w_ip_i(g_{\mathrm{n,i}}) g_{\mathrm{n,i}},
\end{equation}
where $\det J$ is the standard determinant of the Jacobian of the transformation that maps the geometry of the interface element from its global reference frame to the natural reference system.

To evaluate the normal gap $g_\mathrm{n}$ at any point inside the interface element, we need to introduce the nodal displacement vector $\mathbf{d}=(u_1,v_1,\dots,u_4,v_4)^{\text{T}}$, which collects the displacements $u$ and $v$ of the four interface finite element nodes. The relative displacement $\mathbf{g}$ for the nodes $1$-$4$ and $2$-$3$ is then computed by applying a matrix operator $\mathbf{L}$ which calculates the difference between the displacements of nodes $1$ and $2$ with respect to those of nodes $4$ and $3$. The relative displacement within the interface finite element is then given by the linear interpolation of the corresponding nodal values, represented by the multiplication with the matrix $\mathbf{N}$ which collects the shape functions at the element level. Finally, the tangential and the normal gaps are determined by the multiplication with the rotation matrix $\mathbf{R}$ defined by the components of the unit vectors $\mathbf{t}$ and $\mathbf{n}$. In formulae, we have:
\begin{equation}
\mathbf{g}=-\mathbf{R}\mathbf{N}\mathbf{L}\mathbf{d},
\end{equation}
where the operators present the following matrix form:
\begin{subequations}
\begin{align}
\mathbf{L} &=
\left[
\begin{matrix}
-1 &  0 &  0  &  0 &  0 & 0  & 1 &  0\\
 0 & -1 &  0  &  0 &  0 & 0  & 0 &  1\\
 0 &  0 & -1  &  0 &  1 & 0  & 0 &  0\\
 0 &  0 &  0  & -1 &  0 & 1  & 0 &  0
\end{matrix}
\right],\\
\mathbf{N} &=
\left[
\begin{matrix}
N_1 &  0    &  N_2  &  0 \\
    0 & N_1 &  0      &  N_2
\end{matrix}
\right],\\
\mathbf{R} &=
\left[
\begin{matrix}
t_x & t_z\\
n_x   & n_z
\end{matrix}
\right],
\end{align}
\end{subequations}
where $n_x$, $n_z$, $t_x$ and $t_z$ are the components of the unit vectors $\mathbf{n}$ and $\mathbf{t}$ along the $x$ and $z$ directions.

The normal gap is used to compute the normal traction $\pnn$ according to the boundary element method accounting for micro-scale contact interactions. Due to the intrinsic non-linearity of the contact problem, a Newton-Raphson scheme is herein adopted to solve the implicit non-linear algebraic system of equations resulting from the finite element discretization:
\begin{subequations}\label{N-R}
\begin{align}
\mathbf{K}^{(k)} \Delta\mathbf{d}^{(k)} &= -\mathbf{R}^{(k)},\\
\mathbf{d}^{(k+1)} &=\mathbf{d}^{(k)}+\Delta\mathbf{d}^{(k)},
\end{align}
\end{subequations}
where the superscript $k$ denotes the iteration inside the Newton-Raphson loop, and the residual vector $\mathbf{R}_{e}^{(k)}$ and the tangent stiffness matrix $\mathbf{K}_{e}^{(k)}$ associated with the $e-$th interface finite element, assembled to the global residual vector $\mathbf{R}$ and global stiffness matrix $\mathbf{K}$, are:
\begin{subequations}\label{R-K}
\begin{align}
\mathbf{R}_{e}^{(k)} &=-\int_{\Gamma^*_e} \mathbf{L}^{\text{T}}\mathbf{N}^{\text{T}}\mathbf{R}^{\text{T}}\mathbf{p}\,\text{d}\Gamma,\\
\mathbf{K}_{e}^{(k)} &=\int_{\Gamma^*_e} \mathbf{L}^{\text{T}}\mathbf{N}^{\text{T}}\mathbf{R}^{\text{T}}\mathbb{C}\mathbf{R}\mathbf{N}\mathbf{L}\,\text{d}\Gamma,
\end{align}
\end{subequations}
where $\mathbf{p}=(p_\mathrm{t},\pnn)^{\text{T}}=(0,\pnn)^{\text{T}}$ and $\mathbb{C}$ is the linearized interface constitutive matrix:
\begin{equation}
\mathbb{C} =
\left[
\begin{matrix}
\dfrac{\partial p_\mathrm{t}}{\partial \gtt} &  \dfrac{\partial p_\mathrm{t}}{\partial \gnn}\\
\dfrac{\partial \pnn}{\partial \gtt} &  \dfrac{\partial \pnn}{\partial \gnn}
\end{matrix}
\right],
\end{equation}
that, for the frictionless normal contact problem, reads:
\begin{equation}
    \mathbb{C} =
\left[
\begin{matrix}
0 &  0\\
0 &  \dfrac{\partial \pnn}{\partial \gnn}
\end{matrix}
\right],
\end{equation}
and we just need to specify $\partial \pnn/\partial\gnn$ depending on the sign of the normal gap.
For the present multi-scale problem, it has to be remarked that the closed form expression for $\partial \pnn/\partial \gnn$ is not available, and therefore it is computed numerically by a finite difference approximation of two solutions obtained by the application of the boundary element method, one for $\gnn$ and another for the same value of $\gnn$ plus a small perturbation, see the next section.

The integrals in Eqs.~\eqref{R-K} are therefore given as the sum of two terms: 
\begin{subequations}
\begin{align}
\mathbf{R}_{e}^{(k)} & = - \det J\sum_{i=1}^2 w_i \mathbf{L}^{\text{T}}\mathbf{N}^{\text{T}}\mathbf{R}^{\text{T}}\mathbf{p}(x_{\mathrm{g}i}),&\\
\mathbf{K}_{e}^{(k)} & = \det J\sum_{i=1}^2 w_i
\mathbf{L}^{\text{T}}\mathbf{N}^{\text{T}}\mathbf{R}^{\text{T}}\mathbb{C}(x_{\mathrm{g}i})\mathbf{R}\mathbf{N}\mathbf{L},
\end{align}
\end{subequations}
where $w_i=1$ is the weight and $x_{\mathrm{g}1,2}= \mp 1/\sqrt{3}$ are the positions of the two Gauss Points along $\Gamma^*$, where the pressure is going to be evaluated.


\subsection{Boundary element method for micro-scale interactions} \label{sec:BEM_discret}

The unknown value of $\pnn$ at each Gauss Point is herein computed by solving the normal contact problem of a rigid rough surface indenting an elastic half-plane with composite elastic parameters, which is  mathematically the equivalent of solving the contact problem between two deformable rough surfaces \cite{barber03}. 

\begin{figure}
\centering
\subfloat[Rough profiles identified by $e_1(\bxi)$ and $e_2(\bxi)$.\label{fig:original}]
{\includegraphics[width=0.45\textwidth]{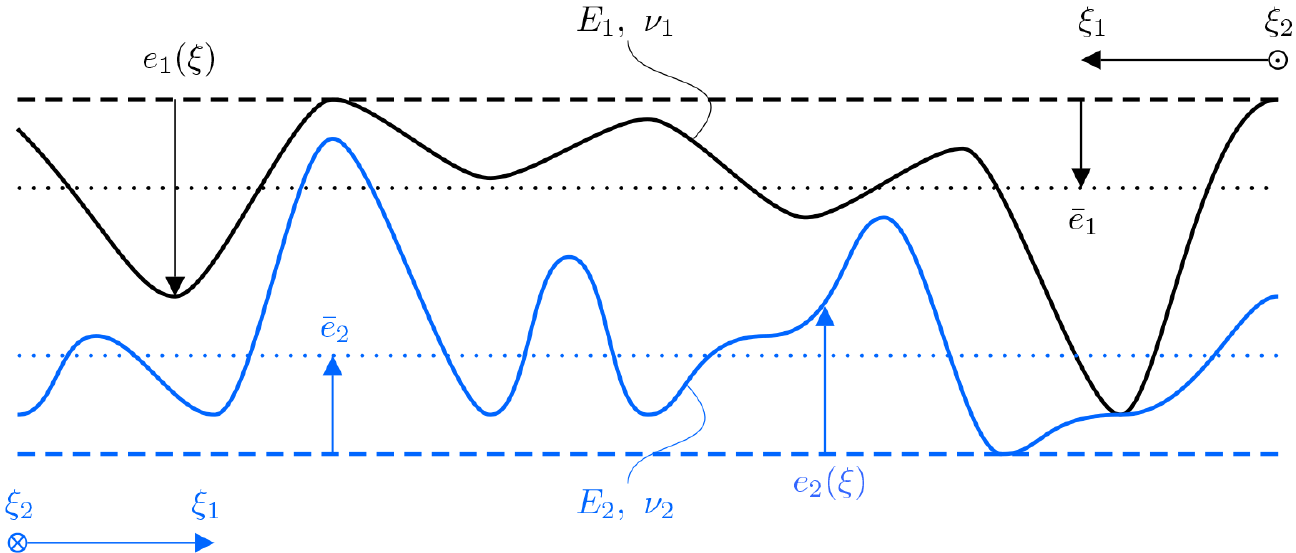}}\\
\subfloat[Composite topography described by $e^*(\bxi)$.\label{fig:composite}]
{\includegraphics[width=0.45\textwidth]{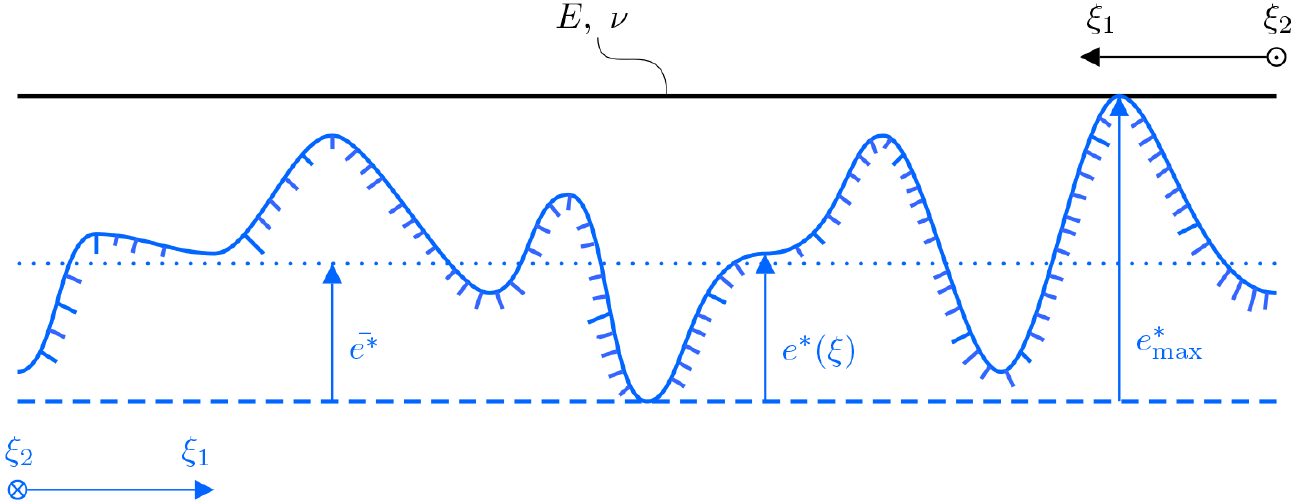}}
\caption{Transformation of two rough profiles~\protect\subref{fig:original} into a flat line, the elastic part, and a profile with composite topography~\protect\subref{fig:composite}, \textit{i.e.} the rigid indenter.}
\label{fig:comp_top}
\end{figure}

Let $e_1(\bxi)$ and $e_2(\bxi)$ be the elevations of two rough surfaces measured from their lowest point, where $\bxi=(\xi_1,\xi_2)^T$ is a position vector referring to the surfaces local reference system (see Fig.~\ref{fig:comp_top}\subref{fig:original}). The elevation of the composite topography can be evaluated as:
\begin{equation}
e^\ast(\bxi)=e_1(\bxi)+e_2(\bxi)-\min[e_1(\bxi)+e_2(\bxi)],
\end{equation}
measured from a new datum set in correspondence of the lowest point, with distance $e^\ast_\mathrm{max}$ from the boundary of the elastic flat half-space, as shown in Fig.~\ref{fig:comp_top}\subref{fig:composite}.  

As illustrated in~\cite{barber_contact,barber_elasticity}, the composite elastic parameters can be computed as:
\begin{subequations}
\label{eq:comp_mod}
\begin{align}
E&=\biggl( \frac{1-\nu_1^2}{E_1}+\frac{1-\nu_2^2}{E_2} \biggr)^{-1}\label{eq:compE},\\
G&=\biggl(\frac{2-\nu_1}{4 G_1}  + \frac{2-\nu_2}{4 G_2} \biggr)^{-1}\label{eq:compG},
\end{align}
\end{subequations}
where $G_i=E_i/[2(1+\nu_i)]$ are the shear modulus of the original bodies and the composite Poisson ratio $\nu$ is related to $G$ and $E$ via $\nu=E/(2G)-1$.

For each Gauss point of the macro-scale model, the following micro-scale contact problem is solved under displacement control, where the far-field displacement corresponds to $\gnn$ from the macro-scale model. For $\gnn=0$, we assume the surfaces touch only at the tallest height of the composite topography, with a resulting zero normal traction. For each $\gnn> 0$, a non-vanishing contact area has to be computed, as well as the corresponding total normal force equivalent to the integral of the normal contact tractions. To do so, the BEM implementation proposed in~\cite{BP15} is employed, in particular the \textit{Warm-Started Non-Negative Least Squares (NNLS)} algorithm is exploited.

According to BEM, the normal displacement at a point of the half-plane characterized by a position vector $\bxi$ is related to the pressure $p(\boldsymbol{\eta})$ exerted at other points by the following relation:
\begin{equation}\label{up}
u(\boldsymbol{\xi})=\int_S H(\boldsymbol{\xi},\boldsymbol{\eta})p(\boldsymbol{\eta})
\,\mathrm{d}\boldsymbol{\eta}
\end{equation}
where $H(\boldsymbol{\xi},\boldsymbol{\eta})$ is the Green function, representing the displacement at point $u(\boldsymbol{\xi})$ caused by a surface pressure $p$ acting at $\boldsymbol{\eta}$, while $S$ is the half-space. For homogeneous, isotropic, linear elastic materials, the Green function has been chosen as (\cite{KJ}, \cite{barber_elasticity}):
\begin{equation}
H(\boldsymbol{\xi},\boldsymbol{\eta})=
\frac{1}{\pi E} \frac{1}{||\boldsymbol{\xi}-\boldsymbol{\eta}||}
\end{equation}
where $E$ denotes the composite Young's modulus of the half-space, while $|| \cdot ||$ represents the Euclidean norm. The total contact force $P$ can be evaluated by integrating the pressure field over the whole interface.
\begin{equation}
P = \int_S p(\boldsymbol{\xi})\,\mathrm{d}\boldsymbol{\xi}
\end{equation}
\begin{figure}
\centering
\includegraphics[width=.45\textwidth]{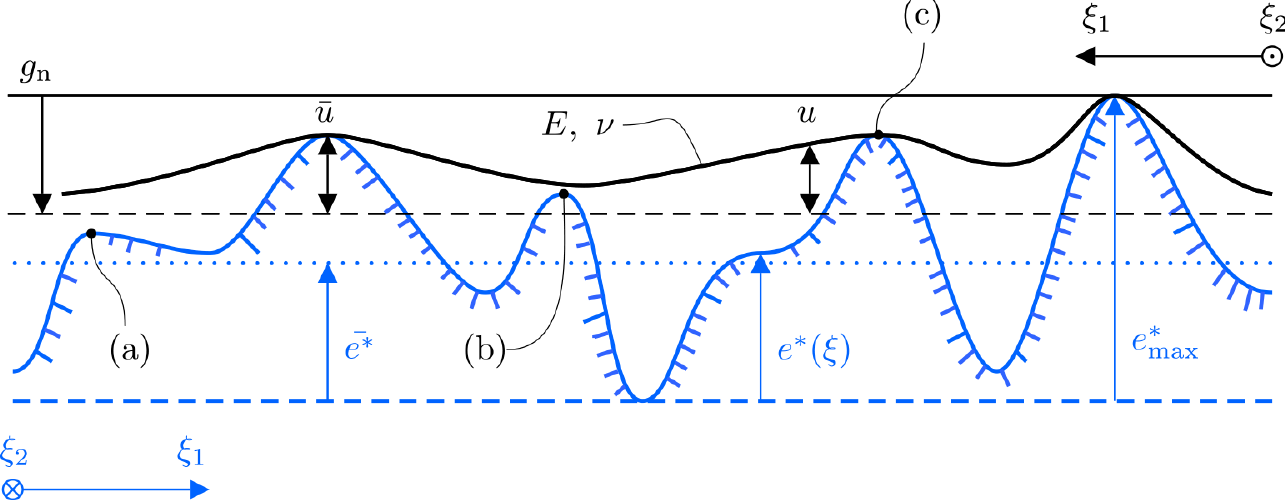}
\caption{Illustration of the contact problem between a rigid rough surface, solid blue line, and an elastic half plane, for a given far field displacement $g_\mathrm{n}$. The rigid body motion of the half plane is indicated by the dashed black lines, while its deformed boundary by the solid black one.}\label{fig:comp3}
\end{figure}

Finally, the mean pressure $p$ is evaluated dividing the total force $P$ by the nominal surface area.  
For a given far-field displacement $\gnn$ in the direction perpendicular to the half-plane, the solution of the normal contact problem $u(\boldsymbol{\eta})$, $p(\boldsymbol{\eta})$ must satisfy Eq.~\eqref{up} together with the unilateral contact constraint:
\begin{subequations}
\begin{align}
w(\boldsymbol{\xi},\gnn)                   &\ge 0, &\\
p(\boldsymbol{\xi})                          &\ge 0, &\\
w(\boldsymbol{\xi},\gnn)p(\boldsymbol{\xi})&= 0,
\end{align}
\end{subequations}
where $w(\boldsymbol{\xi},\gnn)=u(\boldsymbol{\xi})-\overline{u}(\boldsymbol{\xi},\gnn)$ and $\overline{u}(\boldsymbol{\xi},\gnn)$ denotes the indentation of the half-space at the points in contact. A 2D sketch is shown in Fig.~\ref{fig:comp3}, where the deformed configuration of the elastic half-space corresponding to the imposed far-field displacement is represented by the black solid line. The black dashed line represents the rigid body motion of the elastic body under the imposed displacement. The solution of the contact problem requires for the points to be of three types:
\begin{itemize}
\item not in contact from the beginning (a);
\item loosing contact due to elastic interactions (b); 
\item in contact after considering elastic interactions (c).
\end{itemize}

A routine for the solution of this infinite dimensional problem has been implemented by discretizing the rough surface with a square grid with lateral size $l$ and resolution parameter $n$, such that the grid is composed by $N \times N$ cells with $N=2^n+1$ boundary elements per side. The lateral size of each boundary element is $a=l/(2^n+1)$. A random midpoint displacement algorithm has been used to generate the height field $e^\ast_{i,j}$($i=1,...,N$, $j=1,...,N$) of the rough surface, although any data field obtained from experiments can be used in input, without any restriction. For each microscopically rough surface, the mean elevation $\bar{e}^\ast$, the maximum elevation $e_\mathrm{max}^\ast$, and the root mean square roughness $s$ are also available from a preliminary statistical characterization. The discretized matrix form of the problem thus reads: 
\begin{align}
\mathbf{w} &= \mathbf{H}\cdot\mathbf{p}-\overline{\mathbf{u}},\\
\mathbf{w} &\ge \mathbf{0}, \quad\mathbf{p} \ge \mathbf{0}, \quad \mathbf{w}\cdot\mathbf{p} =0,
\end{align}
where $\mathbf{w}$ is the vector of elastic corrections, $\mathbf{p}$ the unknown average contact forces, $\overline{\mathbf{u}}$ the vector of compenetrations and finally $\mathbf{H}$ the matrix collecting the compliance coefficients in its approximated form as in \cite{BCC}:
\begin{equation}
H_{i-k,j-l}=
\begin{cases}
\frac{1}{\pi E}                              
& \text{if $i=k$ and $j=l$},\\
\frac{1}{\pi E}\arcsin{\frac{1}{\|\boldsymbol{\xi}_{i,j}-\boldsymbol{\xi}_{k,l}\|}} 
& \text{if $i\neq k$, $j\neq l$},
\end{cases}
\end{equation}

Due to linear elasticity, $\mathbf{H}$ is symmetric and positive definite. This guarantees that the contact problem has a unique solution for any $\gnn\ge0$. Moreover, the problem corresponds to the Karush-Kuhn-Tucker conditions for optimality of the convex quadratic program:
\begin{align}
&\min_p~\frac{1}{2}\mathbf{p}^T\mathbf{Hp}-\overline{\mathbf{u}}^T\mathbf{p}, &\\
&\text{s.t.}~\mathbf{p}\ge\mathbf{0}.
\end{align}

\subsection{Computation of the contact pressure related to roughness}
The normal contact stiffness and the contact pressure predicted by the boundary element method account for two separate effects: one associated with the roughness of the surface, and another related to the deformation of the half-space \cite{PB11}. The overall compliance of the system is the sum related to roughness and the elastic one. 
In our framework, we need to extract only the effect associated to roughness, since the elastic contribution of the surrounding continuum is already computed in the macro-scale model. Therefore, a correction to the resulting pressure field is required. To this aim, as first step, we need to compute the elastic deformation associated to our micro-scale contact problem and subtract this contribution from the overall system.

If we consider the contact of a flat rigid indenter, with a $l \times l$ square size, acting on an elastic half-plane, an average nominal pressure $\bar{p}$ will cause a uniform displacement $w_0$ equal to:
\begin{equation}
w_0(\overline{p})=\frac{\alpha\overline{p}}{E}l
\label{eq:w0}
\end{equation}
with a mesh size dependent shape factor $\alpha<1$. To compute $\alpha$ for a given mesh resolution, a micro-scale BEM model with a perfectly flat surface has been solved. Since such model only includes linear elastic effects, the resulting gap-pressure relation is linear. The shape factor has been taken as $\alpha=E w_0/l\overline{p}$, where $E$ is the composite Young's modulus, $w_0$ is the imposed far field displacement and also the half-space indentation, $l$ is the lateral size of the square punch and $\overline{p}$ is the mean pressure, evaluated dividing the resulting total load by the nominal area $l^2$. Its values are given in Tab.~\ref{tab:alpha} for different resolution $n$. Different approaches for evaluating the limit value of $\alpha$, as the mesh size resolution approaches the continuum, can be found in \cite[Ch. 4]{barber_contact}, and~\cite{CONWAY1968489,nakamura1993}.

\begin{table}
	\centering
	\begin{tabular}{cccccc}
		\toprule
		$n$ & 1 & 2 & 3 & 4 \\  
		\midrule
		$\alpha$ &  0.778 &  0.806 &  0.826 &  0.841 \\        
		\toprule
		$n$ & 5 & 6 & 7 & 8 \\  
		\midrule
		$\alpha$ &  0.852 &  0.858 &  0.862 &  0.865  \\      
                     \bottomrule
	\end{tabular}
	\caption{Values of the coefficient $\alpha$ computed by solving the problem of a rigid flat indenter in contact with an elastic half space with the BEM algorithm, for different values of the surface resolution parameter $n$}
	\label{tab:alpha}
\end{table}

\begin{figure}
\centering
\includegraphics[width=.45\textwidth,angle=0]{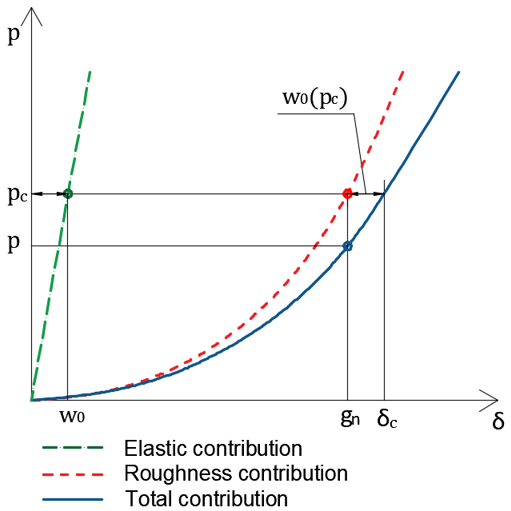}
\caption{Qualitative representation of pressure vs. imposed displacement curve considering the elastic contribution, the roughness contribution, and their combined effect.}
\label{fig:corr}
\end{figure}

Known the values of $\alpha$, we can use the relation between the nominal pressure and the elastic indentation given in Eq.~\eqref{eq:w0}, for computing the gap-pressure curve related only to roughness. This curve can be obtained by evaluating the values of the pressure $p$ for a set of given displacements $\delta$, considering both the elastic and the roughness contributions through the BEM and then applying the relationship \eqref{eq:w0} for computing the roughness related displacement $\delta_r=\delta_\mathrm {t} - w_0(p_\mathrm{t})$ in order to obtain the curve $p=p(\delta_\mathrm{r})$. The result of this correction procedure is graphically shown in Fig.~\ref{fig:corr}, where we can notice that the identified roughness contribution (red dashed curve) is stiffer than the one resulting from the overall system (blue solid curve) and it is obtained by subtracting the elastic contribution (green dashed line) to the BEM curve.

\begin{figure}
\centering
\includegraphics[width=.45\textwidth,angle=0]{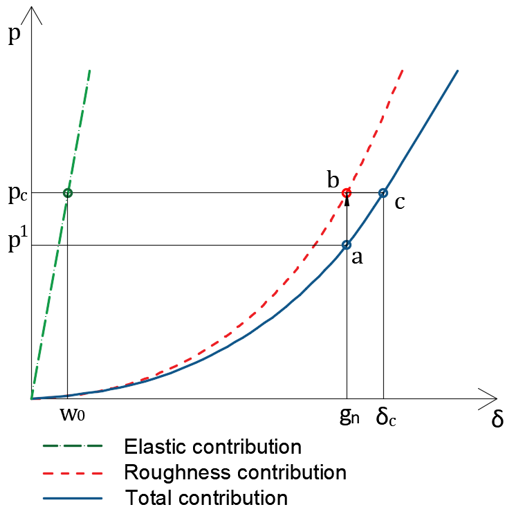}
\caption{Since point (b) is not directly derivable in the element routine, starting from point (a) the iterative procedure evaluates the pressure in (c) which guarantees equilibrium between $w_0$, $\gnn$ and $\delta_\mathrm{c}$ and gives the corrected pressure related to $\gnn$.}
\label{fig:corr2}
\end{figure}

It must be underlined that this subtracting procedure is not directly applicable in the interface element routine, since it requires the evaluation of the entire pressure-gap curve without correction, while the macro-model provides to the micro-model a single displacement $\gnn$ for each Newton-Raphson iteration at each Gauss point. 
Since $\gnn$ is meant to be related to roughness only, a pressure $p_\mathrm{c}$ higher than the one obtained by the unmodified curve ($p^1$) must be found, as shown in Fig.~\ref{fig:corr2}. The required value results from an augmented displacement $\delta_\mathrm{c}=\gnn+w_0$ where the value of $w_0$ can not be evaluated directly, depending on the unknown pressure $p_\mathrm{c}$ and an iterative approach is needed as follows. 

The BEM algorithm takes $\gnn$ as input from the macro-scale model and computes a pressure $p^1 (\gnn)$ that allows for the computation of a correction $w_0^1(p^1)$. The input displacement is then updated as $\delta_{\mathrm{c}}^1=\gnn + w_0^1$ and a new value of the pressure $p^2$ is computed. The relative error on the average pressure is evaluated and eventually the procedure is repeated. At the $i$-th generic iteration, the corrected displacement reads:
\begin{equation}
    \delta_\mathrm{c}^i=\gnn + w_0^i(p^i)
\end{equation}
and it corresponds to a pressure $p^i$. The relative error from an iteration to the next is updated as:
\begin{equation}
    \text{err}=\frac{p^i-p^{i-1}}{p^i}    
    \label{eq:err}
\end{equation}
The iterative procedure stops when the relative error is less then an imposed tolerance and the reached value of pressure is the required value $\pnn$ to be read by the macro-model at the Gauss point.

The value of tolerance has been obtained after a convergence study: the iterative procedure has been tested for a set of imposed displacements, varying the value of the tolerance in order to achieve a good accordance with the corrected gap-pressure curve evaluated with the subtracting procedure. As shown later in Sec.~\ref{sec:examples}, very good accordance has been found between the two curves even for a loose tolerance for all the values of separation taken into account, in line with the results in \cite{revitalized} and \cite{CGP08}.
Furthermore, the given procedure is valid for any desired value of tolerance that can be easily adjusted by the user according to the precision required by the specific case study.

\subsection{Multi-scale coupling}

The coupling between the micro- and the macro-scales has been implemented by exploiting three alternative approaches.

In the first approach, a full integration of FEM and BEM is proposed and it is called FEM BEM Quasi-Newton (FBEM-QN) since an approximation of the Jacobian is used for the iterative update scheme. The interface finite element has been coded as a user element for FEAP, exploiting a Newton-Raphson solution scheme. At each time step and for each Gauss point, the contact pressure $p_\mathrm{n}(\gnn)$ and the contact stiffness $\partial p_\mathrm{n}/\partial \gnn$ are computed by calling the subroutine based on BEM. Such BEM subroutine reads the rough surface height field at the first time step from an input file (the height field is stored in a standard $x$, $y$, $z$ three columns format) and stores it in a history variable for all the next time steps, to avoid continuous access to external files. The BEM subroutine is called once to compute $p_\mathrm{n}$ and then a second time to compute the normal contact stiffness via a finite difference approximation:
\begin{equation}\label{eq:QN}
\dfrac{\partial \pnn}{\partial \gnn} \simeq 
\dfrac{p_{\mathrm{n},k+1}-p_{\mathrm{n},k}}{g_{\mathrm{n},k+1}-g_{\mathrm{n},k}} 
\end{equation}
where $g_\mathrm{n,k}$ is the far-field displacement of the macro-scale model for the current $k$-th Newton-Raphson iteration, and $g_\mathrm{n,k+1}=g_\mathrm{n,k}+\Delta g_\mathrm{n,k}$ is a small perturbation of its value, for which the pressure values $p_\mathrm{n,k}$ and $p_\mathrm{n,k+1}$ are computed by BEM.  

This approach is computationally demanding, and therefore a second approach is also proposed for the numerical evaluation of the normal contact stiffness with the aim of saving CPU time.  In such approach, 
called FEM BEM Cheap Quasi-Newton (FBEM-CQN), the contact stiffness at the current Newton-Raphson iteration is computed by using the displacement and the pressure corresponding to the previous converged time step as the reference values for the application of the finite difference formula. The procedure requires using Eq.~\eqref{eq:QN} only at the first time step and then the following equation is used for the subsequent time steps:\begin{equation}
\frac{\partial\pnn}{\partial\gnn} \simeq \dfrac{p_\mathrm{n,k}^t-\pnn^{t-1}}{g_\mathrm{n,k}^t-\gnn^{t-1}},
\end{equation}
where $t$ and $t-1$ denote, respectively, the current and the previous time steps. This procedure requires storing the values of $\gnn^{t-1}$ and $\pnn^{t-1}$ in another appropriate history variable. 


In the last approach, which is referred to as FEM-BEM semi-analytical (FBEM-SAN), the normal contact problem at the micro-scale is solved off-line according to BEM, based on the generated height field given in input, for a sequence of far-field displacements. The solution of the problem in terms of predicted average contact pressure vs. the imposed far-field displacement is finally fitted with a power-law continuous function of the type:
\begin{equation}
\pnn(\gnn)=a \gnn^b,\label{eq:power}
\end{equation}
which provides a closed-form expression for $p_\mathrm{n}(\gnn)$. Its derivative $\partial p_\mathrm{n}/\partial \gnn$ entering the linearized interface stiffness matrix $\mathbb{C}$ is also available in analytic form. 

The choice of a power-law type fitting function is justified by the argument exposed in~\cite{PB11}. Let's assume to have two rough surfaces in contact, with specific dimensionless contact conductance $\tilde{C}$, dimensionless mean plane separation $\tilde{d}$ and dimensionless nominal contact pressure $\tilde{p}$. Making the hypothesis of incomplete similarity on $\tilde{p}$, a power-law dependence can be postulated between $\tilde{C}$ and $\tilde{p}$, in the form:
\begin{equation}
\tilde{C}=\Phi \tilde{p}^\beta,\label{eq:power_con}
\end{equation}
where $\Phi$ is a coefficient depending on the fractal geometry of the surface and $\beta$ is an exponent that can be obtained by real or numerical experiments. This hypothesis holds for physical systems which are in an intermediate situation between two limit conditions, which in the present setting are the high and low separations regime respectively. Together with the previous hypothesis, the electrical-mechanical analogy established in~\cite{barber03} states that:
\begin{equation}
\tilde{C}=-2\frac{\mathrm{d}\tilde{p}}{\mathrm{d}\tilde{d}}\,.\label{eq:analogy}
\end{equation}
By combining Eq.s~\eqref{eq:power_con} and~\eqref{eq:analogy} the result is an ordinary differential equation with separable variables, with solution, for $\beta \ne 1$: 
\begin{equation}
\frac{\tilde{p}^{1-\beta}}{1-\beta}=-\frac{\Phi}{2}(\tilde{d_0}-\tilde{d}),\label{eq:}
\end{equation}
which is a power-law relation between the nominal pressure and the plane separation. Following this formulation, the function~\eqref{eq:power} has been chosen as fitting function.

A major drawback coming from this approach arises when the state of the system is far from intermediate, i.e. for very high or very low separations. Scatter in the contact pressures is usually observed in the first case, where the contact response is ruled by the statistics of extremes of the lower tail of the asperity elevations distribution, and an artificial smoothing is inevitably introduced by fitting the data with a regular curve. The opposite condition corresponds to very high pressures, where distinct asperities start merging together and form large contact islands. This condition is not reached in the present case, where 
the maximum displacement imposed for acquiring the curve employed in the SAN approach is $3s$, and corresponds to a plane separation still far from the puzzling region of very high pressures.

On the other hand, a great advantage of the FBEM-SAN is its speed. It is expected to be the fastest of the three procedures if the representative rough surface is the same for all the integration points of the macro-scale model (uniform spatial roughness) or when the the same surface topography is used in several load cases. In these conditions, the time required by the BEM to solve the normal contact problem is spent only once, during the off-line stage. However, it is not difficult to imagine different scenarios where the convenience of one method with respect to another is not given for granted. For example, in case of a realistic macro-scale model where roughness is not homogeneous, but depends on the point, the Semi-Analytic method is still applicable, but fitting a different curve for every required Gauss point is necessary. Furthermore, when a different kind of topology is present, e.g. a complex textured rough surface, the power-law expression of the pressure-displacement relation could reasonably fail in predicting the trend of the curve under examination, therefore this kind of interpolation could introduce an undesirable approximation to the problem. In such a different scenario the Semi-Analytic implementation can still be applied, provided that a more suitable and perhaps more complex interpolating function has been determined. The other two approaches are expected to be more competitive when the gap-pressure range involved in the problem is not known from the beginning, and in general when the off-line stage becomes expensive. Testing the efficiency of the FBEM-SAN with respect to the integrated FBEM-QN and CQN in these and other different scenarios would be worth of interest and is left for further investigations.

\section{Numerical examples}\label{sec:examples}

In this section we propose a benchmark test to illustrate the capabilities of the proposed FEM-BEM multi-scale approach and compare the performances of the different solution strategies. 

\begin{figure}
\centering
\subfloat[][]
{\includegraphics[width=0.35\textwidth]{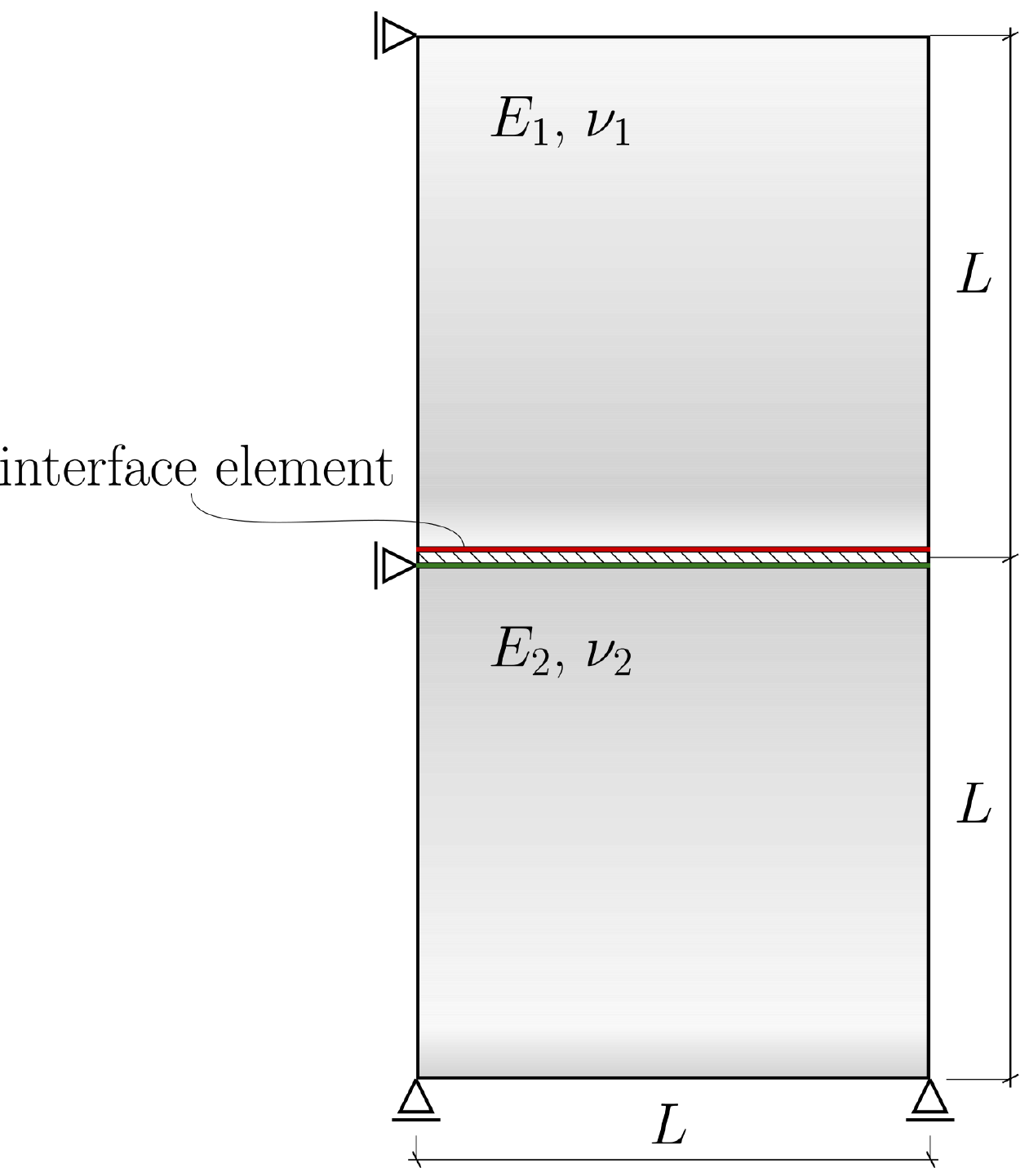}}\\
\subfloat[][]
{\includegraphics[width=0.35\textwidth]{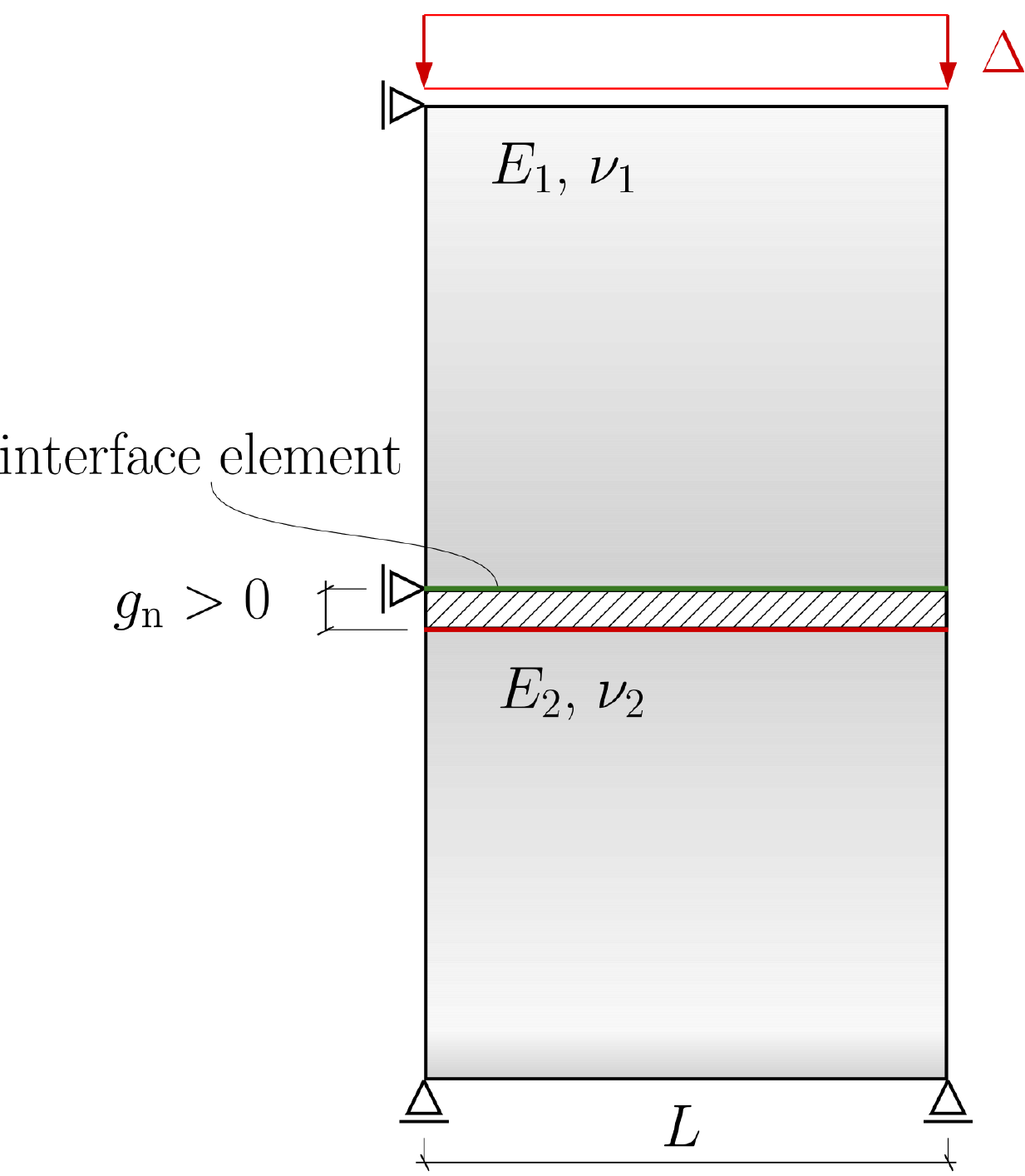}}
\caption{Geometry and boundary conditions of the benchmark test in uniaxial compression.}
\label{fig:blocks}
\end{figure}

Two square blocks of lateral size $L= 10\,$mm are discretized by a single finite element, see Fig.~\ref{fig:blocks}. An interface finite element connects the common boundary of the two bodies.

The two materials have Young's moduli $E_1=E_2=1\,$N$/\mu$m$^2$ 
and Poisson ratios $\nu_1=\nu_2=0.3$, where the subscripts 1 and 2 identify the lower and upper bodies, respectively. 
Choosing the same elastic properties for the two blocks avoids the coupling between the normal and the tangential contact problems, since a frictional constitutive response for the interface is not specified in this test. Using Eqs.~\eqref{eq:compE} and~\eqref{eq:compG}, we end up with a composite Young's modulus $E=0.5495\,$N$/\mu$m$^2$ and a composite Poisson ratio $\nu=-0.3929$ to be used at the interface.

Dirichlet boundary conditions are given by vertical constraints onto the lower side of $\Omega_2$, while a single horizontal constraint is applied at the top-left nodes of both bodies, to avoid rigid body motion. An imposed downward vertical displacement $\Delta$ acts on the upper side of $\Omega_1$, monotonically increasing with a pseudo-time variable to simulate the quasi-static normal contact problem, starting from $0$ up to a maximum value of $3s$, denoting $s$ the root mean square of the surface roughness used to represent the composite topography.
 
The simulations have incorporated three different rough fractal surfaces generated using the \textit{Random Midpoint Displacement} (RMD) algorithm \cite{PB11,peitgen1988science}. The Hurst exponent has been set equal to $H=0.7$, while three resolutions corresponding to $n=6$, $7$ and $8$ have been considered, which implies having $65$, $129$, and $257$ heights per side, respectively. The aim is comparing the computational complexity by increasing the dimension of the contact problem solved by BEM at each Gauss point, and assessing how different coupling strategies affect the accuracy of the contact predictions. 

\begin{figure}
\centering
\includegraphics[width=0.45\textwidth,angle=0]{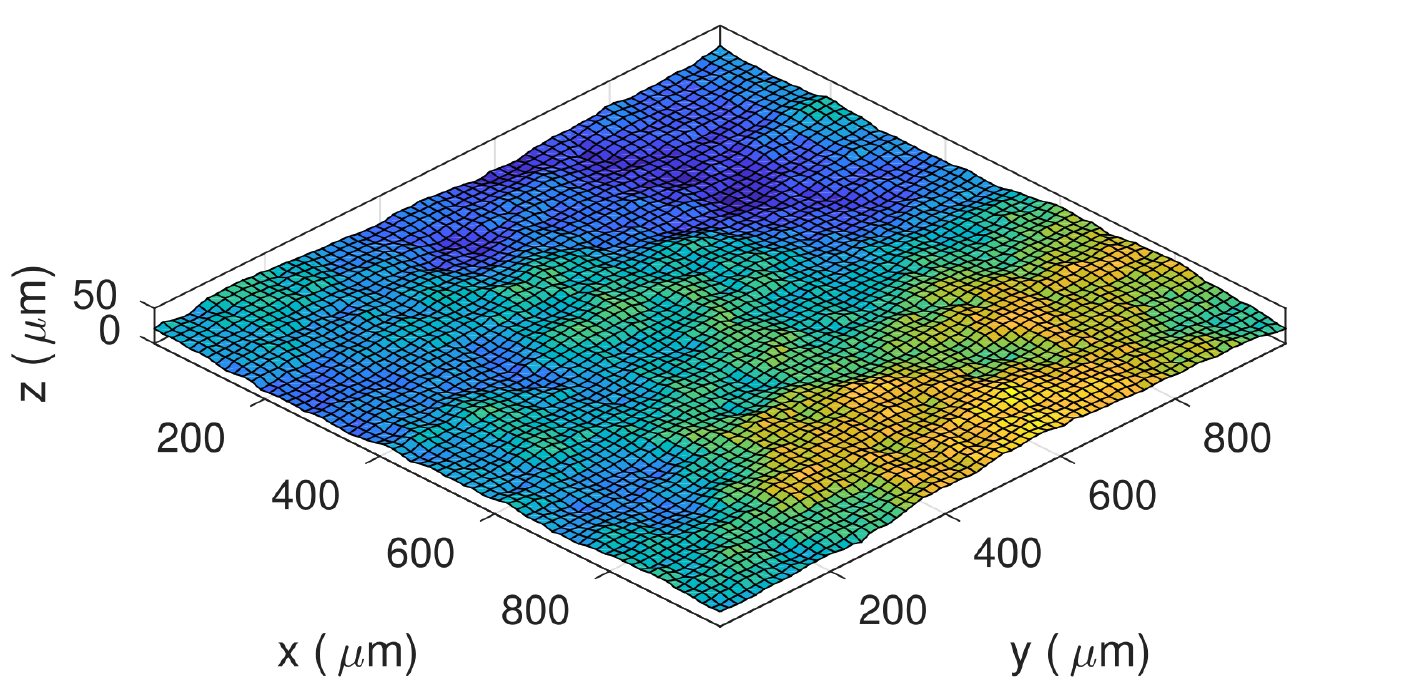}
\caption{Example of a RMD rough surface $(n=6)$.}
\label{fig:surf6}
\end{figure}

For the application of the present method, which hinges on the assumption of scale-separation between the micro and the macro scales, all the rough surfaces input for BEM should be statistically representative of roughness and their lateral size $l$ should be much smaller than the macroscopic lateral size $L$. In the present case, $l=1$ mm, which leads to a ratio $l/L=0.1$. The maximum height of the rough surface is $50\,\mu$m. An example of the generated surface is shown in Fig.~\ref{fig:surf6}. 

\begin{figure}[h]
\centering
\includegraphics[width=0.45\textwidth,angle=0]{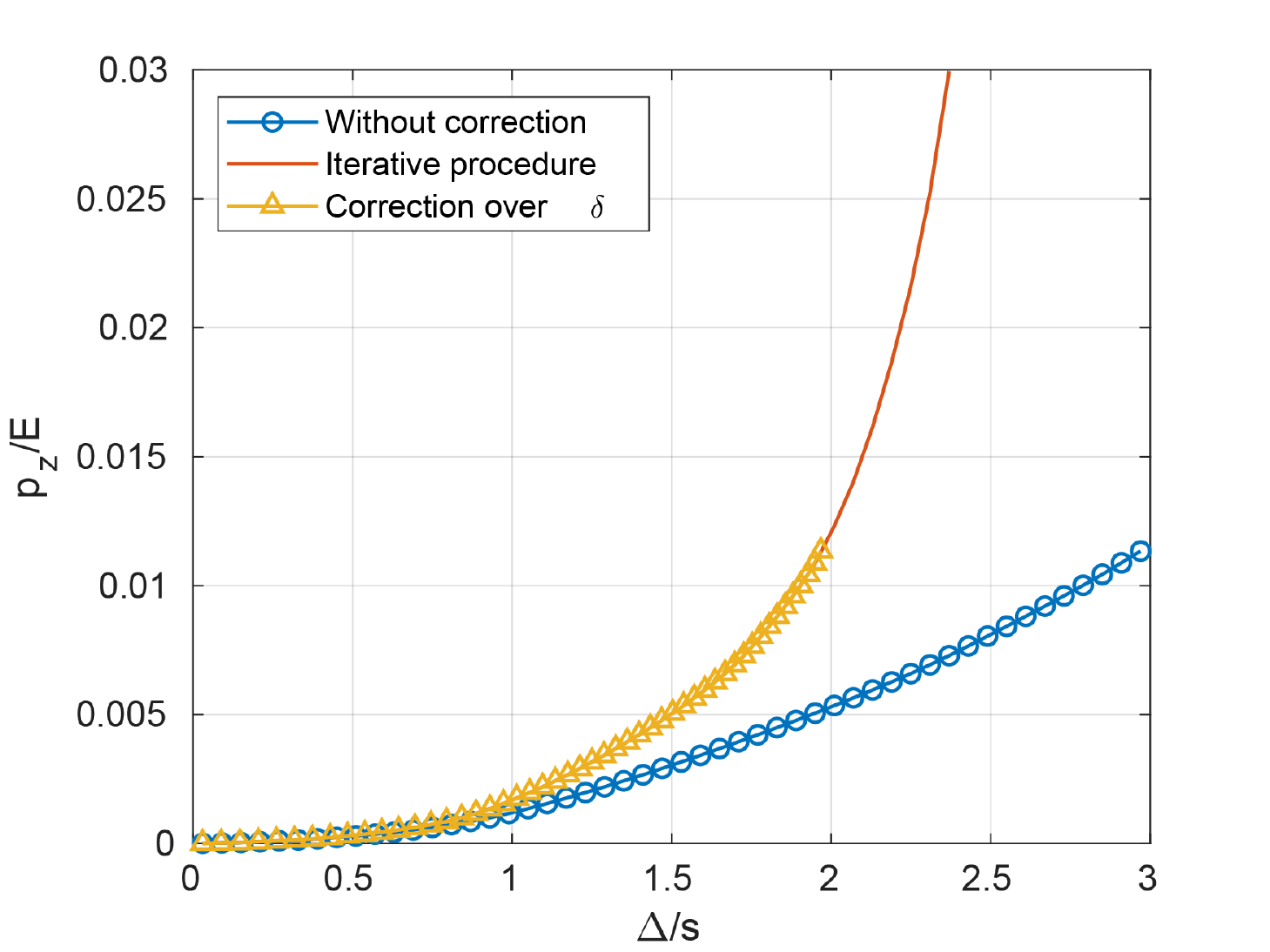}
\caption{Comparison of the gap-pressure curves evaluated using two different procedures with tolerance sets as $1\times 10^{-2}$.}
\label{fig:corr_comp}
\end{figure}

The proposed tolerance value used to control the error in Eq.~\eqref{eq:err} is equal to $1 \times 10^{-2}$ which gives a good accordance between the gap-pressure curves evaluated in the convergence study as shown in Fig.~\ref{fig:corr_comp} for the given dimensionless displacements $\Delta/s$ and the example surface with $n=6$.

The FBEM-QN, FBEM-CQN and FBEM-SAN solution strategies are herein compared in terms of dimensionless force $P/(EA)$ vs. $h^*/s$, where $P$ is the total normal load computed from the sum of the vertical reactions forces at the constrained nodes of the macro-scale finite element model, $E$ is the composite Young modulus, $A$ is the  macro-scale nominal contact area, and $h^*=e_{max}^*-\bar{e}^*-\gnn$ is the actual distance between the flat plane and mean plane of the rough surface. For the FBEM-QN approach, a value of the perturbation $\Delta g_{\mathrm{n},i}=0.01g_{\mathrm{n},i}$ has been chosen.
 
\begin{figure}
\centering
\subfloat[][R-square coefficient.\label{fig:fit_conv}]
{\includegraphics[width=0.45\textwidth]{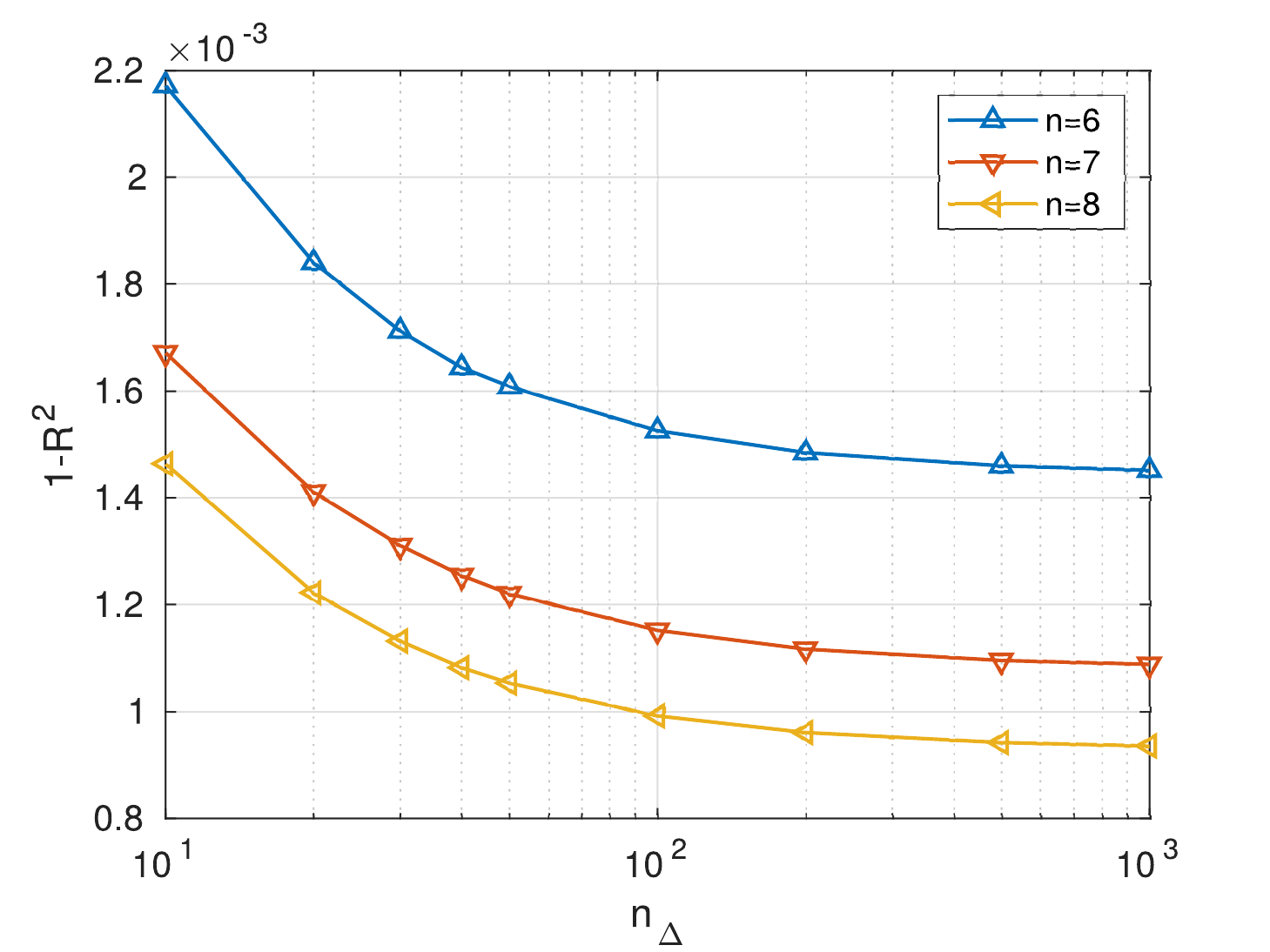}}\\
\subfloat[][Computational time.\label{fig:off_time}]
{\includegraphics[width=0.45\textwidth]{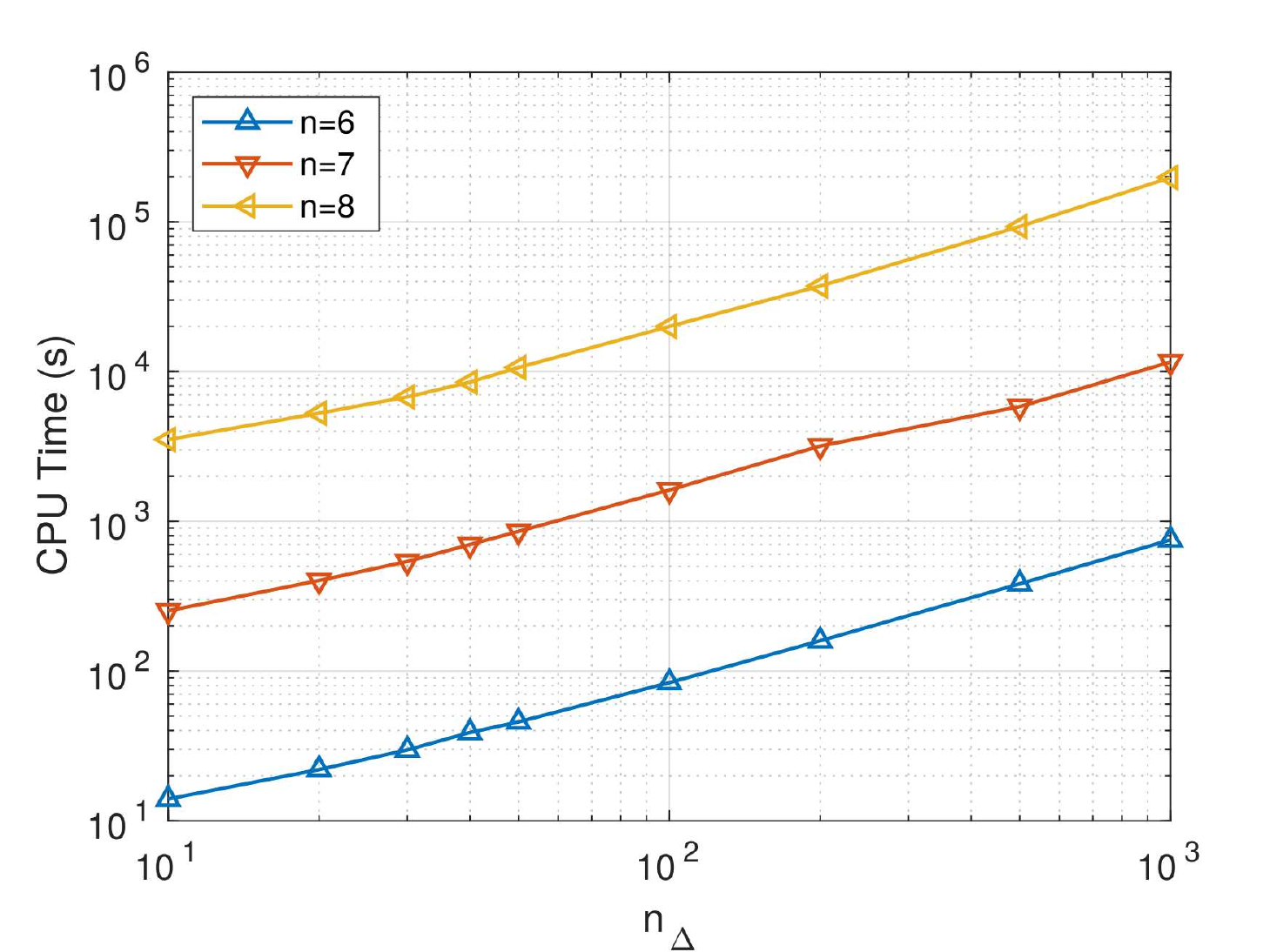}}
\caption{Parametric study over the number of time steps used in the fitting, for the same imposed far field displacement $\Delta$.}
\label{plot:}
\end{figure}

For the FBEM-SAN scheme, the curve used to fit the off-line BEM contact predictions is chosen as a power-law function given by Eq.~\eqref{eq:power} and the fitting has been performed employing MATLAB's built-in \texttt{fitnlm} function\footnote{See \url{https://uk.mathworks.com/help/stats/fitnlm.html} for documentation.} for performing non-linear regressions. The resulting curve coefficients are collected in Tab.~\ref{tab:coeff} for the three different surface resolutions distinguished by the value of $n$, together with the sum of squares due to error ($SSE$), the sum of squares of the regression ($SSR$), the total sum of squares ($SST$) and finally the R-square ($R^2$) coefficients. Improvements in all the estimators can be observed as the resolution gets higher. Another critical point regards the number of time steps $n_\Delta$ to be employed during the off-line computation of the fitting coefficient. Fig.~\ref{fig:fit_conv} and Fig.~\ref{fig:off_time} show, respectively, the variation of $R^2$ and the CPU time required by the off-line stage, with respect to the number of discretization steps. The value of $10^2$ steps, used in the present benchmark example, represents a good trade-off between fitting accuracy and computational time spent during the operation.


\begin{table*}
\centering
\caption{Coefficients of the power-law function $p(\gnn)=a\gnn^b$, together with goodness of fit parameter.}
\begin{tabular}{ccccccc}
\toprule
$n$ & $a~[\mathrm{N}/\mu\mathrm{m}^2]$ & $b$ & $SSE$ & $SSR$ & $SST$ & $R^2$\\ 
\midrule
6 & $1.416\times10^{-06}$ & $2.831$ & $5.677\times10^{-07}$ & $3.773\times10^{-04}$ & $3.722\times10^{-04}$ & $0.9985$\\
7 & $1.240\times10^{-06}$ & $2.862$ & $4.073\times10^{-07}$ & $3.576\times10^{-04}$ & $3.537\times10^{-04}$ & $0.9988$\\
8 & $1.064\times10^{-06}$ & $2.905$ & $3.407\times10^{-07}$ & $3.461\times10^{-04}$ & $3.437\times10^{-04}$ & $0.9990$\\ 
\bottomrule
\label{tab:coeff}
\end{tabular}
\end{table*}

\begin{figure*}
\centering
\subfloat[][FBEM-QN\label{plot:QN}]
{\includegraphics[width=0.47\textwidth]{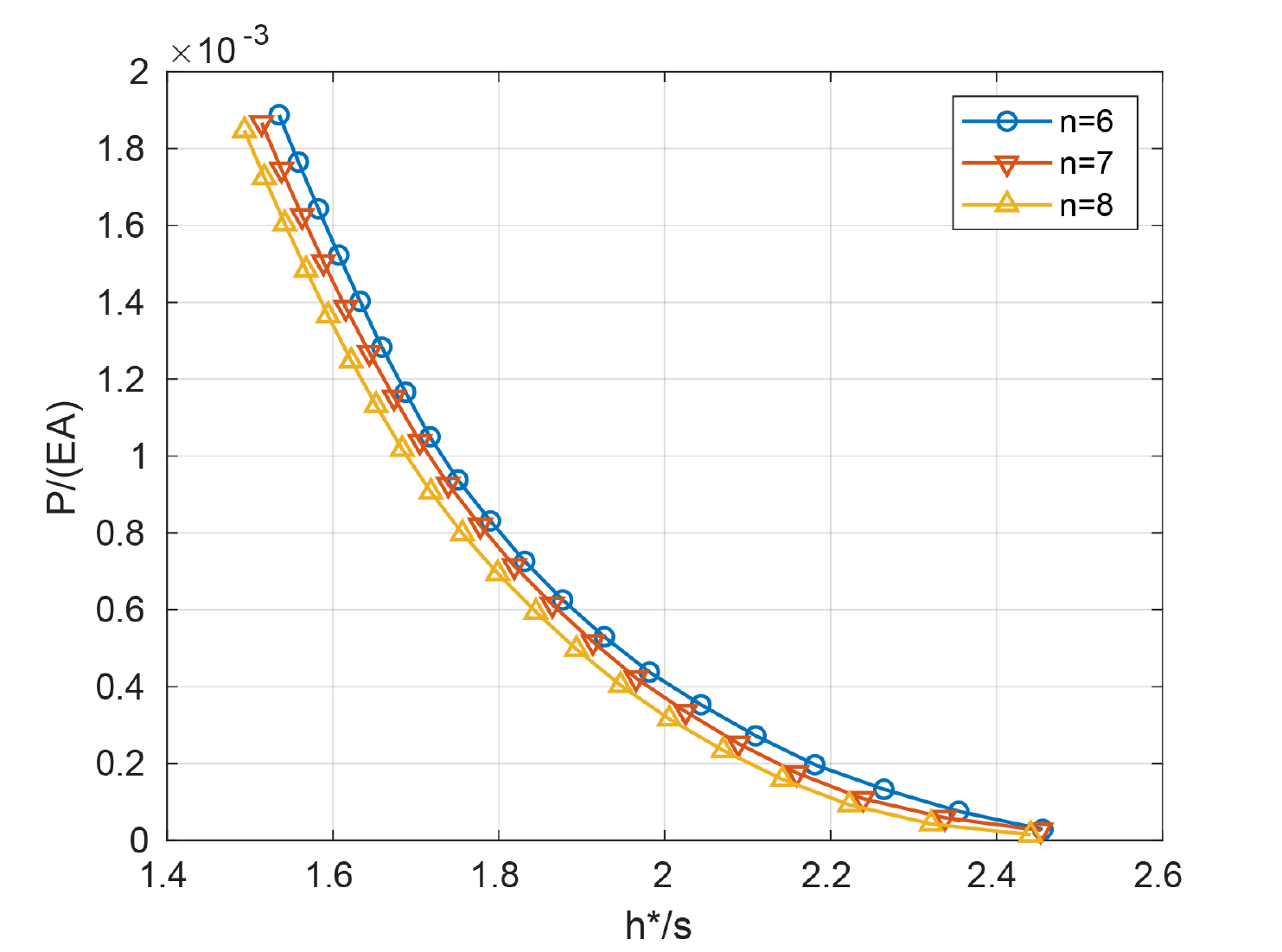}}
\subfloat[][$n=6$\label{plot:6}]
{\includegraphics[width=0.47\textwidth]{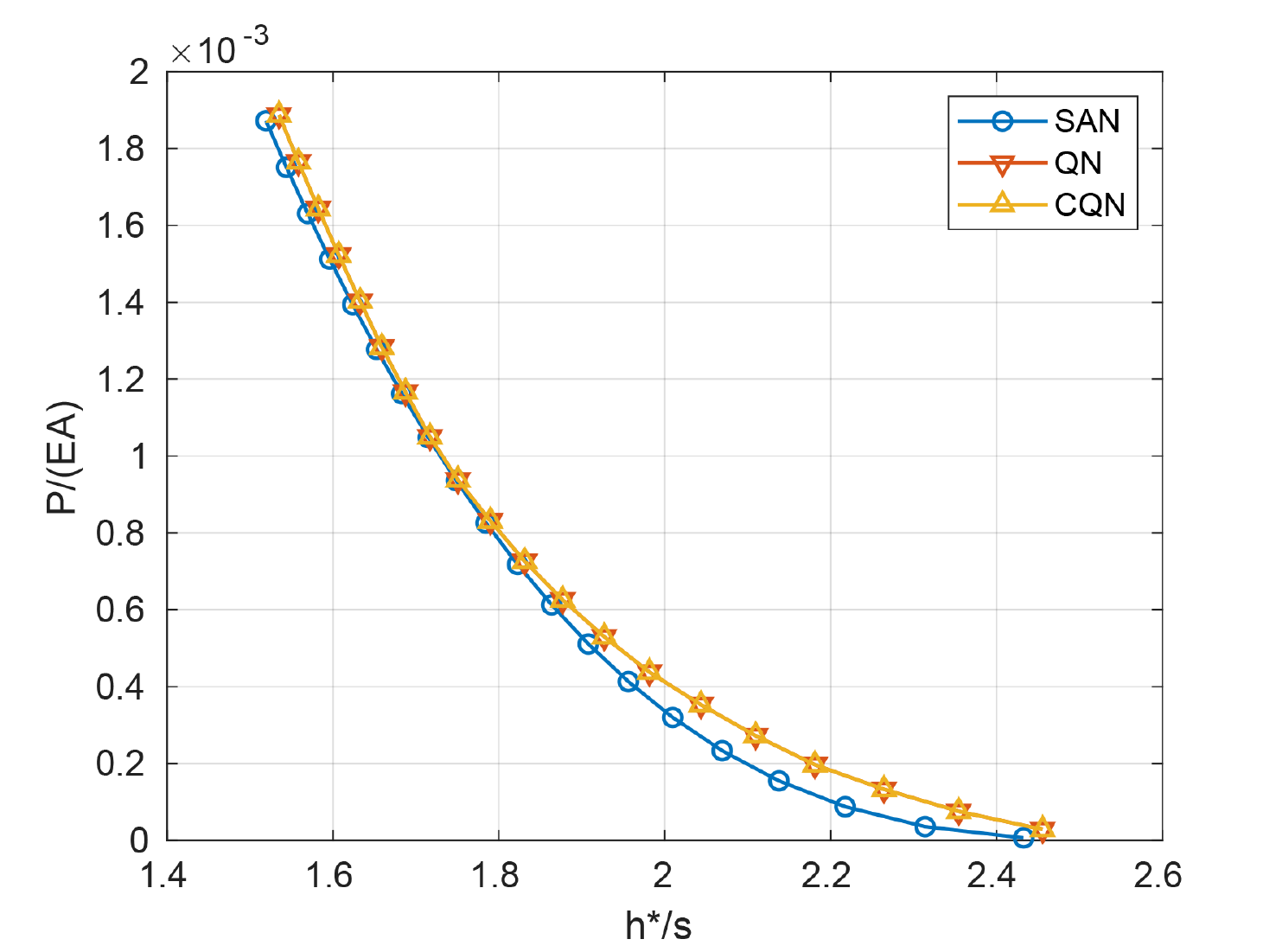}}\\
\subfloat[][FBEM-CQN\label{plot:CQN}]
{\includegraphics[width=0.47\textwidth]{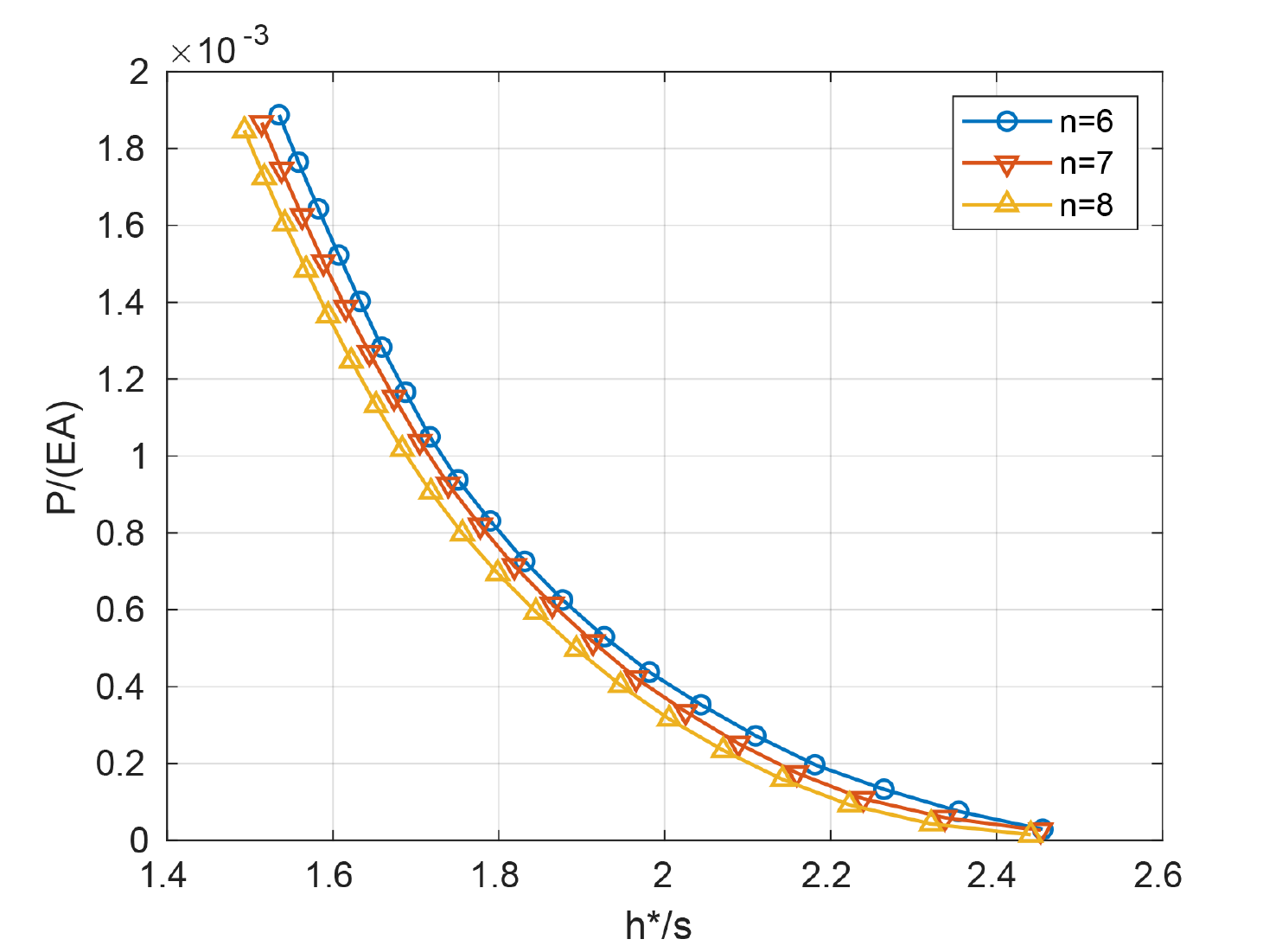}}
\subfloat[][$n=7$\label{plot:7}]
{\includegraphics[width=0.47\textwidth]{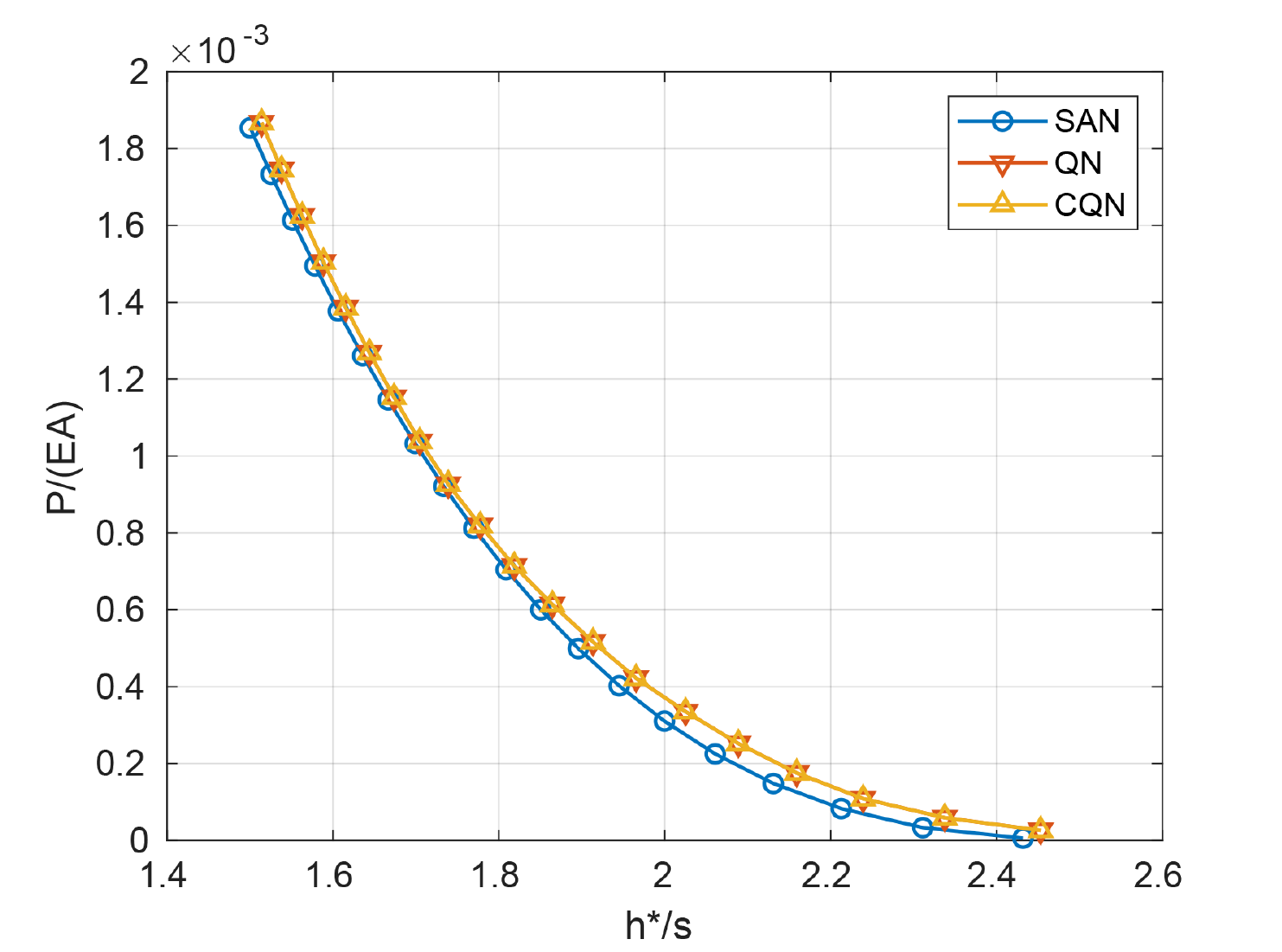}}\\
\subfloat[][FBEM-SAN\label{plot:SAN}]
{\includegraphics[width=0.47\textwidth]{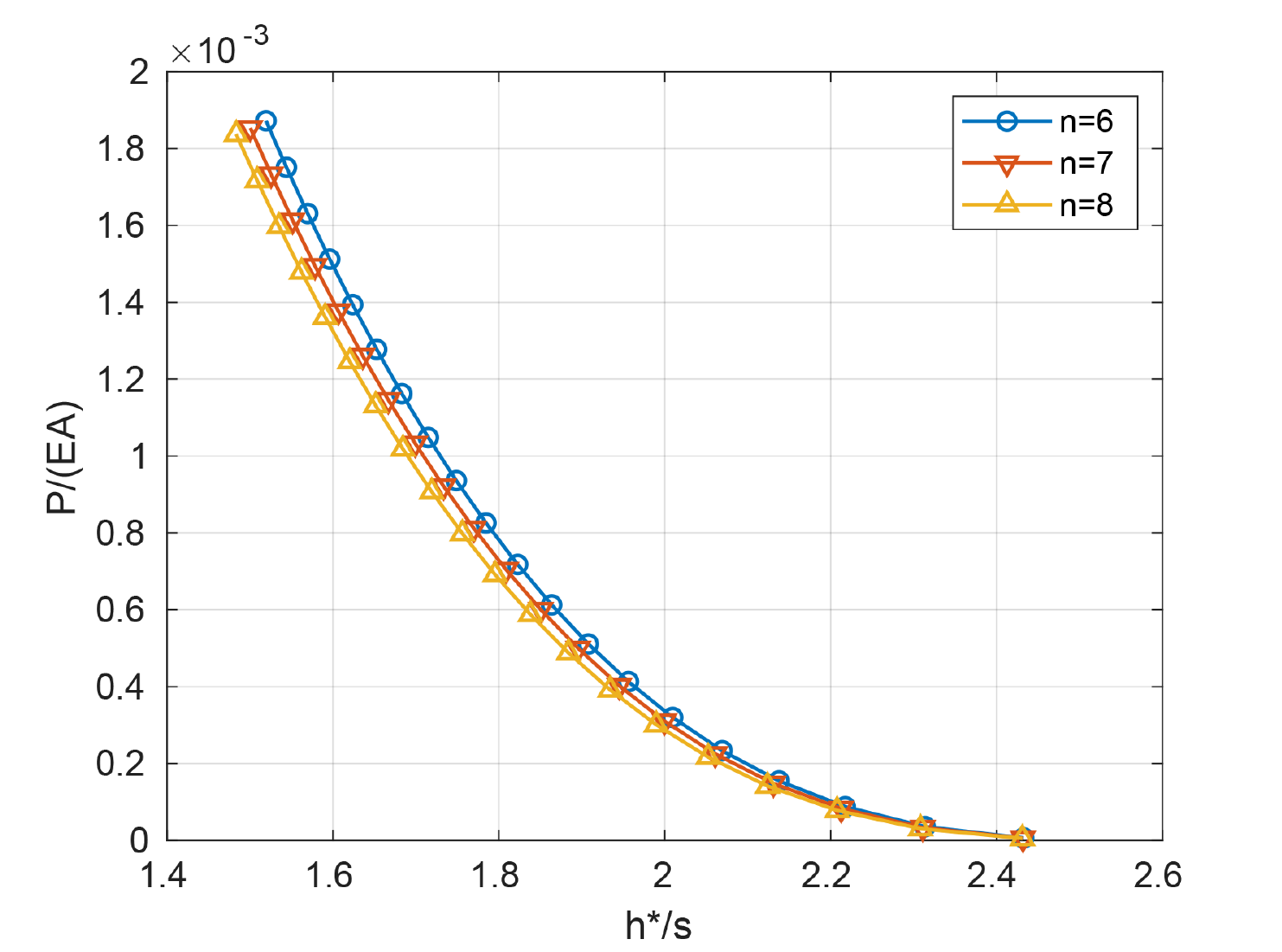}}
\subfloat[][$n=8$\label{plot:8}]
{\includegraphics[width=0.47\textwidth]{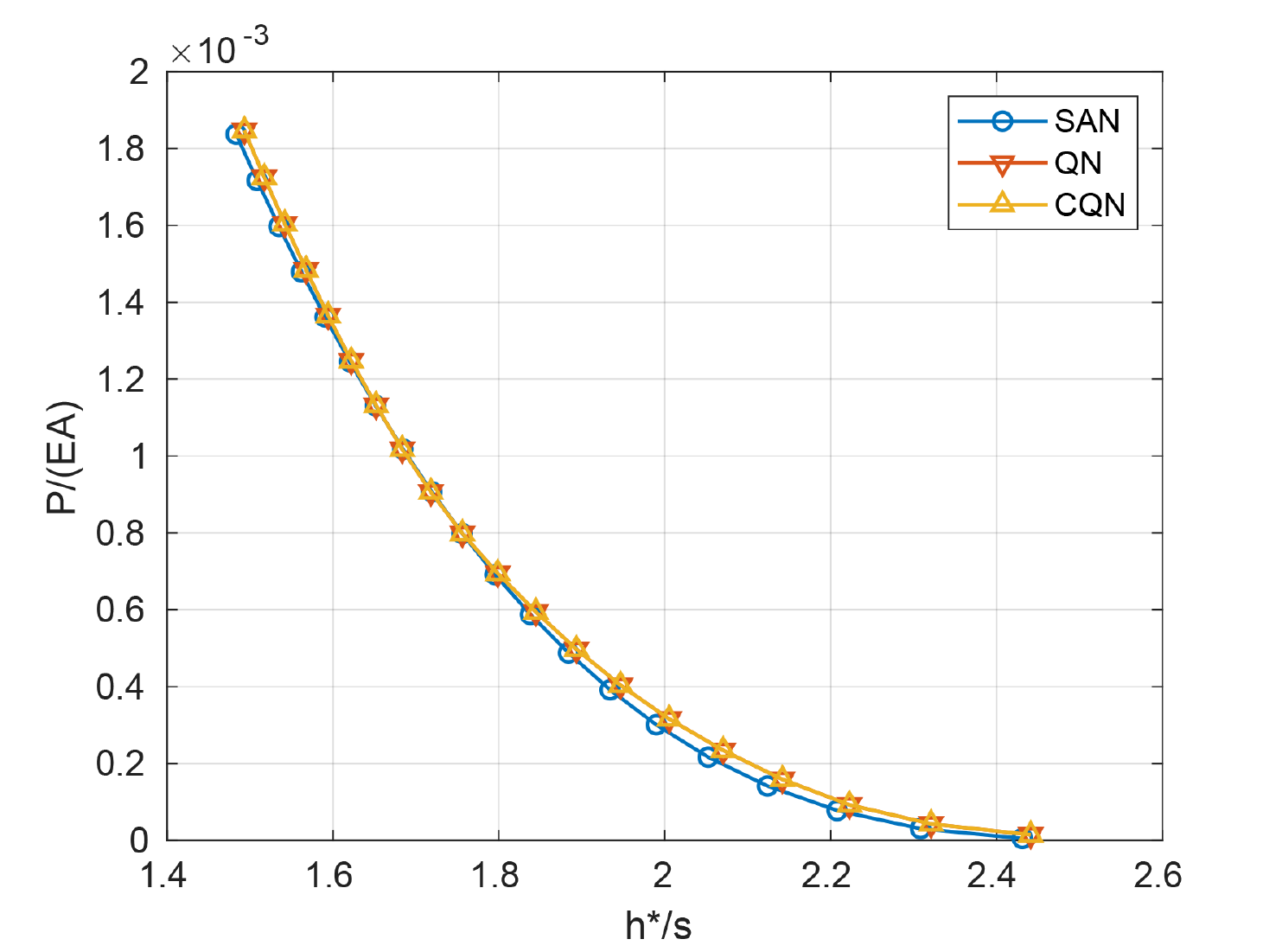}}\\
\caption{Dimensionless contact pressure vs. dimensionless normal gap predictions depending on the solution scheme and the surface resolution.}
\label{plot:h-N}
\end{figure*}

The $P/(EA)$ vs. $h^*/s$ contact predictions are shown for rough surfaces with resolution parameter $n=6$, 7 and 8 in Figs.~\ref{plot:QN}, \ref{plot:CQN} and \ref{plot:SAN}, respectively. The same results are collected for each value of $n$ in Figs.~\ref{plot:6}, \ref{plot:7} and \ref{plot:8} to compare FBEM-QN, FBEM-CQN and FBEM-SAN schemes. Overall, we notice that the three approaches provide almost coincident results for the highest surface resolution (surface with $n=8$), while the semi-analytical scheme leads to slightly different predictions for lower resolutions (surfaces with $n=6$ and $n=7$). As anticipated before, the reason for that is related to the power-law function used to approximate the contact response in the FBEM-SAN scheme, which does not exactly reproduce the actual BEM contact response for coarse meshes or for large separations, being affected by a scatter induced by statistics of extremes of the asperity height distribution.

\begin{figure*}
\centering
\subfloat[][FBEM-QN\label{plot:h-Cmat-QN}]
{\includegraphics[width=0.47\textwidth]{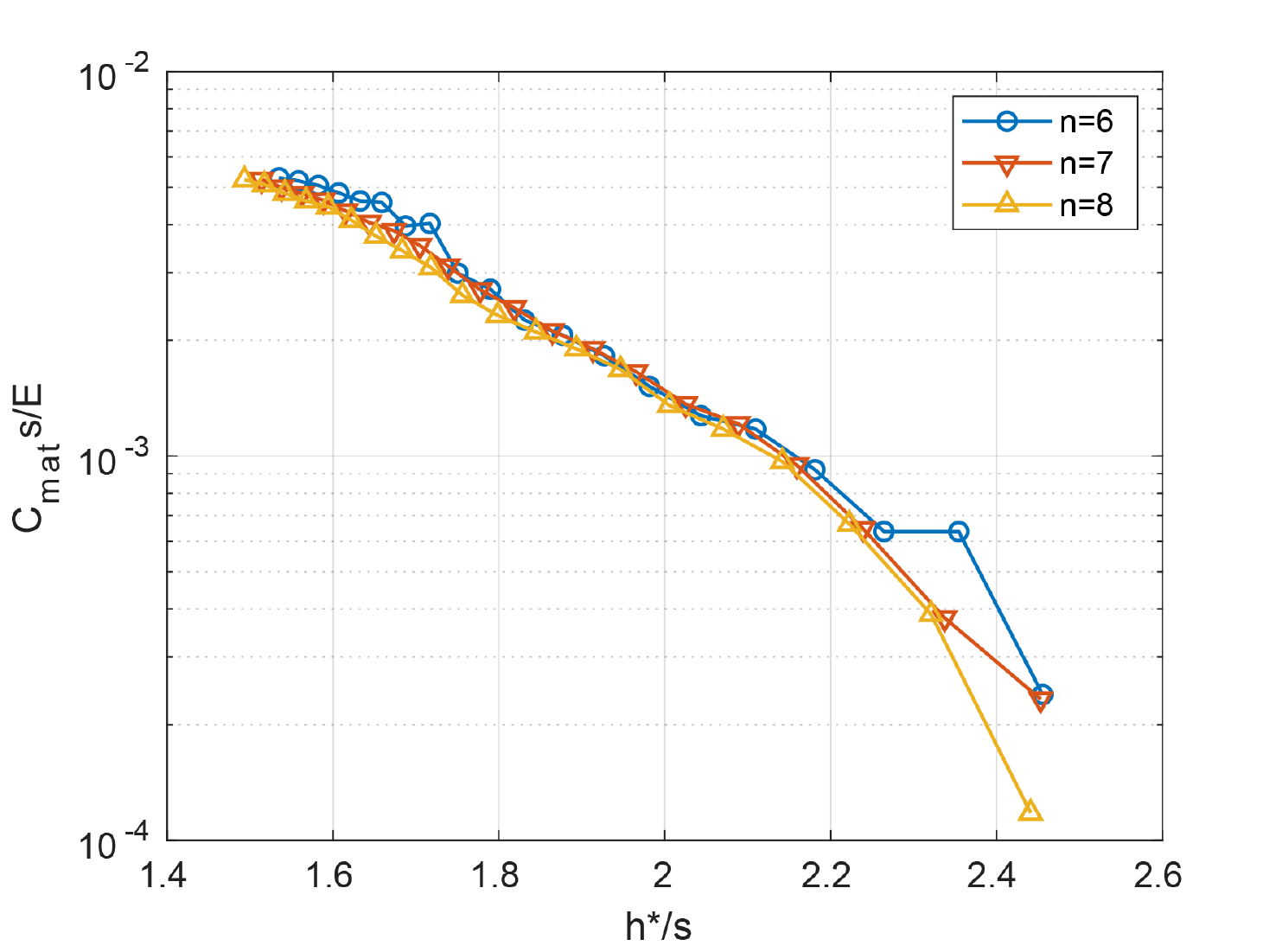}}
\subfloat[][$n=6$\label{plot:h-Cmat-6}]
{\includegraphics[width=0.47\textwidth]{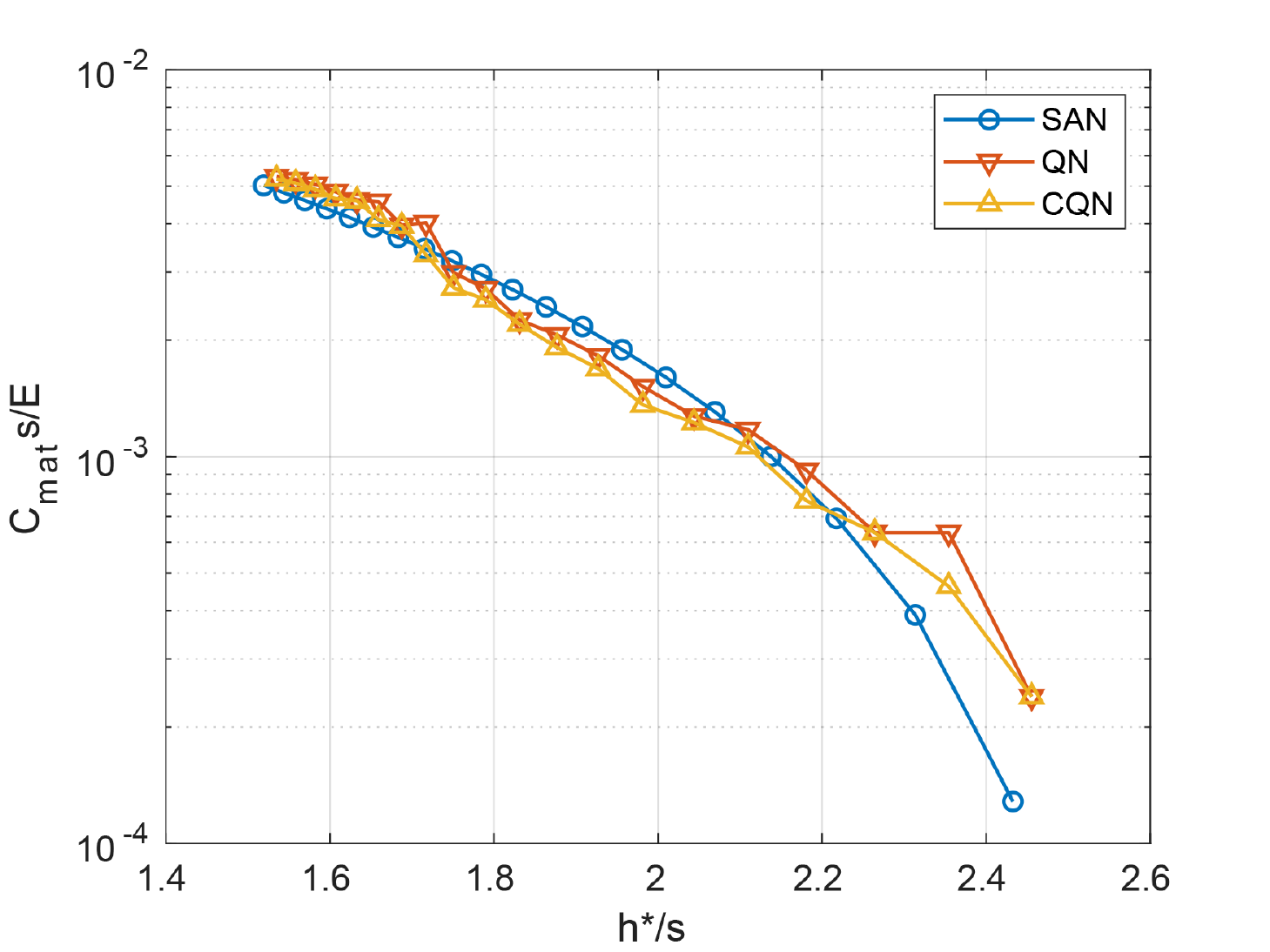}}\\
\subfloat[][FBEM-CQN\label{plot:h-Cmat-CQN}]
{\includegraphics[width=0.47\textwidth]{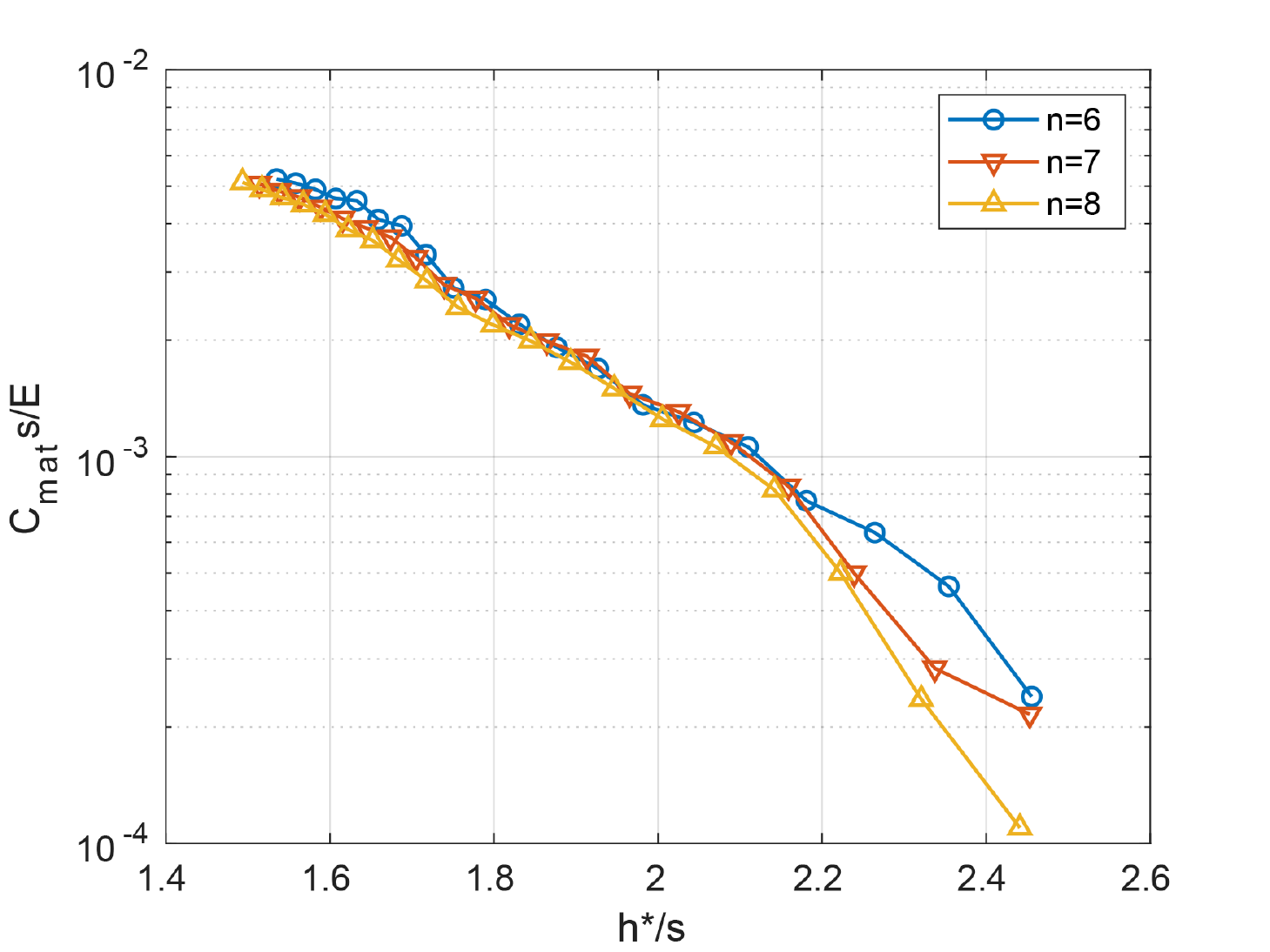}}
\subfloat[][$n=7$\label{plot:h-Cmat-7}]
{\includegraphics[width=0.47\textwidth]{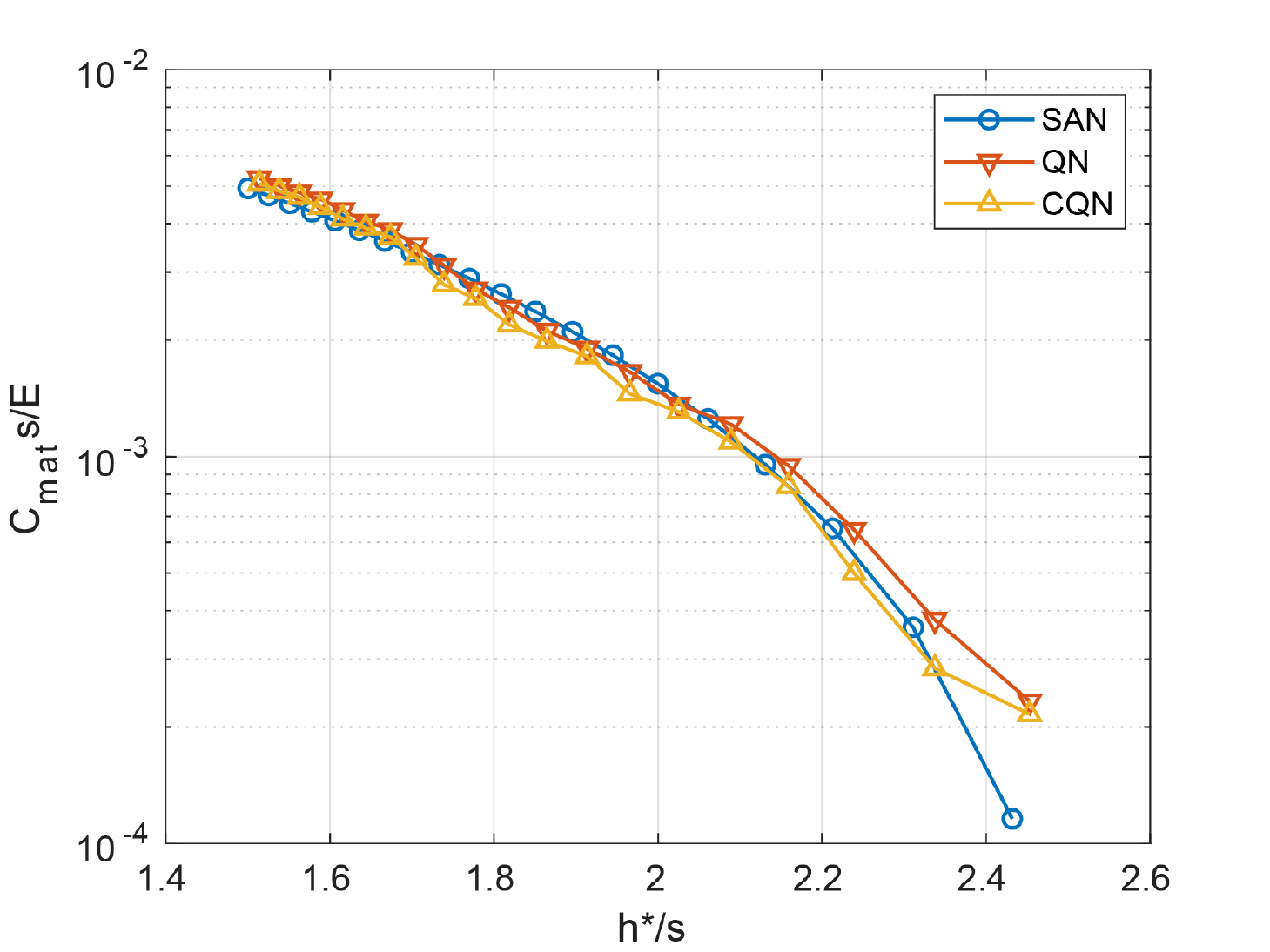}}\\
\subfloat[][FBEM-SAN\label{plot:h-Cmat-SAN}]
{\includegraphics[width=0.47\textwidth]{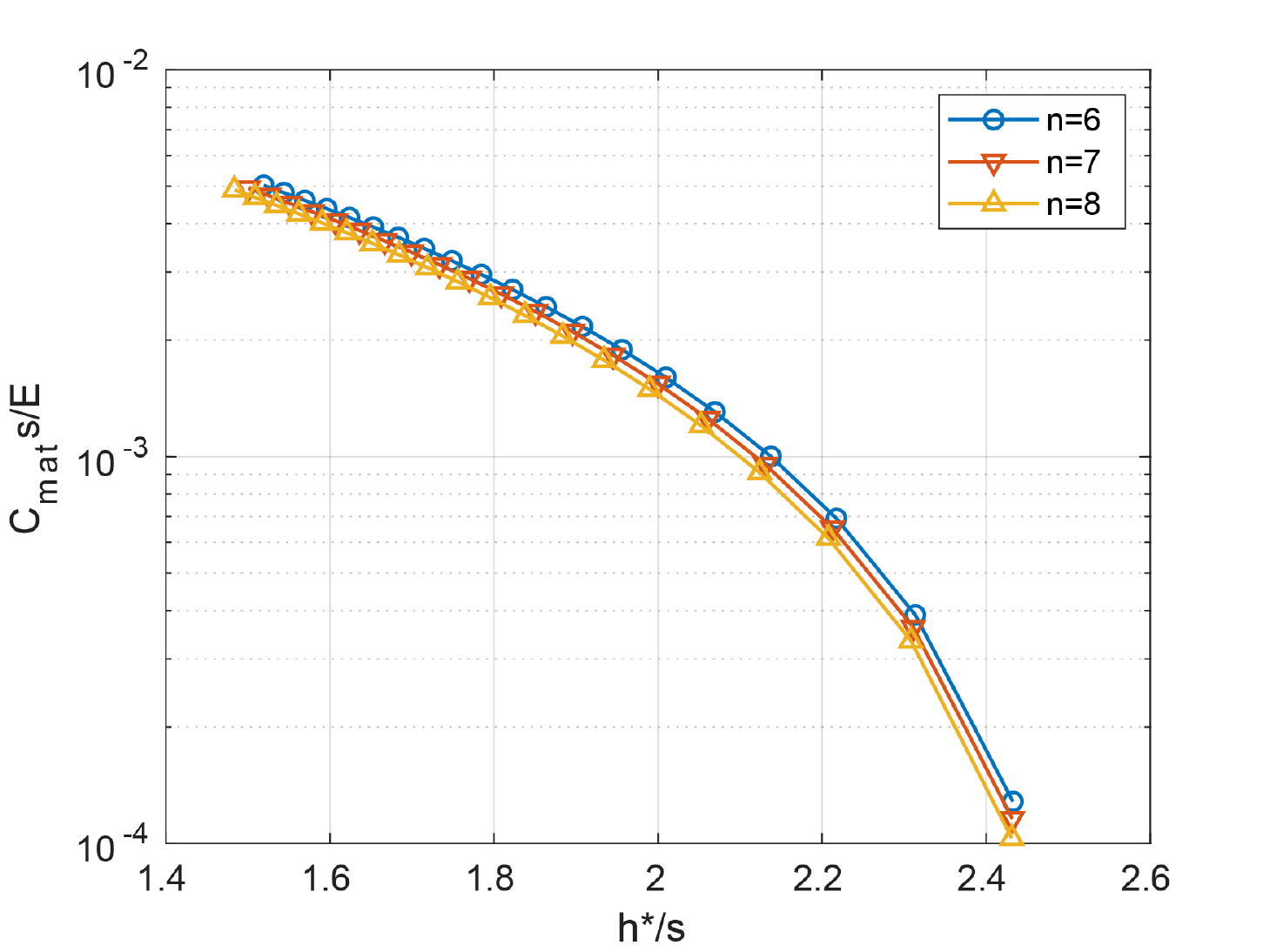}}
\subfloat[][$n=8$\label{plot:h-Cmat-8}]
{\includegraphics[width=0.47\textwidth]{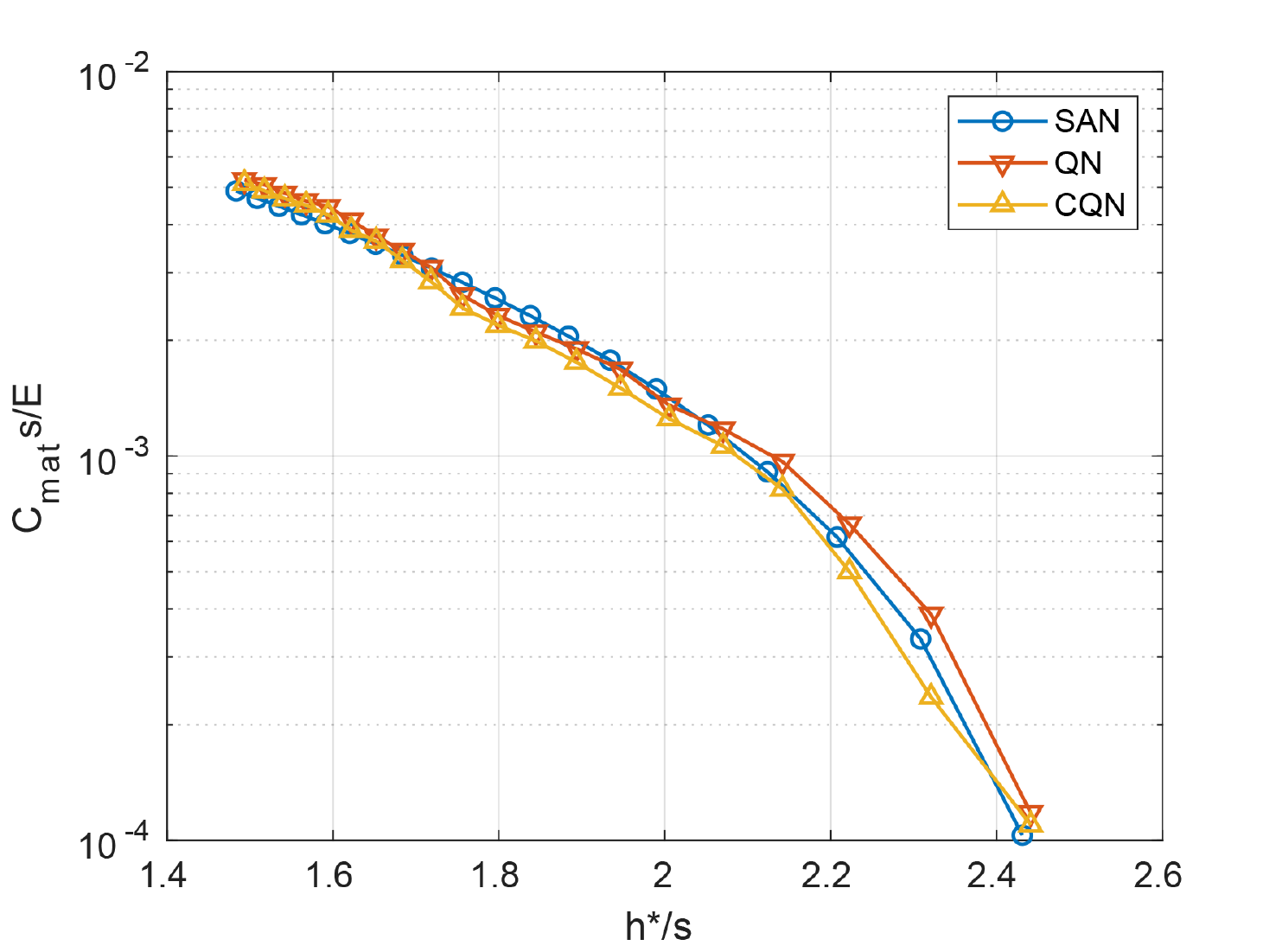}}\\
\caption{Dimensionless contact stiffness vs. dimensionless normal gap predictions depending on the solution scheme and the surface resolution.}
\label{plot:h-Cmat}
\end{figure*}

This trend is even more evident by examining the dimensionless normal contact stiffness $C_\mathrm{mat}s/E$ vs. the dimensionless normal gap $h^*/s$ depending on the resolution parameter $n=6$, 7 and 8 shown in Figs.~\ref{plot:h-Cmat-QN}, \ref{plot:h-Cmat-CQN} and \ref{plot:h-Cmat-SAN}, respectively. The same results are again collected for each $n$ in Figs.~\ref{plot:h-Cmat-6}, \ref{plot:h-Cmat-7} and \ref{plot:h-Cmat-8} to compare FBEM-QN, FBEM-CQN and FBEM-SAN schemes. Overall, we notice that the three approaches provide almost coincident results for the highest surface resolution (surface with $n=8$) and for the low-separations regime. The smoother response predicted by the FBEM-SAN scheme for coarse surfaces and high separations is primarily due to the artificial smoothing of the actual contact response introduced by the power-law best-fitting equation. For large separations, the actual contact behaviour is governed by few asperities in contact and therefore the contact response should present oscillations and a non-smooth behaviour. By increasing the number of contact spots (increasing the pressure or the surface resolution), the collective response tends to be much more stable and smoother, and the power-law best-fit approximation becomes much more reliable.

\begin{figure*}
\centering
\subfloat[][FBEM-QN\label{plot:ite-residual-QN}]
{\includegraphics[width=0.47\textwidth]{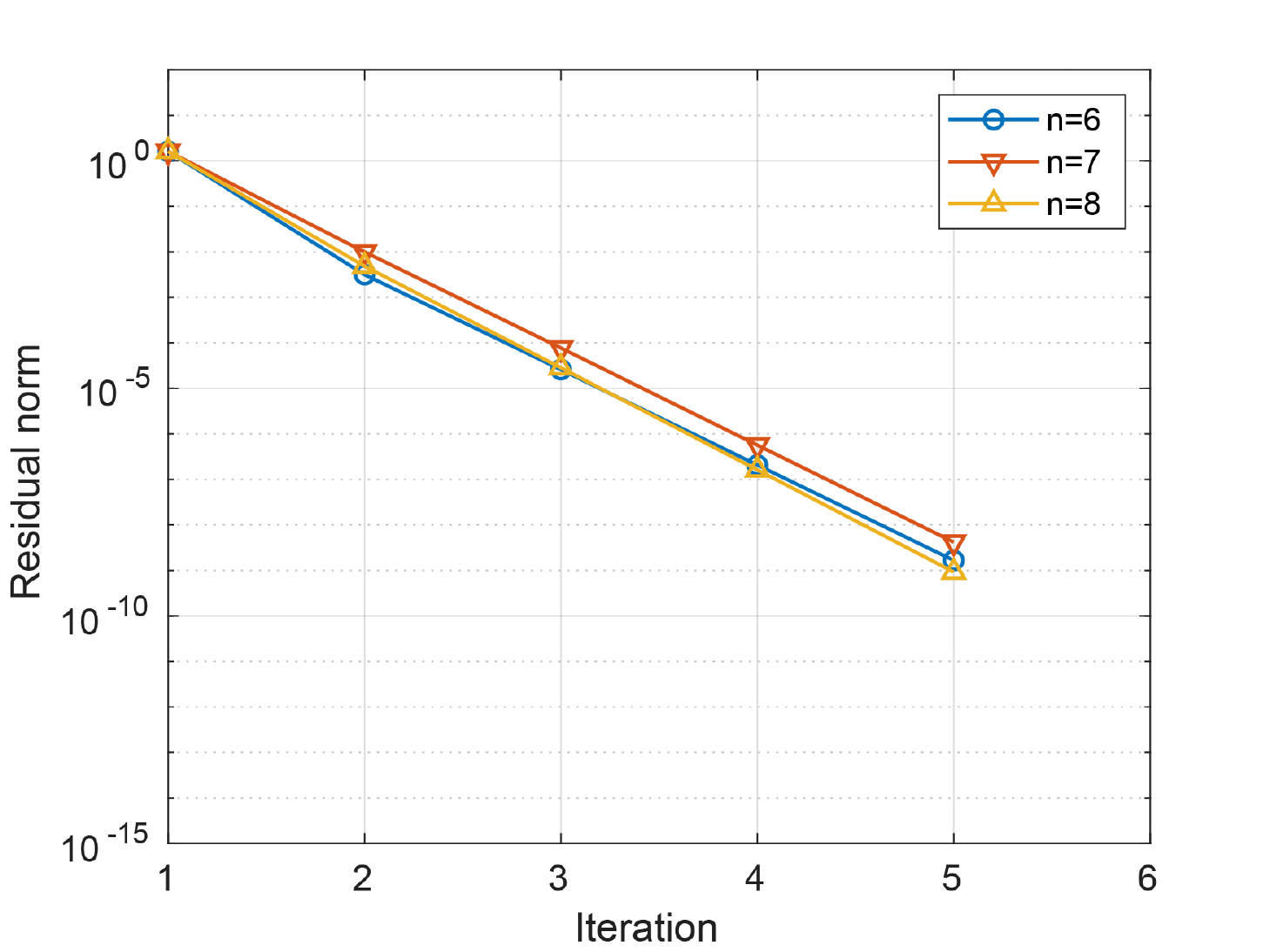}}
\subfloat[][$n=6$\label{plot:ite-residual-6}]
{\includegraphics[width=0.47\textwidth]{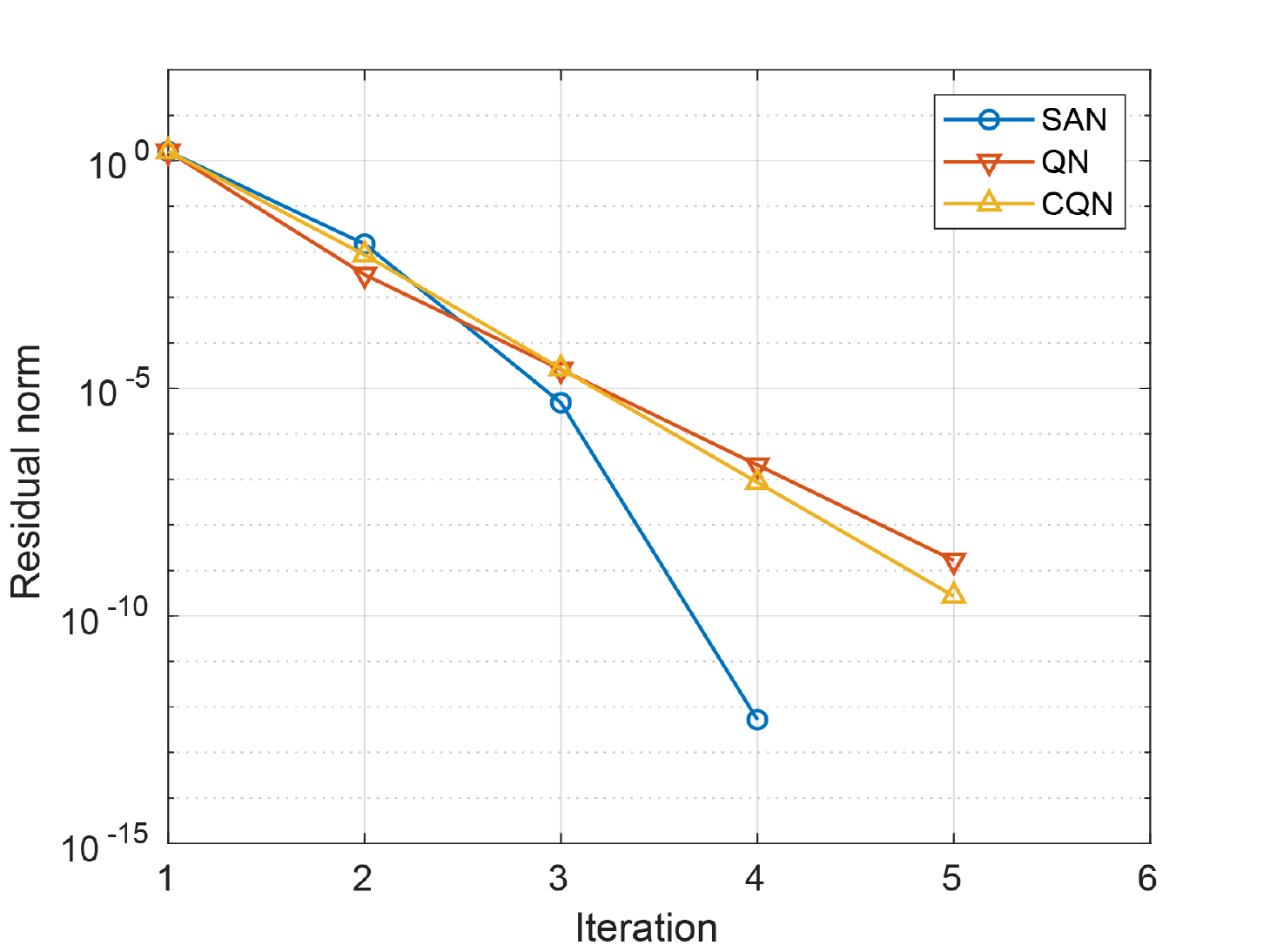}}\\
\subfloat[][FBEM-CQN\label{plot:ite-residual-CQN}]
{\includegraphics[width=0.47\textwidth]{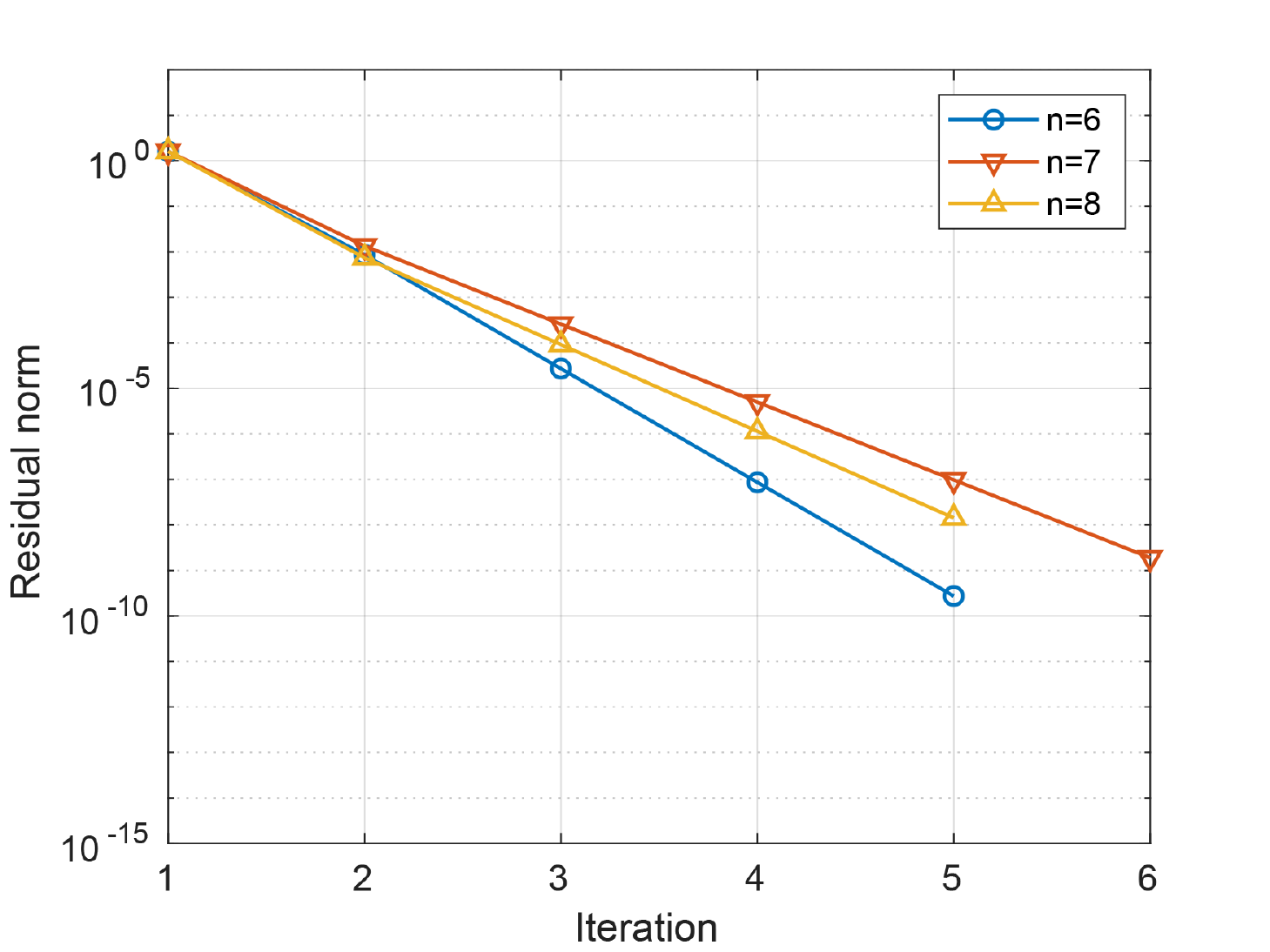}}
\subfloat[][$n=7$\label{plot:ite-residual-7}]
{\includegraphics[width=0.47\textwidth]{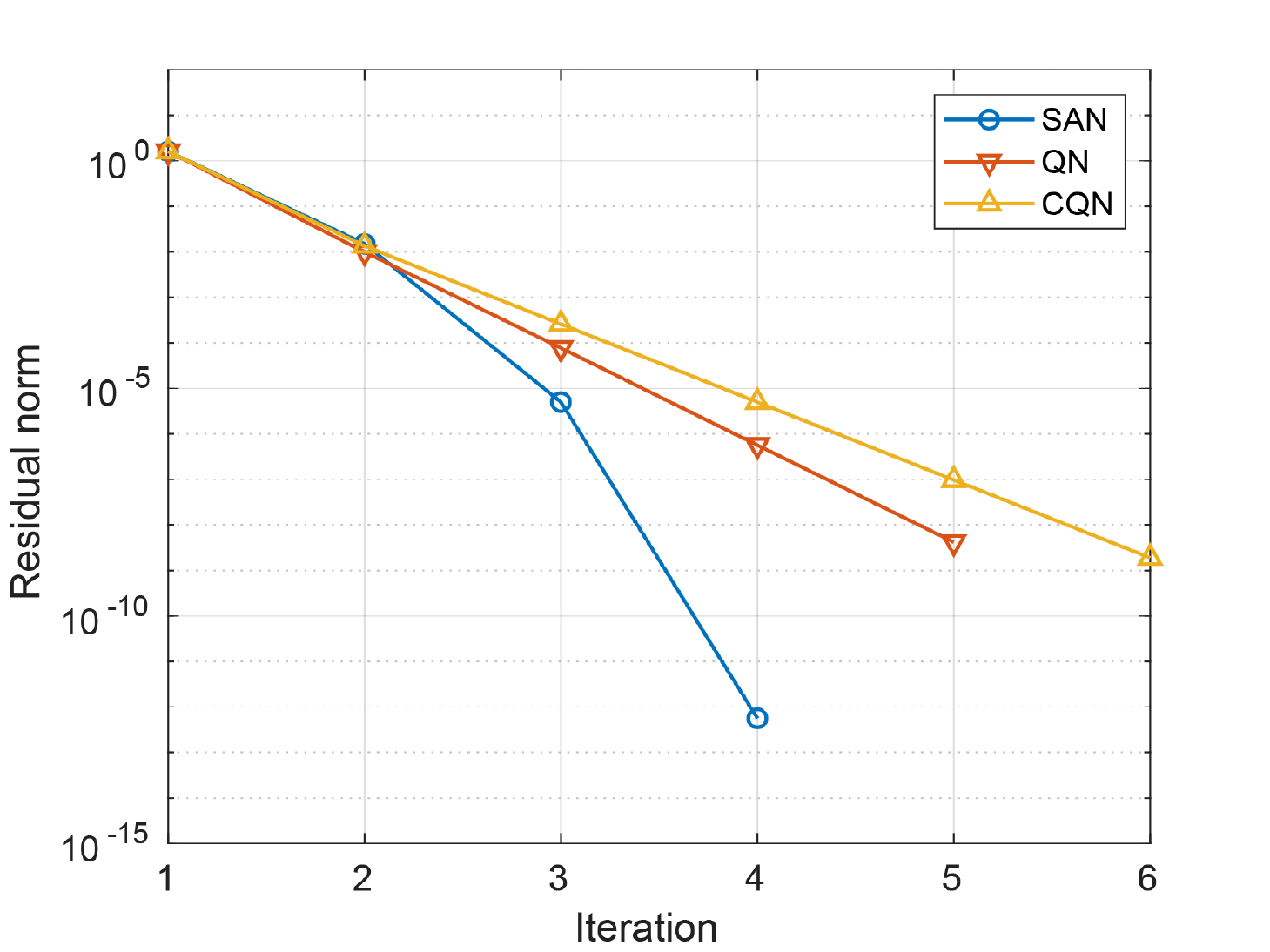}}\\
\subfloat[][FBEM-SAN\label{plot:ite-residual-SAN}]
{\includegraphics[width=0.47\textwidth]{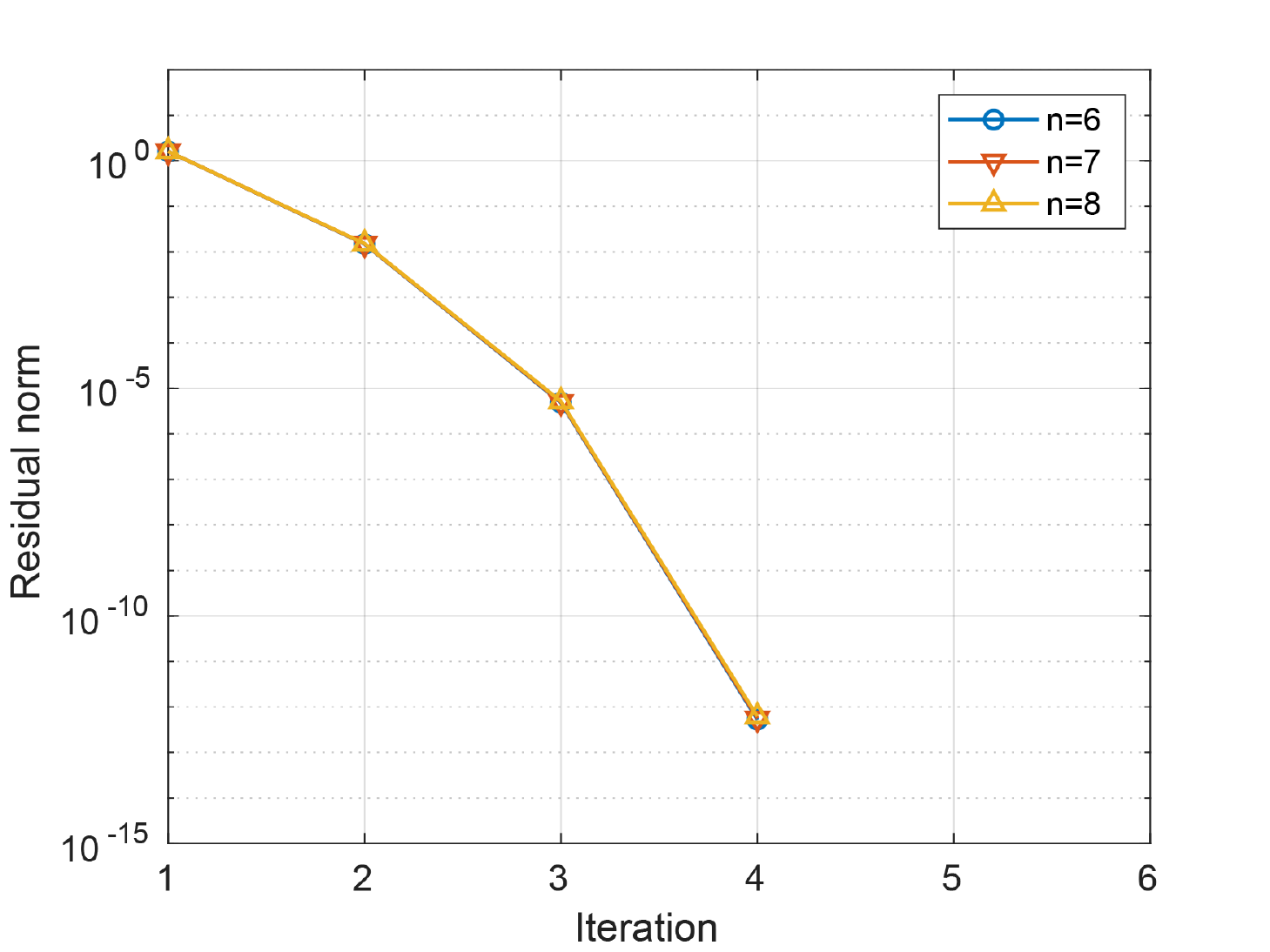}}
\subfloat[][$n=8$\label{plot:ite-residual-8}]
{\includegraphics[width=0.47\textwidth]{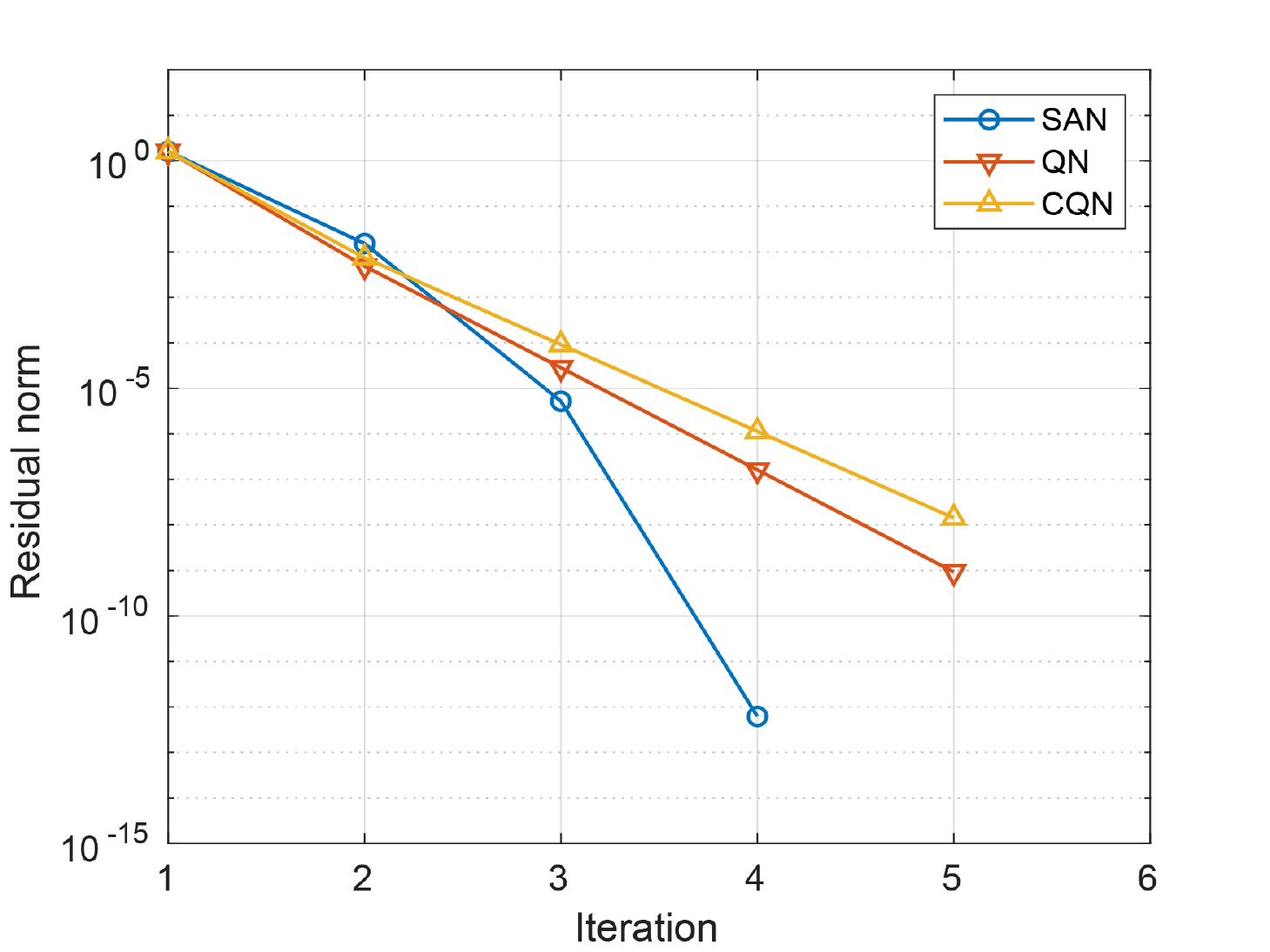}}\\
\caption{Residual norm vs. iteration step depending on the solution scheme and the surface resolution.}
\label{plot:ite-res}
\end{figure*}

The evolution of the residual norm vs. the number of iterations of the numerical scheme used to solve the set of nonlinear algebraic equations  is highlighted in Figs.~\ref{plot:ite-residual-6}, \ref{plot:ite-residual-7} and \ref{plot:ite-residual-8}, for the FBEM-QN, FBEM-CQN and FBEM-SAN solution strategies applied to surfaces with different resolution parameter $n$. Furthermore, figures \ref{plot:ite-residual-QN}, \ref{plot:ite-residual-CQN} and \ref{plot:ite-residual-SAN} compare the convergence rate of the three numerical strategies for the same surface resolution. These results correspond to the last time-step ($\Delta= 3 s$), with a convergence tolerance of $1\times 10^{-9}$. As expected, the FBEM-SAN displays a quadratic convergence, regardless of the resolution, since the tangent stiffness is computed exactly from the derivative of the pressure-separation relation, which is given in analytic form. The FBEM-QN and FBEM-CQN display a slower convergence rate than FBEM-SAN, requiring at least one iteration more then the semi-analytic approach, due to the numerical approximation of the tangent stiffness matrix. 


\begin{figure*}
\centering
\subfloat[][FBEM-QN\label{plot:h-CPU-QN}]
{\includegraphics[width=0.47\textwidth]{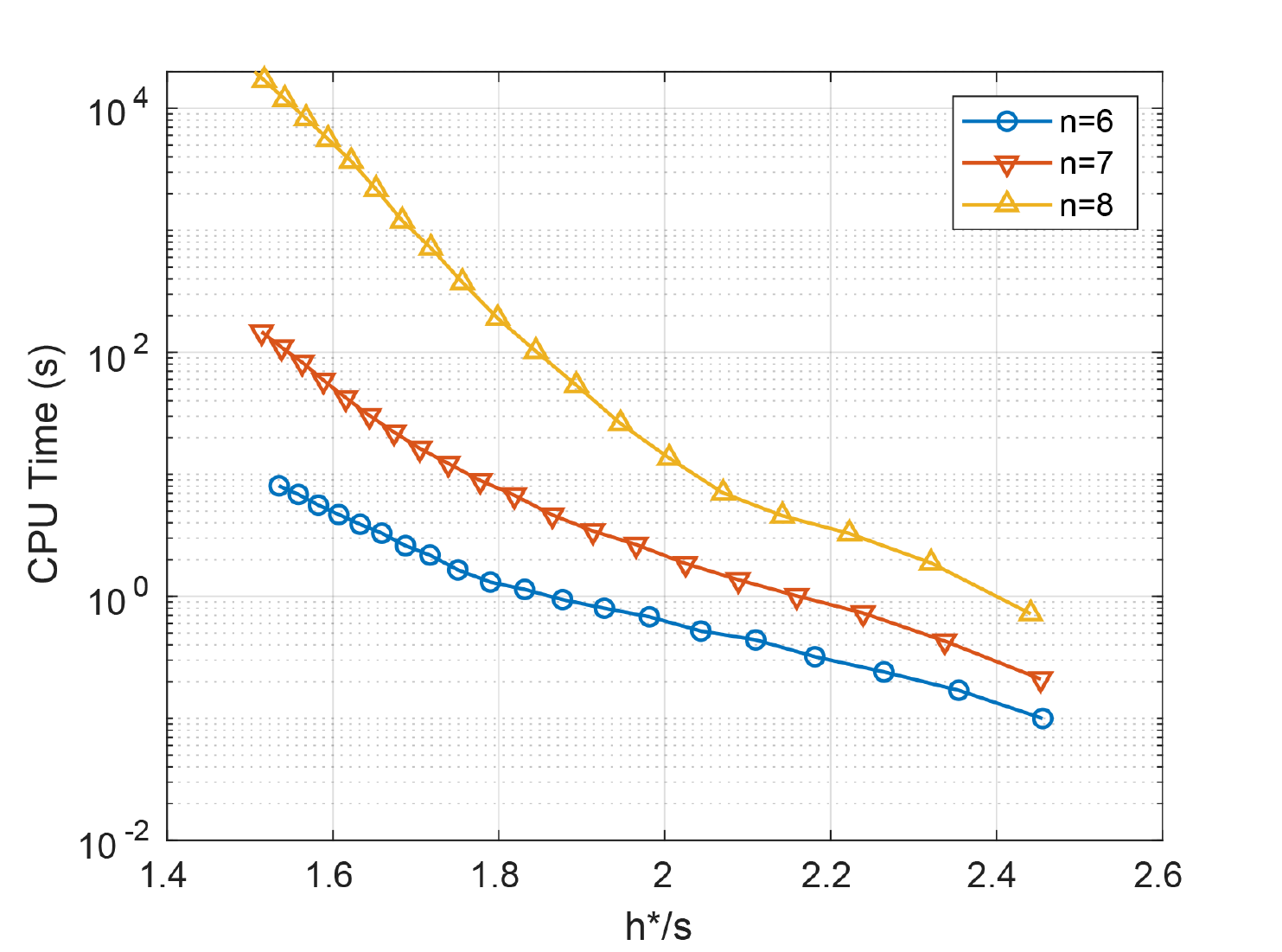}}
\subfloat[][$n=6$\label{plot:h-CPU-6}]
{\includegraphics[width=0.47\textwidth]{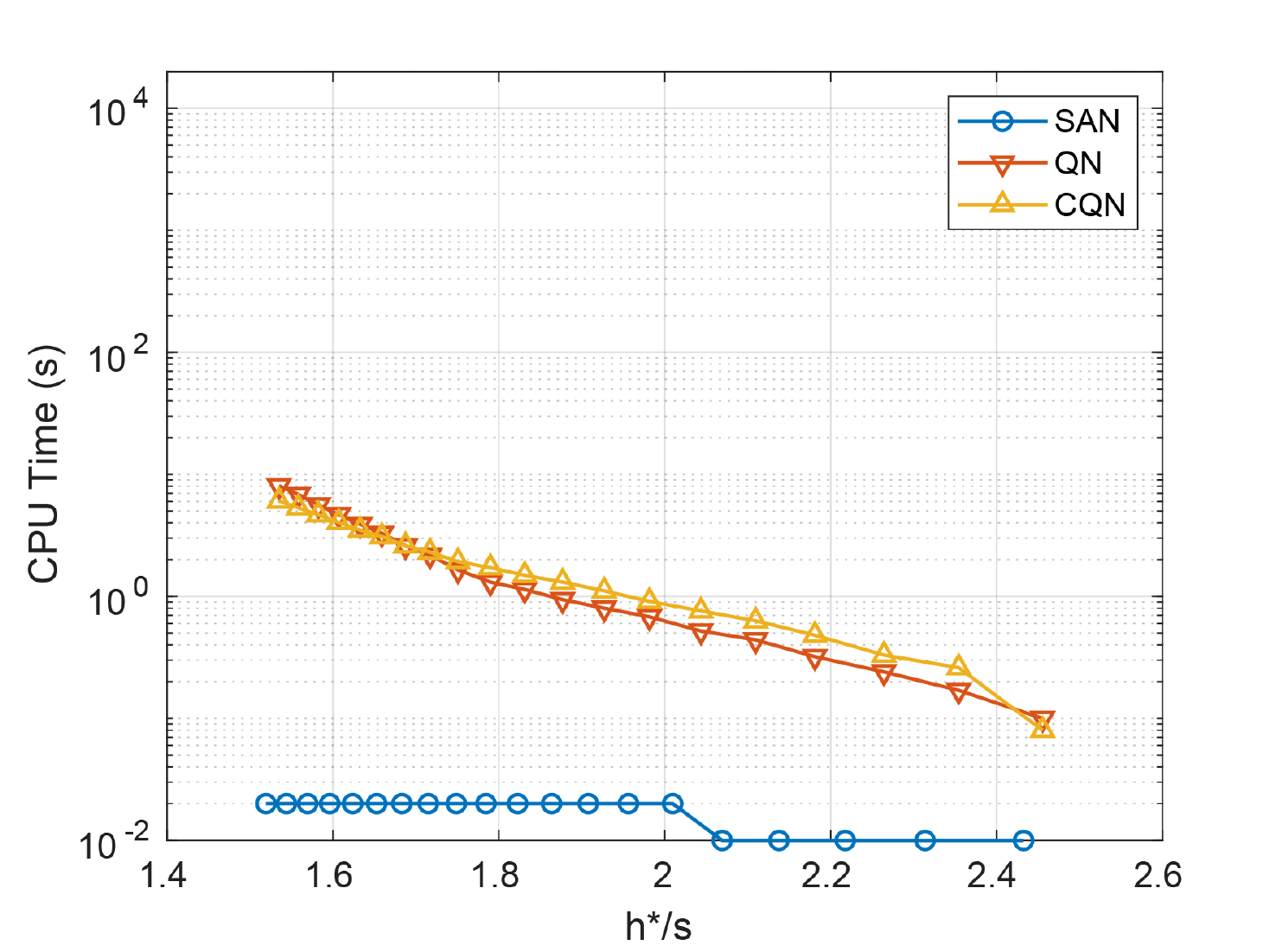}}\\
\subfloat[][FBEM-CQN\label{plot:h-CPU-CQN}]
{\includegraphics[width=0.47\textwidth]{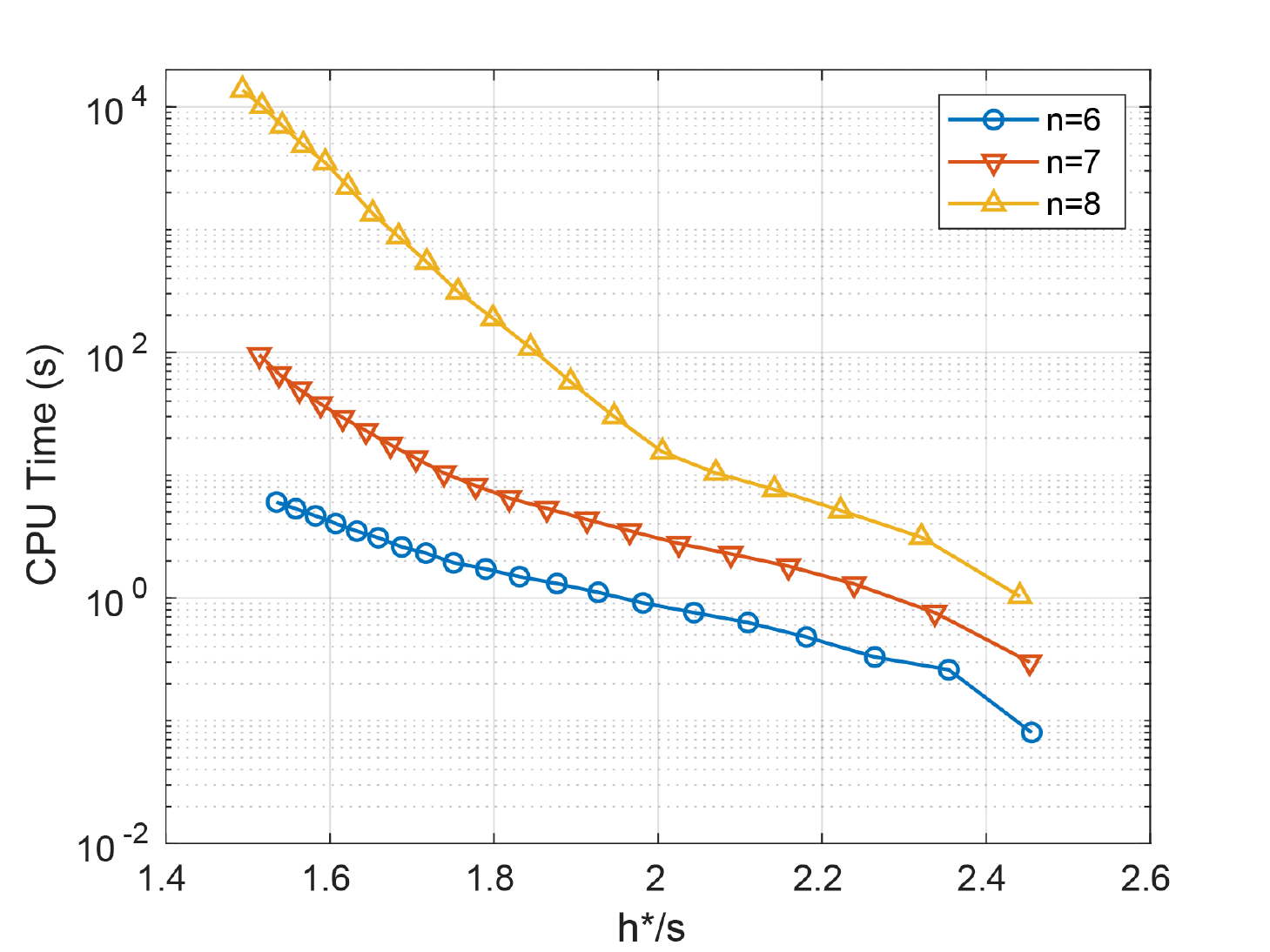}}
\subfloat[][$n=7$\label{plot:h-CPU-7}]
{\includegraphics[width=0.47\textwidth]{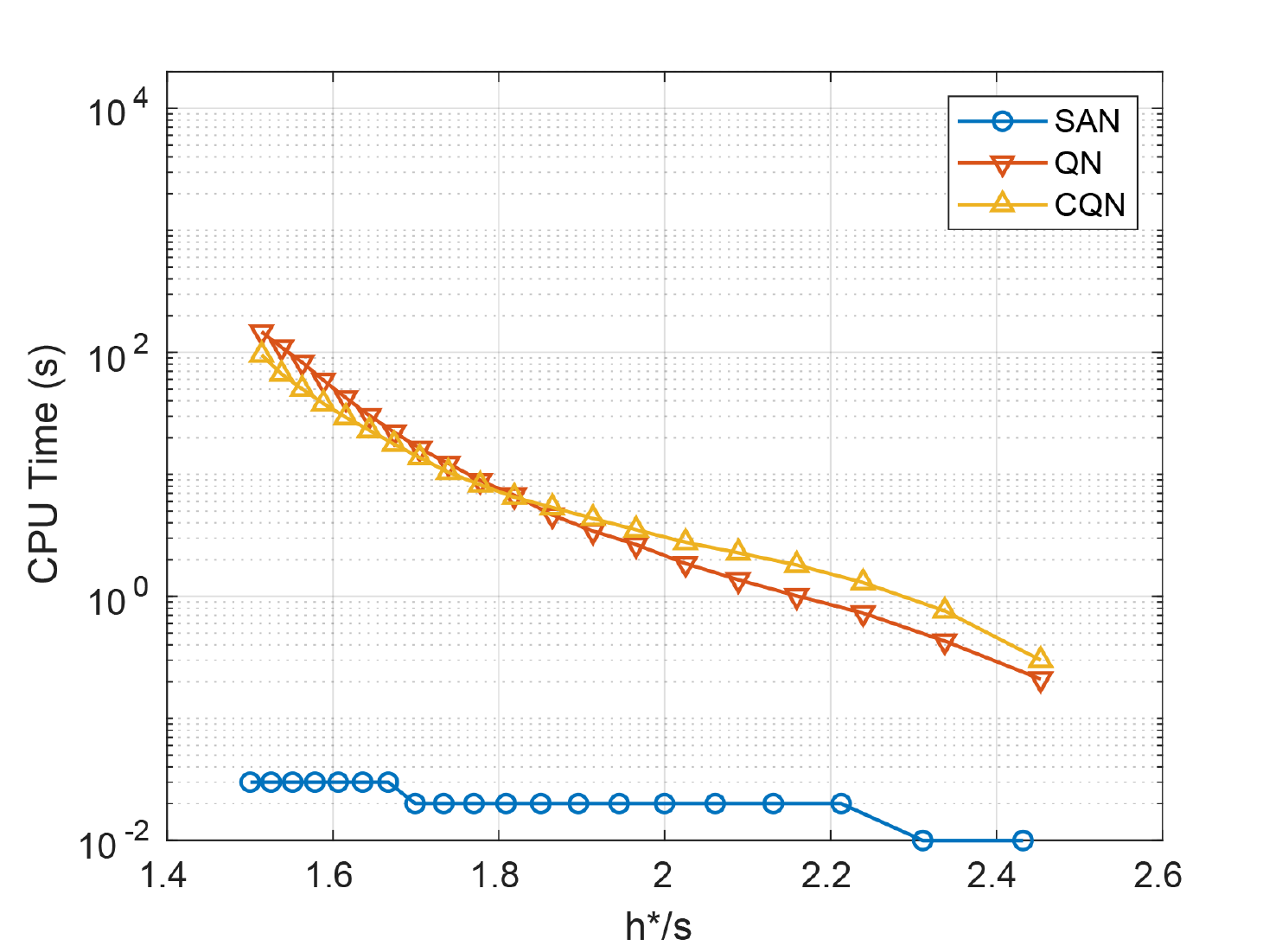}}\\
\subfloat[][FBEM-SAN\label{plot:h-CPU-SAN}]
{\includegraphics[width=0.47\textwidth]{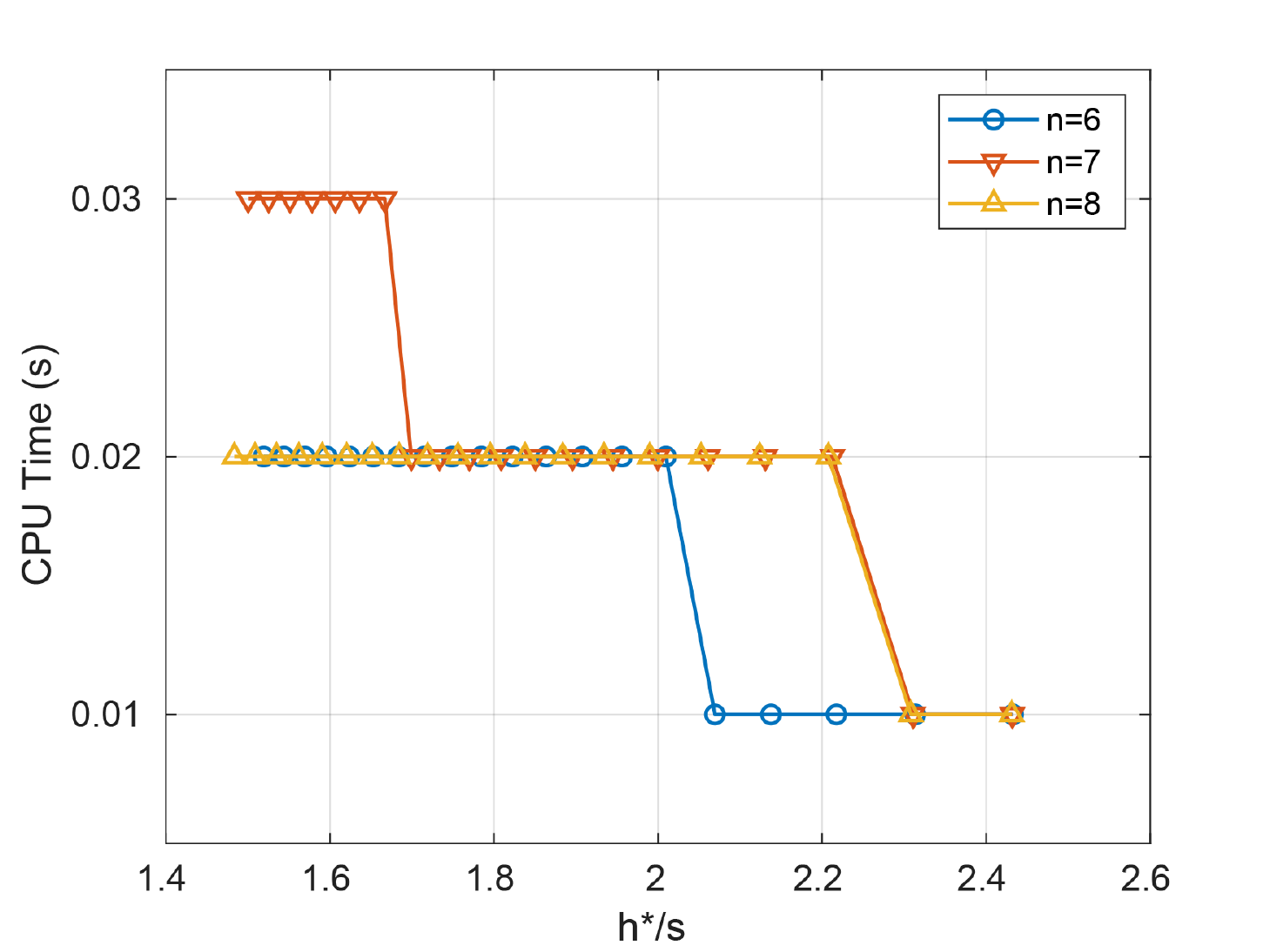}}
\subfloat[][$n=8$\label{plot:h-CPU-8}]
{\includegraphics[width=0.47\textwidth]{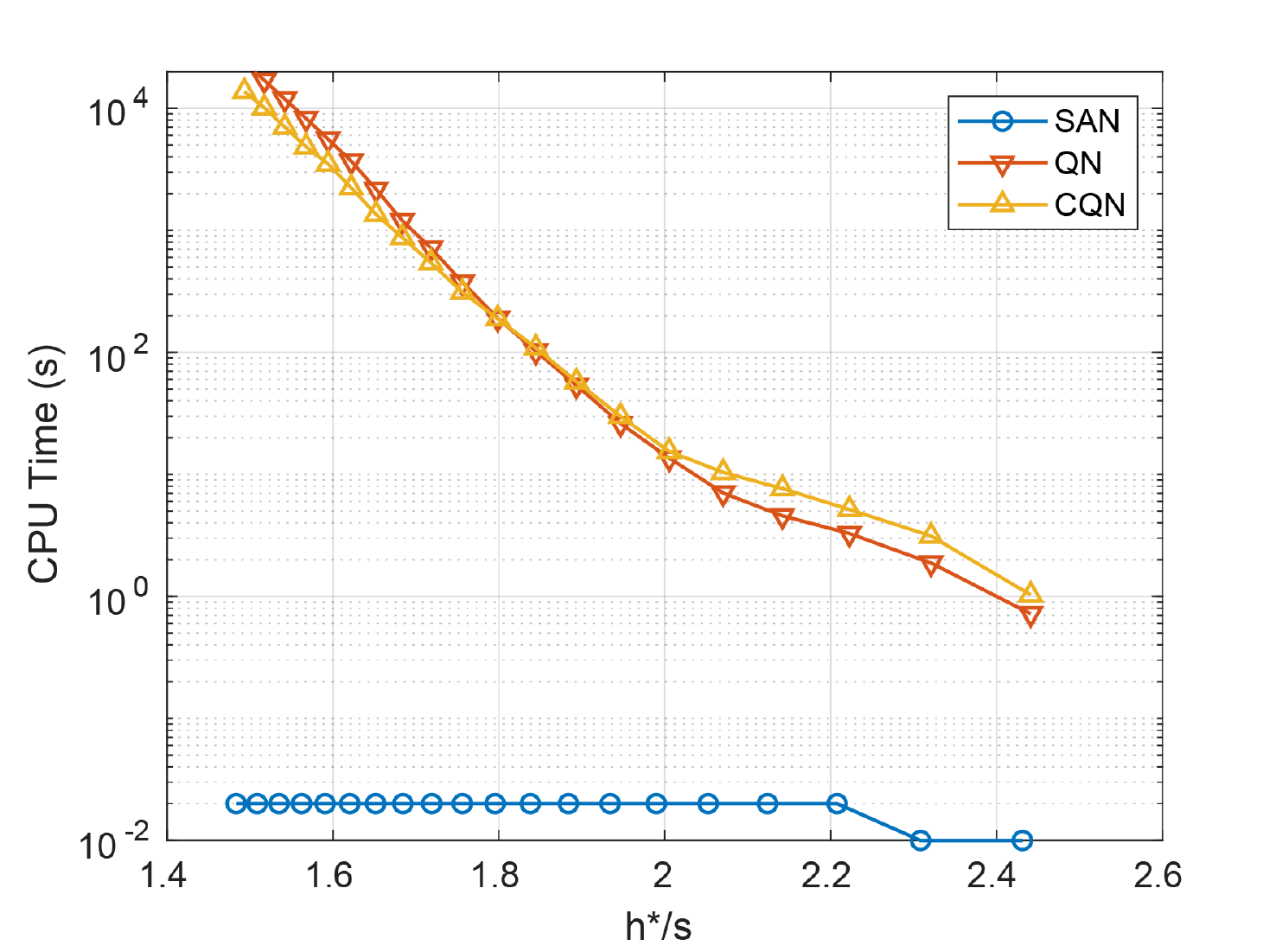}}\\
\caption{CPU time (s) vs. dimensionless normal gap predictions depending on the solution scheme and the surface resolution.}
\label{plot:h-CPU}
\end{figure*}

The CPU time required to solve the contact problem is shown in Figs.~\ref{plot:h-CPU-QN}, \ref{plot:h-CPU-CQN} and \ref{plot:h-CPU-SAN} for the FBEM-QN, FBEM-CQN and FBEM-SAN solution strategies and in Figs.~\ref{plot:h-CPU-6}, \ref{plot:h-CPU-7} and \ref{plot:h-CPU-8} for the three different resolutions.
The CPU time for the FBEM-SAN strategy includes only the time required for FEM to solve the macro-scale contact problem  without the time for the off-line execution of BEM, since this preparatory step is very case specific and depends not only on the maximum value of pressure required, but also on the accuracy requested to the fitting operation, as already shown in Fig.~\ref{fig:off_time}.

The FBEM-SAN is much faster than the other two strategies especially for intermediate and low separations, when the time required for the micro-scale BEM computations spent to predict the contact pressure and the contact stiffness in the QN and in the CQN schemes is significant. Both the integrated approaches becomes more expensive as the number of contact points increases, for the higher resolution or the decreasing separation between the surfaces. These last two strategies show almost the same CPU time, with slight differences: the QN is faster at the beginning, for high separation while the CQN allows to save time in the low separation range.



\begin{figure*}
	\centering
	\subfloat[][Dimensionless overall reaction force against the imposed displacement.]
	{\includegraphics[width=0.5\textwidth]{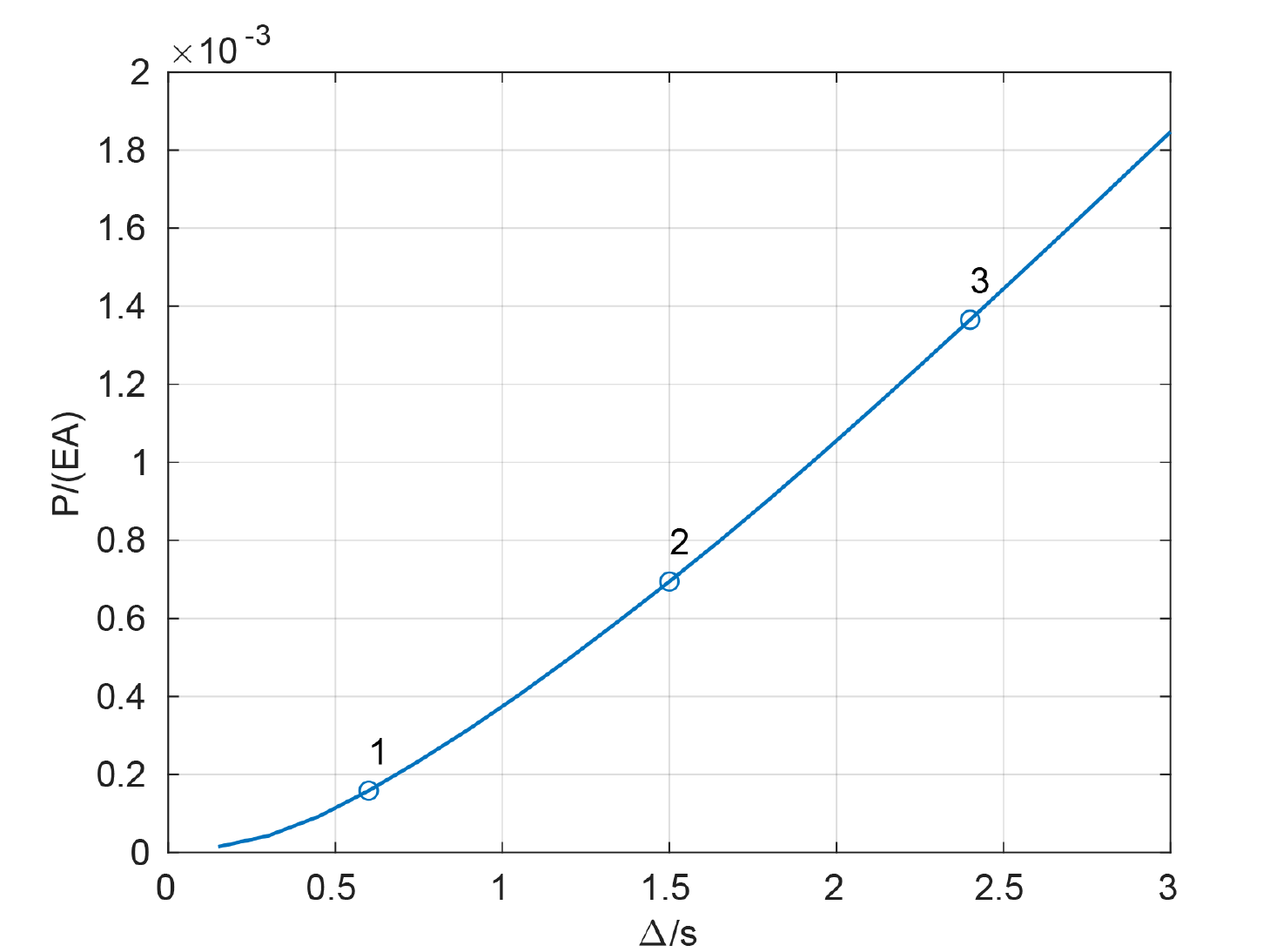}}
	\subfloat[][Point 1, $A/A_\mathrm{n}=0.38\%$.]
	{\includegraphics[width=0.5\textwidth]{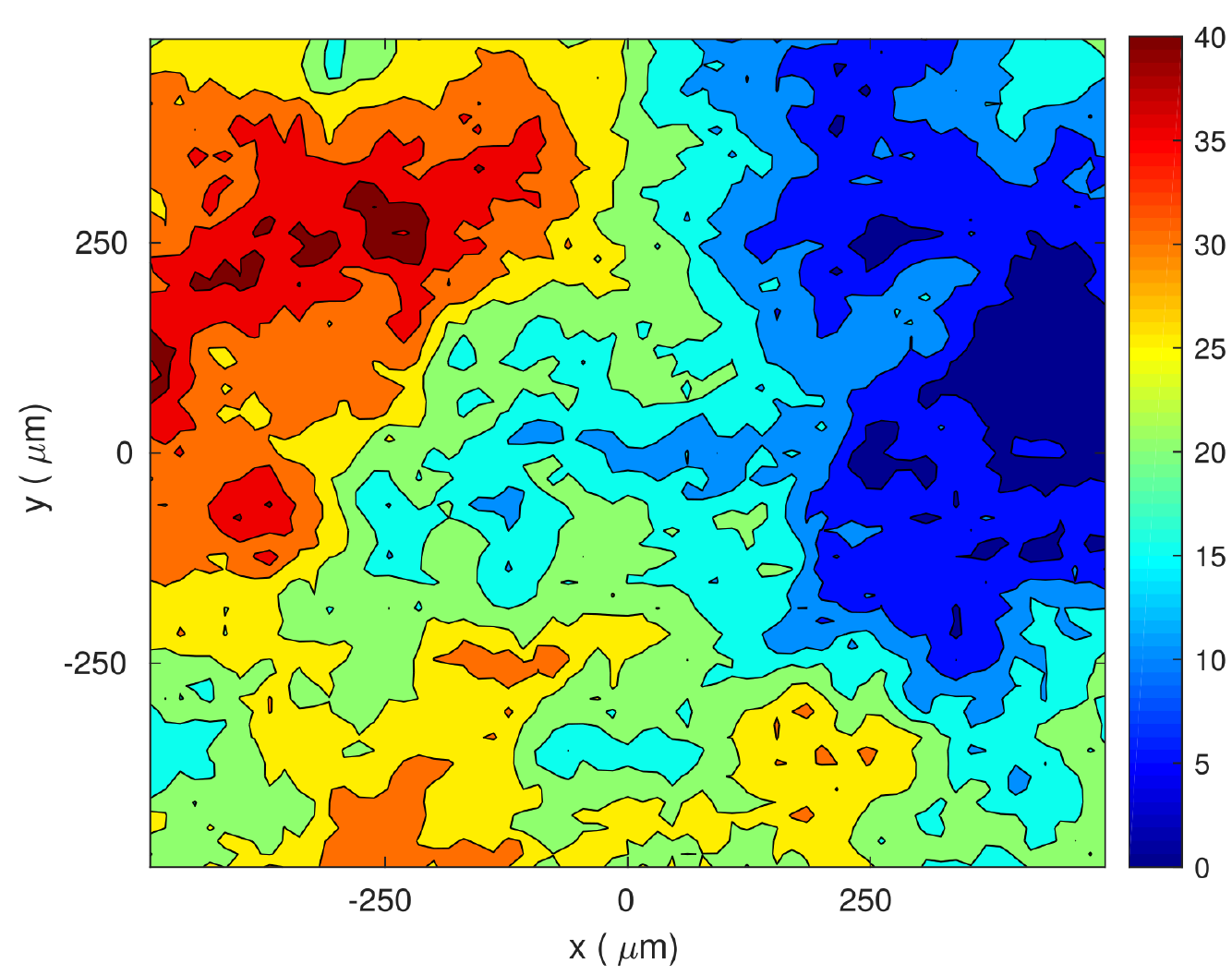}}\\
	\subfloat[][Point 2, $A/A_\mathrm{n}=1.66\%$.]
	{\includegraphics[width=0.5\textwidth]{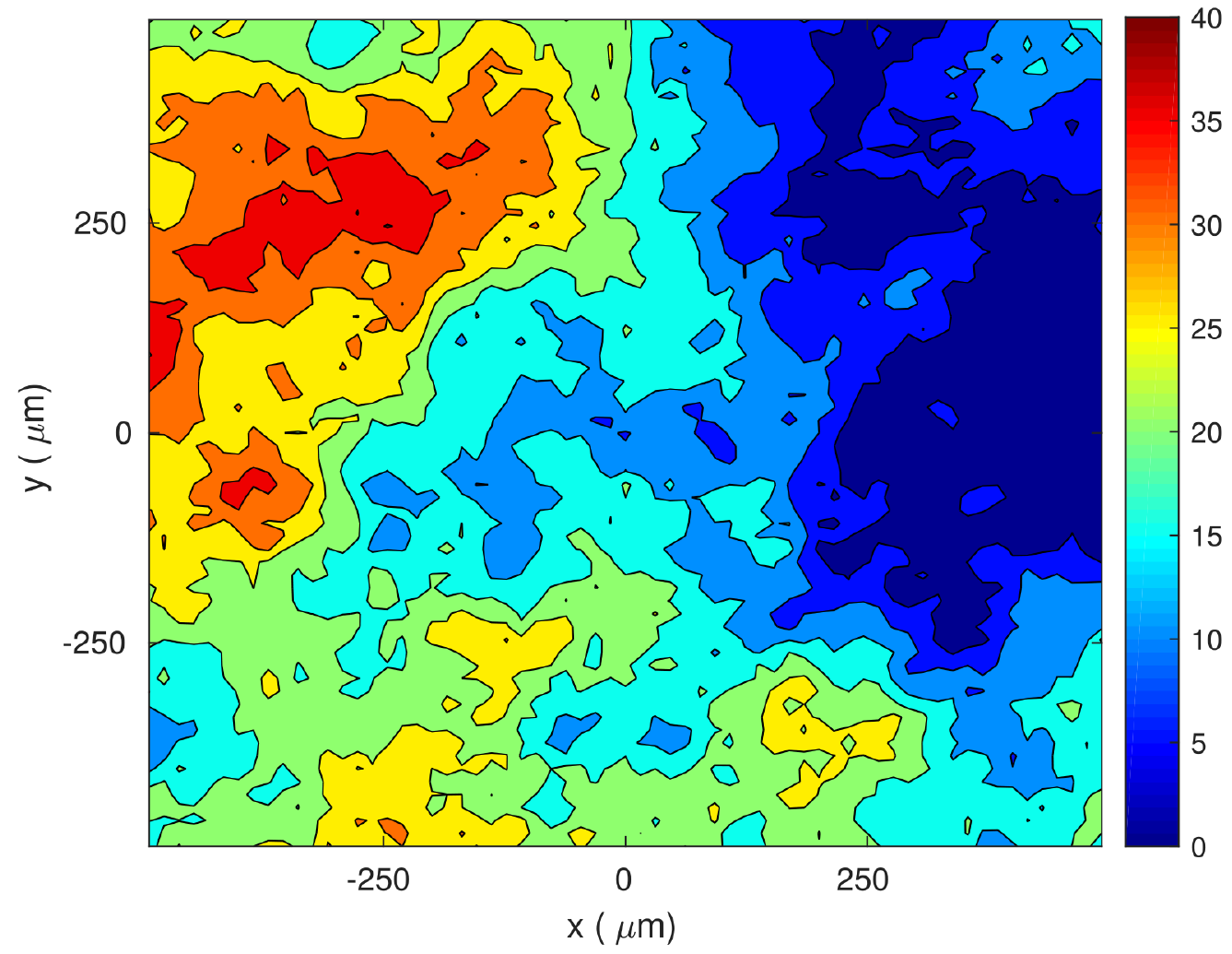}}
	\subfloat[][Point 3, $A/A_\mathrm{n}=3.12\%$.]
	{\includegraphics[width=0.5\textwidth]{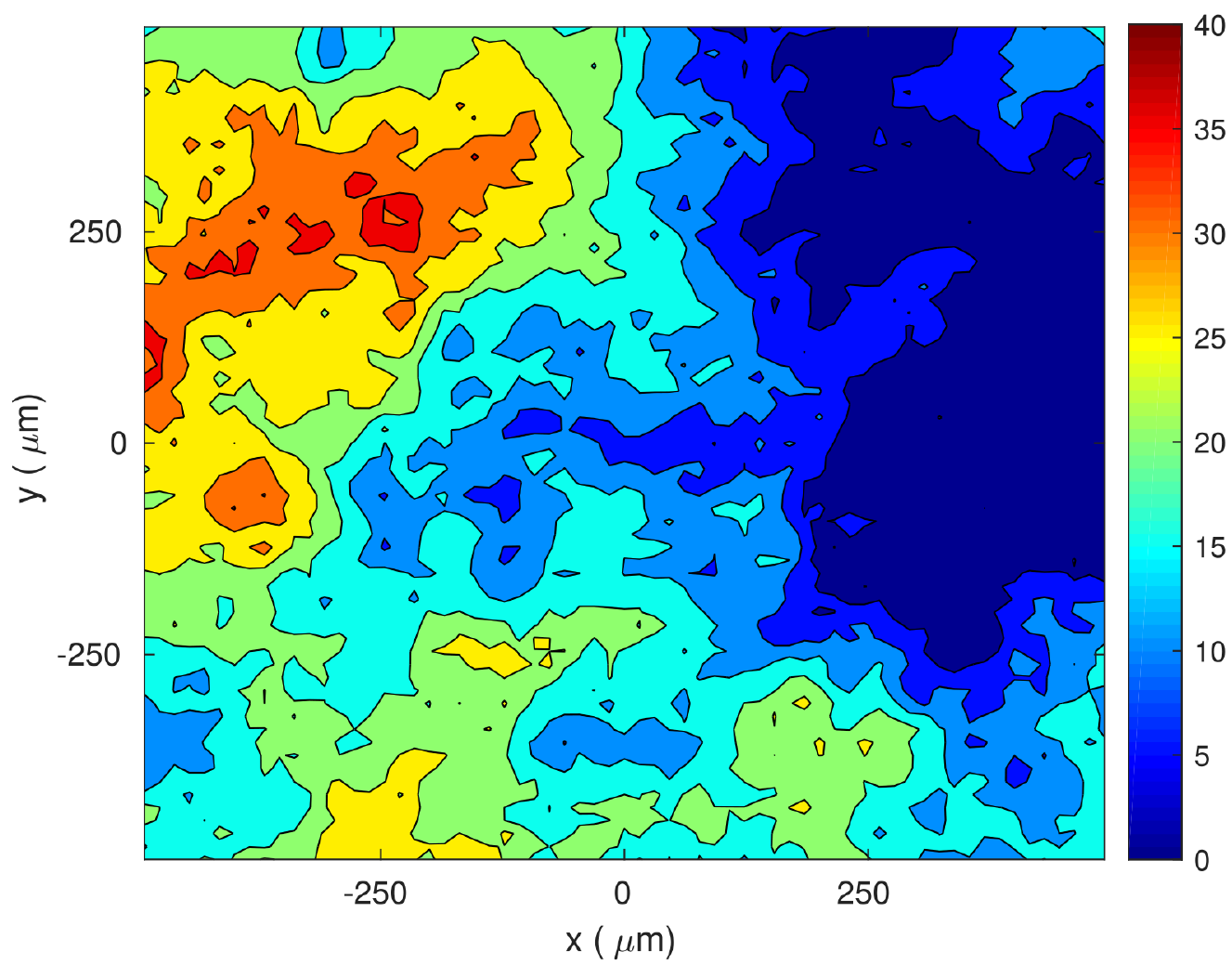}}\\
	\caption{Evolution of the free volume of the real geometry at the micro-scale for three different levels of imposed displacement, for the $n=7$, FBEM-QN case. For every one of the three contour plots the ratio between the actual contact area $A$ and the nominal one $A_\mathrm{n}$ is provided, while the dark blue islands show the contact area, the deepest valley are marked in red. }
	\label{fig:contourf}
\end{figure*}

In addition to the examined computational performance of the three approaches, it is important to notice that FBEM-QN and FBEM-CQN allow to extract local information about the micro-scale contact problem. As an example, the pressure field and the free volume evolution can be easily extracted at each time step from the model without any additional effort (see Fig.~\ref{fig:contourf}). Their values can be very useful for multi-field problems involving heat transfer or reaction-diffusion phenomena and for simulations including wear and friction where knowing the contact islands and the pressure distribution plays a key role. On the other hand, in the FBEM-SAN approach, the information about the percentage of contact area can be easily recovered using an additional interpolating function during the off-line stage in order to obtain the relation between the total contact area and the average contact pressure.

\section{Conclusion}

A multi-scale FEM-BEM contact mechanics formulation has been proposed to address contact problems involving a nominally smooth surface in the macro-scale and a microscopically rough topology in the micro-scale. The assumption of scale separation is put forward, which assumes that a statistically representative rough surface can be defined at each point along the macroscopical contact surface. Coupling of the two scales is enforced by passing the normal gap at each integration point of the interface finite element to the boundary element code to solve the microscopical normal contact problem. In return to the macro-scale, the normal contact traction and tangent stiffness matrix are provided, to be used in a Newton-Raphson algorithm and its variants.

As compared to the previous methods proposed in the literature \cite{Zava1}, the approach does not rely on a closed-form solution at the micro-scale associated to a specific microscopical contact model, which implies assumptions on roughness statistics. BEM is in fact applied to any height field without introducing any simplifying assumption on the topology. Therefore, different statistically rough surfaces can also be provided in input depending on the position along the macro-scale contact surface. In spite of this higher versatility, a drawback arises from the much higher computational cost, which is especially due to the numerical computation of the tangent normal contact stiffness, requiring an additional BEM computation for a perturbed normal gap. Possible acceleration strategies, all the way down to an off-line BEM computation and interpolation of the microscopical contact solution to derive a closed-form equation analogous to that provided by micromechanical contact theories, are explored. This last approach has some disadvantages being less accurate in the high separation regions, where the contact spots are governed by the extremes of the summit distribution, and care should be taken to perform an interpolation covering all the range of pressures expected, while, on the other hand, the process is automatic for the other two routines. Quite the opposite, the computational cost for the semi-analytic procedure is indeed lower than the other two routines and this strategy can be exploited for its speed in problems where the interpolation of the pressure-displacement solution can be easily computed off-line and used multiple times in the macro-scale model.

The fully integrated FEM-BEM routines are more suitable for dealing with highest resolution surfaces and low pressure regimes or, more generally, for problems where the interpolation of the micromechanical results adds undesired approximation to the model. Furthermore, these procedures provide more details on the micro-scale contact problem directly from the macroscopic model as the actual micro-pressure distribution and the actual contact area, useful for extending the method to multi-field problems. The semi-analytic approach, on the other hand, could be extended in order to provide the percentage of the contact area using another interpolation function, useful information in case of problems involving frictions.

As a final remark, in spite of the additional computational cost associated to concurrent FEM-BEM coupling, this approach presents the highest versatility and it can be very efficient to deal with wear phenomena affecting the micro-scale computations. In the case of wear, for instance, the height field of the rough surface stored in each integration point can be progressively updated by the prediction of a wear law. 
Similarly, in the case of surface erosion by a fluid, local surface roughness properties can be updated along with the simulation. Those aspects, as well as frictional energy dissipation, are open issues worth investigating in future research studies.

\begin{acknowledgements}
The authors would like to acknowledge funding from the MIUR-DAAD Joint Mobility Program 2017 to the project "Multi-scale modelling of friction for large scale engineering problems". The project has been granted by the Italian Ministry of Education, University and Research (MIUR) and by the Deutscher Akademischer Austausch Dienst (DAAD) through funds of the German Federal Ministry of Education and research (BMBF).
\end{acknowledgements}

\bibliography{ref}

\begin{thebibliography}{10}
\expandafter\ifx\csname url\endcsname\relax
  \def\url#1{\texttt{#1}}\fi
\expandafter\ifx\csname urlprefix\endcsname\relax\def\urlprefix{URL }\fi
\expandafter\ifx\csname href\endcsname\relax
  \def\href#1#2{#2} \def\path#1{#1}\fi

\bibitem{LR05}
B.~Luan, M.~Robbins, The breakdown of continuum models for mechanical contacts,
  Nature 435 (2005) 929--932.

\bibitem{RMF02}
J.~Raja, B.~Muralikrishnan, S.~Fu, Recent advances in separation of roughness,
  waviness and form, Precision Engineering 26 (2002) 222--235.

\bibitem{PH_SI_JMES}
M.~Paggi, D.~Hills, Special issue on \protect{EUROMECH} 575, Proceedings of the
  Institution of Mechanical Engineers, Part C: Journal of Mechanical
  Engineering Science 230~(9) (2016) 1373--1373.

\bibitem{PH_SI_JSA}
M.~Paggi, D.~Hills, Editorial of the special issue on the \protect{EUROMECH}
  colloquium 575, The Journal of Strain Analysis for Engineering Design 51~(4)
  (2016) 239--239.

\bibitem{vakis}
A.~Vakis, V.~Yastrebov, J.~Scheibert, L.~Nicola, D.~Dini, C.~Minfray,
  A.~Almqvist, M.~Paggi, S.~Lee, G.~Limbert, J.~Molinari, G.~Anciaux,
  S.~{Echeverri Restrepo}, A.~Papangelo, A.~Cammarata, P.~Nicolini,
  R.~Aghababaei, C.~Putignano, S.~Stupkiewicz, J.~Lengiewicz, G.~Costagliola,
  F.~Bosia, R.~Guarino, N.~Pugno, G.~Carbone, M.~Mueser, M.~Ciavarella,
  Modeling and simulation in tribology across scales: an overview, Tribology
  International\href {http://dx.doi.org/10.1016/j.triboint.2018.02.005}
  {\path{doi:10.1016/j.triboint.2018.02.005}}.

\bibitem{rabino}
E.~Rabinowicz, Friction and Wear of Materials, Wiley, New York, 1965.

\bibitem{KJ}
K.~Johnson, Contact Mechanics, Cambridge University Press, Cambridge, UK, 1985.

\bibitem{goriacheva}
I.~Goryacheva, Contact Mechanics in Tribology, Vol.~61, Springer Netherlands,
  Dordrecht, 1998.

\bibitem{perssonbook}
B.~Persson, Sliding Friction, Physical Principles and Applications, Springer,
  2000.

\bibitem{vlpopov}
V.~Popov, Contact Mechanics and Friction, Springer, Berlin, Heidelberg, 2010.

\bibitem{popov_hess}
V.~Popov, M.~Hess, Method of Dimensionality Reduction in Contact Mechanics and
  Friction, Springer, Berlin, Heidelberg, 2015.

\bibitem{barber_contact}
J.~Barber, Contact Mechanics, Springer International Publishing, 2018.

\bibitem{exp4}
J.~McCool, Comparison of models for the contact of rough surfaces, Wear 107
  (1986) 37--60.

\bibitem{ZBP04}
G.~Zavarise, M.~Borri-Brunetto, M.~Paggi, On the reliability of microscopical
  contact models, Wear 257 (2004) 229--245.

\bibitem{GW66}
J.~Greenwood, J.~Williamson, Contact of nominally flat surfaces, Proceedings of
  the Royal Society of London A: Mathematical, Physical and Engineering
  Sciences 295 (1966) 300--319.
\newblock \href {http://dx.doi.org/10.1098/rspa.1966.0242}
  {\path{doi:10.1098/rspa.1966.0242}}.

\bibitem{revitalized}
M.~Ciavarella, V.~Delfine, G.~Demelio, A ``re-vitalized'' greenwood and
  williamson model of elastic contact between fractal surfaces, Journal of the
  Mechanics and Physics of Solids 54 (2006) 2569--2591.
\newblock \href {http://dx.doi.org/10.1016/j.jmps.2006.05.006}
  {\path{doi:10.1016/j.jmps.2006.05.006}}.

\bibitem{GW2006}
J.~Greenwood, A simplified elliptic model of rough surface contact, Wear 261
  (2006) 191--200.

\bibitem{PC10}
M.~Paggi, M.~Ciavarella, The coefficient of proportionality $\kappa$ between
  real contact area and load, with new asperity models, Wear 268 (2010)
  1020--1029.
\newblock \href {http://dx.doi.org/10.1016/j.wear.2009.12.038}
  {\path{doi:10.1016/j.wear.2009.12.038}}.

\bibitem{CGP08}
M.~Ciavarella, J.~Greenwood, M.~Paggi, Inclusion of ``interaction'' in the
  greenwood and williamson contact theory, Wear 265 (2008) 729--734.
\newblock \href {http://dx.doi.org/10.1016/j.wear.2008.01.019}
  {\path{doi:10.1016/j.wear.2008.01.019}}.

\bibitem{MB90}
A.~Majumdar, B.~Bhushan, Role of fractal geometry in roughness characterization
  and contact mechanics of surfaces, ASME Journal of Tribology 112 (1990)
  205--216.

\bibitem{BCC}
M.~Borri-Brunetto, A.~Carpinteri, B.~Chiaia, Scaling phenomena due to fractal
  contact in concrete and rock fractures, International Journal of Fracture 95
  (1999) 221--238.
\newblock \href {http://dx.doi.org/10.1023/A:1018656403170}
  {\path{doi:10.1023/A:1018656403170}}.

\bibitem{CARPINTERI20052901}
A.~Carpinteri, M.~Paggi, Size-scale effects on the friction coefficient,
  International Journal of Solids and Structures 42 (2005) 2901--2910.

\bibitem{Persson1}
B.~Persson, O.~Albohr, U.~Tartaglino, A.~Volokitin, E.~Tosatti, On the nature
  of surface roughness with application to contact mechanics, sealing, rubber
  friction and adhesion, Journal of Physics: Condensed Matter 17 (2005) R1.

\bibitem{A81}
T.~Andersson, The boundary element method applied to two-dimensional contact
  problems with friction, in: Boundary Element Methods, Vol.~3, 1981, pp.
  239--258.

\bibitem{M94}
K.~Man, Contact mechanics using boundary elements, topics in engineering,
  Vol.~22, Southampton, Boston, 1994.

\bibitem{wriggers2}
P.~Wriggers, J.~Reinelt, Multi-scale approach for frictional contact of
  elastomers on rough rigid surfaces, Computer Methods in Applied Mechanics and
  Engineering 198 (2009) 1996--2008.
\newblock \href {http://dx.doi.org/10.1016/j.cma.2008.12.021}
  {\path{doi:10.1016/j.cma.2008.12.021}}.

\bibitem{nelias}
J.~Leroux, B.~Fulleringer, D.~Nélias, Contact analysis in presence of
  spherical inhomogeneities within a half-space, International Journal of
  Solids and Structures 47 (2010) 3034--3049.
\newblock \href {http://dx.doi.org/10.1016/j.ijsolstr.2010.07.006}
  {\path{doi:10.1016/j.ijsolstr.2010.07.006}}.

\bibitem{finite}
C.~Putignano, G.~Carbone, D.~Dini, Mechanics of rough contacts in elastic and
  viscoelastic thin layers, International Journal of Solids and Structures
  69-70 (2015) 507 -- 517.
\newblock \href {http://dx.doi.org/10.1016/j.ijsolstr.2015.04.034}
  {\path{doi:10.1016/j.ijsolstr.2015.04.034}}.

\bibitem{PEI}
L.~Pei, S.~Hyun, J.~Molinari, M.~Robbins, Finite element modeling of
  elasto-plastic contact between rough surfaces, Journal of the Mechanics and
  Physics of Solids 53 (2005) 2385--2409.
\newblock \href {http://dx.doi.org/10.1016/j.jmps.2005.06.008}
  {\path{doi:10.1016/j.jmps.2005.06.008}}.

\bibitem{HPMR}
S.~Hyun, L.~Pei, J.-F. Molinari, M.~Robbins, Finite-element analysis of contact
  between elastic self-affine surfaces, Phys. Rev. E 70 (2004) 026117.
\newblock \href {http://dx.doi.org/10.1103/PhysRevE.70.026117}
  {\path{doi:10.1103/PhysRevE.70.026117}}.

\bibitem{MPJR}
M.~Paggi, J.~Reinoso, A variational approach with embedded roughness for
  adhesive contact problems, Mechanics of Advanced Materials and
  Structures\href {http://dx.doi.org/10.1080/15376494.2018.1525454}
  {\path{doi:10.1080/15376494.2018.1525454}}.

\bibitem{Zava1}
G.~Zavarise, P.~Wriggers, E.~Stein, B.~Schrefler, Real contact mechanisms and
  finite element formulation--a coupled thermomechanical approach,
  International Journal for Numerical Methods in Engineering 35 (1992)
  767--785.

\bibitem{Sapora2014}
A.~Sapora, M.~Paggi, A coupled cohesive zone model for transient analysis of
  thermoelastic interface debonding, Computational Mechanics 53 (2014)
  845--857.
\newblock \href {http://dx.doi.org/10.1007/s00466-013-0934-8}
  {\path{doi:10.1007/s00466-013-0934-8}}.

\bibitem{LENARDA2018374}
P.~Lenarda, A.~Gizzi, M.~Paggi, A modeling framework for electro-mechanical
  interaction between excitable deformable cells, European Journal of Mechanics
  - A/Solids 72 (2018) 374--392.

\bibitem{popp2018contact}
A.~Popp, P.~Wriggers, Contact Modeling for Solids and Particles, CISM
  International Centre for Mechanical Sciences, Springer International
  Publishing, 2018.

\bibitem{Seitz2018}
A.~Seitz, W.~A. Wall, A.~Popp, Nitsche's method for finite deformation
  thermomechanical contact problems, Computational Mechanics\href
  {http://dx.doi.org/10.1007/s00466-018-1638-x}
  {\path{doi:10.1007/s00466-018-1638-x}}.

\bibitem{Seitz2018bis}
A.~Seitz, W.~A. Wall, A.~Popp, A computational approach for
  thermo-elasto-plastic frictional contact based on a monolithic formulation
  using non-smooth nonlinear complementarity functions, Advanced Modeling and
  Simulation in Engineering Sciences 5~(1) (2018) 5.
\newblock \href {http://dx.doi.org/10.1186/s40323-018-0098-3}
  {\path{doi:10.1186/s40323-018-0098-3}}.

\bibitem{HIERMEIER2018532}
M.~Hiermeier, W.~A. Wall, A.~Popp, A truly variationally consistent and
  symmetric mortar-based contact formulation for finite deformation solid
  mechanics, Computer Methods in Applied Mechanics and Engineering 342 (2018)
  532 -- 560.
\newblock \href {http://dx.doi.org/https://doi.org/10.1016/j.cma.2018.07.020}
  {\path{doi:https://doi.org/10.1016/j.cma.2018.07.020}}.

\bibitem{Farah2017}
P.~Farah, W.~A. Wall, A.~Popp, An implicit finite wear contact formulation
  based on dual mortar methods, International Journal for Numerical Methods in
  Engineering 111~(4) (2017) 325--353.
\newblock \href {http://dx.doi.org/10.1002/nme.5464}
  {\path{doi:10.1002/nme.5464}}.

\bibitem{barber03}
J.~Barber, Bounds on the electrical resistance between contacting elastic rough
  bodies, Proceedings of the Royal Society of London A 459 (2003) 53--66.
\newblock \href {http://dx.doi.org/10.1098/rspa.2002.1038}
  {\path{doi:10.1098/rspa.2002.1038}}.

\bibitem{barber_elasticity}
J.~Barber, Elasticity, Springer, Dordrecht, 3rd Edition, 2010.

\bibitem{BP15}
A.~Bemporad, M.~Paggi, Optimization algorithms for the solution of the
  frictionless normal contact between rough surfaces, International Journal of
  Solids and Structures 69--70 (2015) 94--105.

\bibitem{PB11}
M.~Paggi, J.~Barber, Contact conductance of rough surfaces composed of modified
  rmd patches, International Journal of Heat and Mass Transfer 54 (2011)
  4664--4672.
\newblock \href {http://dx.doi.org/10.1016/j.ijheatmasstransfer.2011.06.011}
  {\path{doi:10.1016/j.ijheatmasstransfer.2011.06.011}}.

\bibitem{CONWAY1968489}
H.~Conway, K.~Farnham, The relationship between load and penetration for a
  rigid, flat-ended punch of arbitrary cross section, International Journal of
  Engineering Science 6~(9) (1968) 489 -- 496.
\newblock \href {http://dx.doi.org/10.1016/0020-7225(68)90001-3}
  {\path{doi:10.1016/0020-7225(68)90001-3}}.

\bibitem{nakamura1993}
M.~Nakamura, Constriction resistance of conducting spots in an electric contact
  surface, WIT Transactions on Modelling and Simulation 3 (1993) 10.
\newblock \href {http://dx.doi.org/10.2495/BT930121}
  {\path{doi:10.2495/BT930121}}.

\bibitem{peitgen1988science}
H.~Peitgen, D.~Saupe, M.~Barnsley, The Science of fractal images,
  Springer-Verlag, 1988.

\end{thebibliography}
\bibliographystyle{elsarticle-num}

\end{document}